\documentclass[preprint,12pt]{elsarticle}

\usepackage{lineno,hyperref}
\usepackage{bm}
\usepackage{amsfonts}
\usepackage{amssymb}
\usepackage{amsmath}
\usepackage{isomath}

\usepackage{adjustbox}

\usepackage{scalerel,stackengine}
\stackMath
\newcommand\reallywidehat[1]{%
\savestack{\tmpbox}{\stretchto{%
  \scaleto{%
    \scalerel*[\widthof{\ensuremath{#1}}]{\kern-.6pt\bigwedge\kern-.6pt}%
    {\rule[-\textheight/2]{1ex}{\textheight}}
  }{\textheight}%
}{0.5ex}}%
\stackon[1pt]{#1}{\tmpbox}%
}

\usepackage{stmaryrd} 
\usepackage[usenames]{color}
\usepackage{epsfig}
\usepackage{graphicx}
\usepackage{psfrag}

\graphicspath{{Figures/}}
\usepackage{float}
\floatstyle{boxed}
\newfloat{algorithm}{htb}{loa}
\floatname{algorithm}{Algorithm}
\usepackage{algorithm,algpseudocode}
\usepackage{caption}
\usepackage{subcaption}
\usepackage{color}
\usepackage{xcolor}
\usepackage{multirow}
\usepackage{natbib}
\modulolinenumbers[5]

\usepackage[left = 2cm, right=2.5cm, top=2cm, bottom =2cm]{geometry}

\begin{document}
\begin{frontmatter}

\title{ Finite Operator Learning: Bridging Neural Operators and Numerical Methods for Efficient Parametric Solution and Optimization of PDEs }








\author{ Shahed Rezaei$^{1,*}$, Reza Najian Asl$^{2,*}$, Kianoosh Taghikhani$^{1}$, Ahmad Moeineddin$^{3}$,\\  Michael Kaliske$^{3}$, Markus Apel$^{1}$ }
\address{$^1$ACCESS e.V., Intzestr. 5, 52072 Aachen, Germany}
\address{$^2$ Chair of Structural Analysis, Technical University of Munich, Arcisstr. 21, 80333 München, Germany}
\address{$^3$Institute for Structural Analysis, Technische Universität Dresden, \\Georg-Schumann-Str. 7, 01187 Dresden, Germany}
\address{$^*$ corresponding authors: s.rezaei@access-technology.de, reza.najian-asl@tum.de\footnote{The first two authors contributed equally to this work.}}

\begin{abstract}
  We introduce a method that combines neural operators, physics-informed machine learning, and standard numerical methods for solving PDEs. The proposed approach unifies aforementioned methods and we can parametrically solve partial differential equations in a data-free manner and provide accurate sensitivities. These capabilities enable gradient-based optimization without the typical sensitivity analysis costs, unlike adjoint methods that scale directly with the number of response functions. 
  \textcolor{black}{Our Finite Operator Learning (FOL) approach originally employs feed-forward neural networks to directly map the discrete design space to the discrete solution space, and can alternatively be combined with existing physics-informed neural operator techniques to recover continuous solution fields, while avoiding the need for automatic differentiation when formulating the loss terms.} 
  The discretized governing equations, as well as the design and solution spaces, can be derived from any well-established numerical techniques. In this work, we employ the Finite Element Method (FEM) to approximate fields and their spatial derivatives. \textcolor{black}{Thanks to the finite-element formulation, Dirichlet boundary conditions are satisfied by construction, and Neumann boundary conditions are naturally included in the FE residual through the weak form.} Subsequently, we conduct Sobolev training to minimize a multi-objective loss function, which includes the discretized weak form of the energy functional, boundary conditions violations, and the stationarity of the residuals with respect to the design variables.
  Our study focuses on the heat equation and the mechanical equilibrium problem. \textcolor{black}{First, we primarily address the property distribution in heterogeneous materials, where Fourier-based parametrization is employed to significantly reduce the number of design variables.} 
  Second, we explore changes in the source term in such PDEs. Third, we investigate the solution under different boundary conditions. In the context of gradient-based optimization, we examine the tuning of the microstructure's heat transfer characteristics.  
  Our technique also simplifies to an efficient matrix-free PDE solver that can compete with standard available solvers. 
  This is demonstrated by solving a nonlinear thermal and mechanical PDE on a complex 3D geometry.
\end{abstract} 

\end{frontmatter}


\section{Introduction} 

Simulations of physical phenomena are recognized as the third pillar of science \cite{Weinzierl2021}. 
While we focus on developing new models to accurately describe reality, a pressing and arguably more significant challenge lies in the efficient (i.e., rapid) and precise (i.e., high-accuracy) evaluation of these models for optimization purposes. This challenge is particularly noteworthy in the context of developing digital twins or shadows \cite{BERGS202181}. 

\subsection{Current trends in computational methods}
Over the past decades, numerous numerical techniques have been developed for various applications, gaining significant attention for their remarkable predictive capabilities in solving model equations. Among them, one can name the finite difference method (FDM), finite volume method (FVM), and FEM \cite{Faroughi22, Liu2022}. Despite their great predictive abilities, these methods remain time-consuming in recent applications including multiphysics and coupled sets of nonlinear equations \cite{Ruan2022, REZAEI2023103758}. Another issue is that these methods cannot solve the equations in a parametric way. 
In other words, by any change in the input parameters, one has to repeat the calculations.

Machine learning (ML) techniques present a promising solution for overcoming the previously mentioned challenges, offering a considerable speed advantage over classical solvers once effectively trained \cite{Montans2023}. The efficiency of ML algorithms holds particular relevance in multi-scaling analysis (refer to \cite{fernandez2020, Peng2021} and references therein). While training time is a factor, it represents a one-time investment compared to the substantial speedup achieved \cite{mianroodi2022lossless}. 

To ensure reliability, adequate training data encompassing relevant scenarios is essential.
Moreover, integrating well-established physical constraints enhances prediction capabilities \cite{REZAEI2022PINN}. 
Among various options, a well-established method to perform the latter task is the physics-informed neural networks (PINNs) as introduced by \citet{RAISSI2019}. 
Additionally, in scenarios where the underlying physics of the problem is completely known, it becomes possible to train the neural network without any initial data \cite{SIRIGNANO20181339, RAISSI2019, TORABIRAD2020109687, REZAEI2022PINN}. 
However, PINN based models are still under developments and one of their problems is convergence for low-frequency components \cite{Markidis2021}. 
In response to these shortcomings, extensions to the PINN framework have been proposed. Some advancements are rooted in the domain decomposition method \cite{JAGTAP2020113028, jagtap2020extended, Moseley2023}, while others focus on novel architectures and adaptive weights to augment accuracy \cite{HAGHIGHAT2021, WANG21, McClenny22}. Recent investigations have proposed the reduction of differential orders to construct loss functions and utilizing the energy form of equations \cite{REZAEI2022PINN, E2018, SAMANIEGO2020112790, FUHG2022110839}.


\color{black}
\subsection{Idea behind parametric learning and neural operators}
Despite the above progress, one main issue with solving partial differential equations using standard PINNs is that the solutions are still limited to specific boundary value problems. Furthermore, in many engineering applications, the training time of the PINN for the forward problem still cannot compete with those of numerical methods such as the FEM. 

One strategy is leveraging Operator Learning (OL), which entails mapping two function spaces to each other. Some well-established methods for operator learning include but are not limited to DeepOnet \cite{Lu2021}, Fourier Neural Operator (FNO) \cite{li2021fourier}, Graph Neural Operators \cite{li2020neuraloperatorgraphkernel} and Wavelet Neural Operator \cite{TRIPURA2023115783}. 
Despite their pioneering contributions to the field, each of these methods has its pros and cons. The standard structure of DeepONet, for instance, is discretization specific \cite{ZHONG2024117274} and must be adapted depending on the nature of the solution field \cite{HAGHIGHAT2024116681}. On the hand, FNO shows great interpolation power for different grid densities, but it performs best with rectangular domain shapes and periodic boundary conditions. 
For an in-depth exploration of this topic, refer to \cite{Faroughi22, hildebrand2023comparison}.


\textcolor{black}{The concept of PINN can also be integrated with operator learning. In \cite{Wang_Paris2021}, a physics-informed DeepONet framework was utilized to solve engineering problems. 
The idea of FNO is similarly enhanced by incorporating physical constraints, which improves prediction accuracy, as discussed in \cite{li2023physicsinformed}. }
\citet{KORIC2023123809} investigated data-driven and physics-based DeepONets for solving heat conduction equations incorporating a parametric source term.
\citet{ZHU201956} utilized a convolutional encoder-decoder neural network to train physics-constrained deep learning models without labeled data.
\citet{GAO2021110079} introduce a novel physics-constrained CNN learning architecture based on elliptic coordinate mapping.
\citet{ZHANG2021100220} introduce a physics-informed neural network tailored for analyzing digital materials. 
\color{black}
The reviewed works illustrates the potential of physics-informed operator learning as an emerging computational tool. Nevertheless several challenges remain. First, the design parametric space is vast. Reducing the number of free input parameters is crucial, as dealing with an almost infinite parameter space hinders the practical application of operator learning. Second, training based on physical constraints rather than data introduces additional computational costs due to reliance on automatic differentiation algorithms. Finally, their application to extremely heterogeneous domains in 2D and 3D settings, where sharp solutions are expected, has not been sufficiently investigated

Interestingly, researchers have begun addressing some of these challenges. To tackle input space reduction, \citet{kontolati2023learning} propose mapping high-dimensional datasets to a low-dimensional latent space of salient features using suitable autoencoders and pre-trained decoders. Similarly, \citet{ZHANG2023116214} introduce encoder-decoder architectures to learn solutions to differential equations in weak form, capable of generalizing across domains, boundary conditions, and coefficients.
\color{black}
To increase training efficiency, another trend in the literature focuses on approximating derivatives more efficiently using classical numerical methods. This methodology transforms the loss function from a differential equation to an algebraic one. However, it is important to note that this approach may introduce discretization errors, influencing the network's outcomes.
\citet{Fuhg2023} introduced the deep convolutional Ritz method, a parametric PDE solver relying on minimizing energy functionals. 
See also \cite{REN2022114399, ZHAO2023105516, ZHANG2023111919, Hansen_2024}. 

\subsection{Inverse design and optimization through ML}
\label{sec:intro_ML_opt}
Besides the prediction purposes of numerical simulations, a major goal and subsequent question is how to design the structure or material based on specific objectives (i.e., desired constraints).
Strategically designing material microstructures can optimize properties such as strength, and conductivity, resulting in lightweight, robust, and thermally stable materials. However, this optimization process is complex due to the vast design space, high simulation costs, and the expense of sensitivity analysis. 
Researchers have explored the applicability of neural networks to numerical design analysis and optimization in various ways. For example, \cite{Najmon2024,Lee2024,ZHENG2021113894,LIN2018103} employed NNs in a fully data-driven manner, while \cite{He2023,chandrasekhar2021tounn,ZHANG2021114083,zehnder2021ntopo} utilized them for parameterization and as PDE solvers. For a critical review of the research on the application of artificial intelligence in topology optimization, the reader is encouraged to see \cite{Woldseth2022}.

In PDE-constrained optimization, there are primarily two approaches used to solve the optimization problem: the Simultaneous Analysis and Design (SAND) approach and the Nested Analysis and Design (NAND) approach \citep{haftka1985simultaneous}. 
In the SAND approach, state and design variables are updated simultaneously, offering a richer exploration space but with high computational demands and non-physically feasible intermediate solutions. In contrast, the NAND technique fully satisfies state equations within a nested loop, producing valid intermediate designs and is more suitable for large-scale problems. Consequently, NAND is preferred for PDE constrained gradient-based optimization. However, deep neural networks favor the SAND approach due to the ease of integrating optimization responses into the loss function, along with PDE and boundary condition losses.

Besides the standard machine learning framework, research has also begun on combining ideas from PINNs into optimization problems. \citet{Zhang2024} present a weak-form PINN for the topology optimization of heterogeneous materials. 
\citet{mowlavi2023topology} combined PINN with topology optimization to detect hidden geometries in the noninvasive imaging processes. Their approach allows the network to predict physical fields such as displacements, stresses, and the density field. 
\citet{Tillmann2023} incorporated the free-form deformation parametrization into PINNs to enable rapid design space exploration in fluid shape optimization. 
\citet{JEONG2023116401} provided a framework for structural topology optimization using two PINNs. First, the Deep Energy Method (DEM) is trained to approximate the deformation of the design domain. Then, the sensitivity-analysis PINN (S-PINN) is used to differentiate the objective function with respect to design variables using automatic differentiation. 
\citet{He2023} explored a specific class of topology optimization using the deep energy method for predicting the displacement field. In this case, structural compliance is minimized, which is self-adjoint. Consequently, sensitivities are readily available. 

\color{black}
In summary, neural-network-based optimal designs can perform on par with fully FEM-based optimization. However, many studies report higher computational costs for physics-informed DL approaches. This is largely due to the frequent adoption of the SAND formulation, where the optimization variables are embedded directly into the loss function of the state problem, making the implementation straightforward but often computationally expensive. For PDE-constrained gradient-based optimization, the NAND strategy is typically preferred, as it reduces computational effort and produces valid intermediate solutions. 
We note, however, that this efficiency advantage may depend on the physics being modeled; for certain non-self-adjoint or coupled problems, such as solid–fluid interaction, recent NN-based SAND formulations have demonstrated competitive performance with FEM-based NAND approaches \cite{YOUSEFPOUR2025117698}.

\color{black}

\subsection{Current contributions}
Our proposed approach, coined as \underline{F}inite \underline{O}perator \underline{L}earning (FOL), parametrically solves discretized Partial Differential Equations (PDEs) and provides accurate solution-to-design sensitivity. 
In this framework, \textit{finite} indicates that the discrete solution field is hardwired to the network architecture and constrained through the physical loss function formulated using classical numerical methods for PDEs. 
\textcolor{black}{In this work, we utilized FEM-based approach, working with the discretized weak form of the problem. By leveraging the interpolation capabilities of shape functions, we avoid relying on automatic differentiation. 
To address input reduction, we propose Fourier based parameterization, significantly reducing the computational complexity of the training process.}

\begin{figure}[H]
  \centering
  \includegraphics[width=0.99\linewidth]{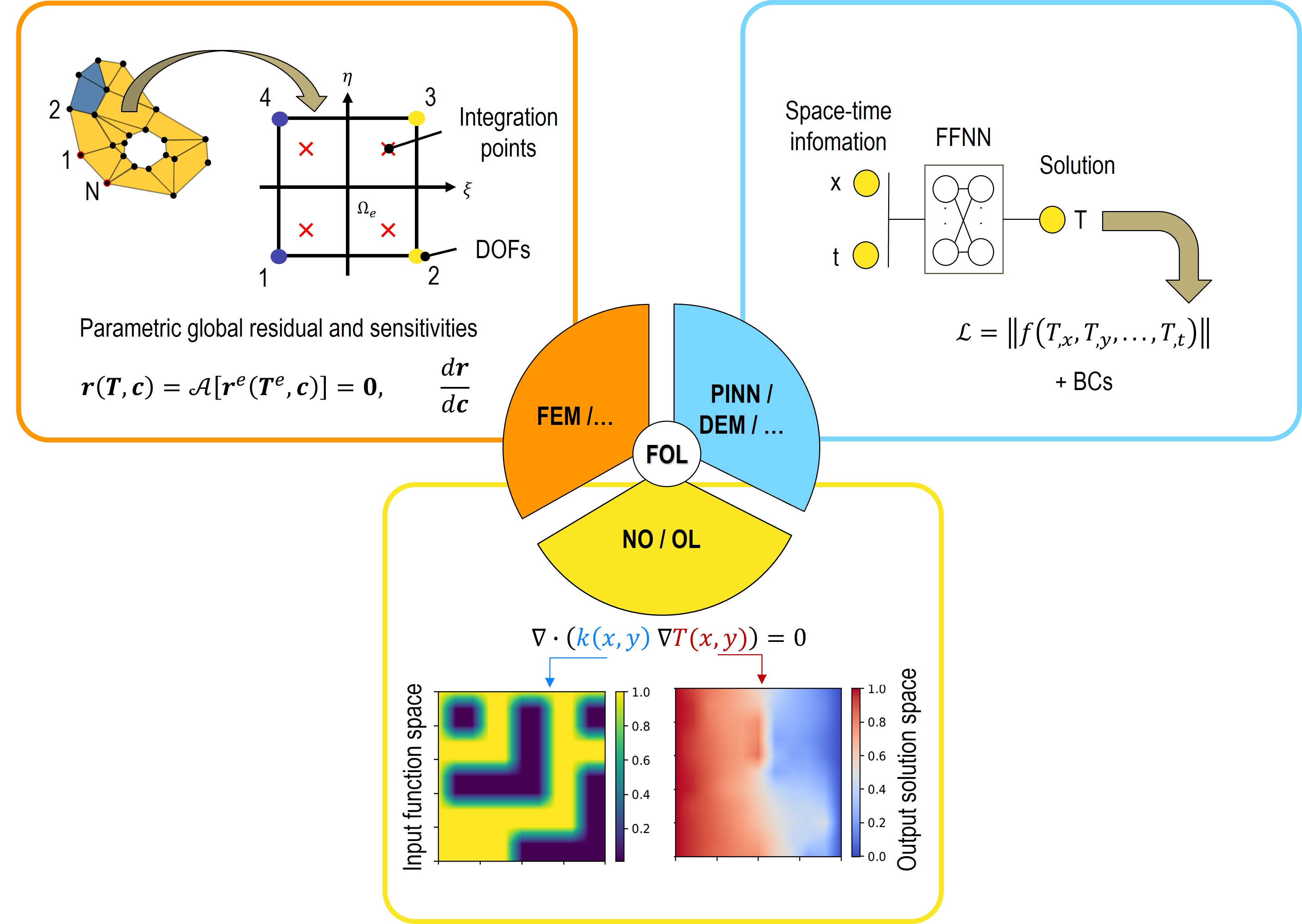}
  \caption{Overall pipeline of the proposed Finite Operator Learning (FOL) framework.}
\label{fig:intro}
\end{figure}

Mathematically, FOL maps any discrete design or parameter space (input layer) to any discrete solution space (output layer) while satisfying physical constraints, specifically minimum energy, zero residuals, and the stationarity of these residuals w.r.t the design variables. 
\textcolor{black}{The latter results in the so-called Sobolev training \cite{VLASSIS2021113695,son2021sobolev,avrutskiy2020enhancing,czarnecki2017sobolev}. \textcolor{black}{To illustrate this potential, we demonstrate the applicability of FOL to the analysis and optimization of steady-state heat equations by using the Fourier parametrization technique \cite{doosti2021topology,lee4711477topology}.}
See also Fig.~\ref{fig:intro} for the overal idea of the proposed FOL approach.}

\textcolor{black}{At the end, it is important to highlight the distinction of FOL from related strategies in the literature, namely Hierarchical deep-learning neural networks (HiDeNN) \cite{SAHA2021113452, ZhangChengLi2021} and gradient PINN (gPINN) \cite{YU2022114823}. HiDeNN constructs hierarchical networks with weights and biases dependent on nodal positions, enabling r-adaptivity and highly accurate local interpolation, while gPINN adds gradient-based losses with respect to spatial coordinates to improve training. 
In contrast, FOL targets operator learning in a parametric space, incorporating losses based on derivatives of the solution field with respect to design or material parameters, and providing scalable, matrix-free parametric predictions. These differences highlight that FOL is complementary rather than competing with these approaches, potentially enabling next-generation parametric solvers with both high accuracy and efficiency.}

The rest of this paper is structured as follows. Section \ref{seq:problem_definition} presents the design analysis and optimization problem, including the heat transfer governing equations, Fourier parametrization, and gradient-based optimization. In Section \ref{sec:FOL}, we present mathematically FOL as a physics-based operator learning for primal and sensitivity analysis. Sections  \ref{sec:res_part1} and \ref{sec:opt_res} critically examine the performance of the proposed framework compared to the reference method, FEM, in terms of the accuracy of predicted primals and gradients, optimality, and computational cost. 
Finally, we conclude with our achievements and discuss potential future developments.

\section{Problem definition and methodology}
\label{seq:problem_definition}
In this work, the design problem is formulated using the NAND approach (see Section \ref{sec:intro_ML_opt}) which involves three key steps: analysis of the underlying governing equations, design parameterization, and design variation, typically aimed at optimization. 
Each of these steps will be thoroughly examined in the following subsections. Mathematically, the goal is to optimize a function $J$ that depends on physical design variables, denoted by $k$, both directly and indirectly through state variables, denoted by $T$. 
These state variables are governed by a set of partial differential equations represented by $\mathcal{R}$. 
We initially formulate the problem in its continuous form and subsequently apply the finite element technique to solve it numerically. 
\color{black}
The described design and analysis problem can be mathematically formulated in its continuous form as
\begin{equation}
  \label{eq:math_problem_def}
  \begin{aligned}
  & \underset{c \in \mathbb{R}}{\text{min}}
  & & J(k,T), \\
  & \text{subject to} 
  & & \mathcal{A} (k,c) = 0, \;\;\;\; {k} \in \mathbb{R},
  \\
  & & & \mathcal{R}(k,T) = {0},  \;\;\; \text{in} \;\; \Omega, 
  \\
  & & & h_{j} = 0, \; \, \; j = 1, \ldots, m_{h}, \\
  & & & g_{j} \leq 0, \; \; \, j = 1, \ldots, m_{g}.
  \end{aligned}
\end{equation}

\color{black}

In the stated formulation, parametrization is addressed by the operator $\mathcal{A}$. This operator functionally maps the true control variables, represented as $c$, to the physical design variables $k$. We also note that if no specific parametrization is involved, the parametrization operator acts as the identity, resulting in $c = k$.   
\textcolor{black}{Additionally, $m_{h}$ equality constraints $h_{j}$ and $m_{g}$ inequality constraints $g_{j}$ are included for completeness.}
\textcolor{black}{In what follows, we will describe each component of Eq.~\ref{eq:math_problem_def} using specific examples from engineering problems. }

\subsection{Governing equations}
\label{sec:governing_equation}
\color{black}
In the current work, we focus on two main types of PDEs commonly encountered in engineering applications from a computational mechanics perspective: thermal diffusion and mechanical equilibrium. 
It is worth noting that the same derivation principles can be applied to analogous fields such as electrostatics or magnetostatics. 
Additionally, this approach extends to areas like chemical diffusion \cite{REZAEI2021104612} (under certain assumptions) and Darcy flow. 
See also the Poisson problem. 

Without losing generality, we write the main derivations and formualtion to explore the problem of steady-state diffusion in a heterogeneous solid, and we refer readers to \cite{REZAEI2022PINN} for a summary of the equations for mechanical equilibrium. 
Moreover, when it comes to the optimization problem, we intend to optimize the conductivity distribution for a given desired microstructure property.

The governing equation for this problem is based on the heat balance which reads
\begin{align}
\label{StrongfromThermal}
\mathcal{R}\left(k,T\right) = \text{div}(-k\,\nabla T) + Q &= 0~~~~~~ \text{in}~ \Omega, \\
\label{BCsthermalD}
T &= \bar{T}~~~~~\text{on}~ \Gamma_D, \\
\label{BCsthermalN}
\bm{q}\cdot \bm{n} = q_n &= \bar{q}~~~~~~\text{on}~ \Gamma_N.
\end{align} 
In the above relations, $Q$ is the heat source term. 
Moreover, Dirichlet and Neumann boundary conditions are introduced in Eq.~\ref{BCsthermalD} and Eq.~\ref{BCsthermalN}, respectively. 
In this context, the heat flux $\bm{q}$ is defined based on Fourier's law
\begin{align}
\label{Fourier}
\bm{q} = -k(x,y,z)\,\nabla T,
\end{align} 
where $k(x,y,z)$ is the phase-dependent heat conductivity coefficient. 
\color{black}

By introducing $\delta T$ as a test function and with integration by parts, the weak form of the steady-state diffusion problem reads
\begin{align}
\label{eq:weakformthermal}
\int_{\Omega}\,k(x,y,z)\,\nabla^T T\,\delta(\nabla T)~dV\,+\,\int_{\Gamma_N}\bar{q}~\delta T~dA\,-\int_{\Omega}\,Q\,\delta T~dV\,=\,0.
\end{align}

Finally, using the standard finite element method, the discrete residual form of the steady-state diffusion equation can be expressed as:
\begin{align}
\label{eq:ResidualFormThermal}
\boldsymbol{r}(\boldsymbol{k},\boldsymbol{T}) \,=\, \boldsymbol{K}\left(\boldsymbol{k}\right) \boldsymbol{T} -\boldsymbol{f}&= \boldsymbol{0}~~~~~~ \boldsymbol{r} \in \mathbb{R}^r, \; \text{in} \;\; \Omega, \\
\label{eq:discrete_BCsthermalD}
\boldsymbol{T} &= \bar{T}~~~~~\text{on}~ \Gamma_D, \\
\label{BCsthermalN_2}
\bm{q}\cdot \bm{n} = q_n &= \bar{q}~~~~~~\text{on}~ \Gamma_N.
\end{align}
Here, $\boldsymbol{K}$ represents the stiffness matrix, $\boldsymbol{r}$ is the residual vector, $\boldsymbol{T}$ is the solution vector, and $\boldsymbol{f}$ denotes the force vector. It is important to note that the heat conductivity field, represented by $\boldsymbol{k}$, is also discretized at nodes and serves as the physical design variable subject to change and optimization. For the derivation of the residual form, please refer to the \ref{app:A}.

\subsection{Parameterization}
A key aspect of parametric learning and optimization is the underlying parameter space. An efficient design space should be capable of exploring complex designs while avoiding unnecessary design parameters. In this regard, there are two main approaches in the literature: filter-based techniques, such as implicit and explicit filtering \cite{najian2023implicit}, and filter-free techniques, such as CAD-based \cite{kiendl2014isogeometric} and Fourier-based \cite{white2018toplogical} parameterizations. While the first approach is well-established and widely used in practice, the second approach gains increasing attention due to its significant reduction of the design variables, making it computationally more efficient. 

In the following, we begin with a parameterization-free representation of the discrete conductivity field, which varies spatially. We then introduce the Fourier transform as a robust method to control this spatially varying physical space using a limited number of variables.

\subsubsection{Parameterization-free design space}
\textcolor{black}{To better clarify this concept, we now focus on a two-phase material where the properties of each phase remain constant. The parametric space currently consists of the distribution of thermal conductivity values within the microstructure.
Therefore, the microstructure's morphology is altered, allowing for arbitrary shapes and volume fraction ratios. The same strategy can be applied to other parameters of interest (i.e., source and boundary terms).} 
\begin{figure}[H] 
  \centering
  \includegraphics[width=0.99\linewidth]{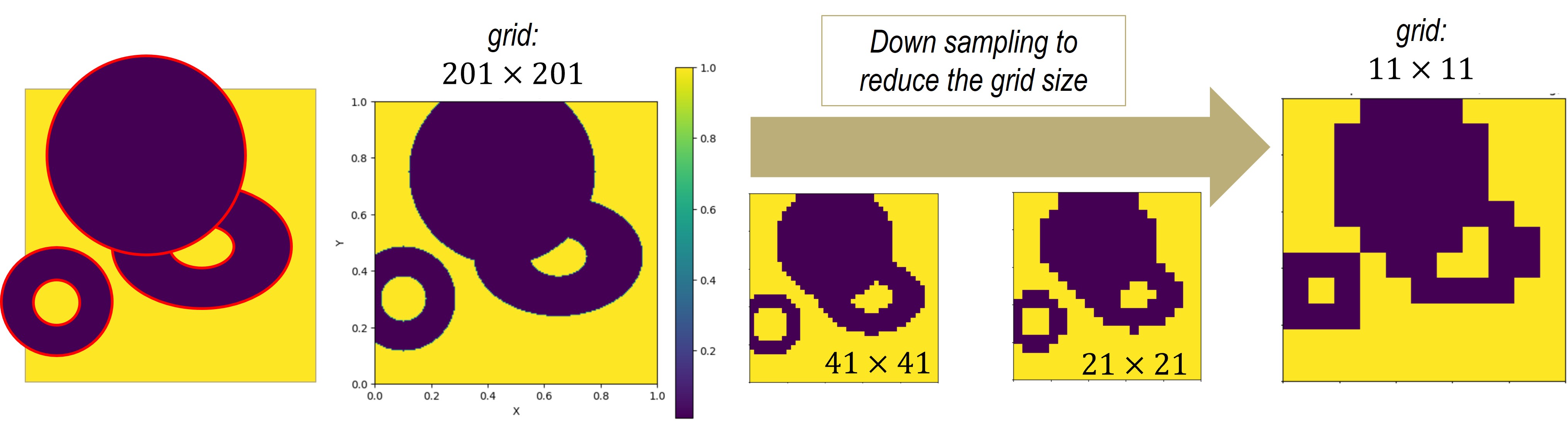}
  \caption{Downsampling the grid size involves using convolutional layers combined with max-pooling, resulting in a reduced space of $11 \times 11 = 121$ grid points. The training will be conducted based on an 11 by 11 grid size. The value of thermal conductivity lies between $k_{min}=0.01$ and $k_{max}=1.0~[W/m^2]$.}
  \label{fig:shapes}
\end{figure}
Based on experience with composite materials, we propose the following strategy for creating samples: 1) select a set of random center points based on a user-defined number, 2) choose random values for the inner radius, outer radius, and rotation angle of the ellipse, with ranges for these values all user-defined, and 3) assign a (low) conductivity value to the grid nodes inside the defined region. 
We then perform a downsampling process via max-pooling on the high-resolution microstructure images, as depicted in Fig.~\ref{fig:shapes}. For more details on down and upsampling strategy, see \cite{koopas2024introducing}.

\textcolor{black}{It is noteworthy that even for a selected reduced grid (with $11 \times 11$ nodes) and considering a two-phase material, there can be \(2^{121} \approx 2.6 \times 10^{36}\) independent samples which is almost unfeasible to cover. 
Hence, the user should determine the relevant parameter space specific to each problem. Another strategy is to reduce the design space by representing it with a different set of inputs that capture the main features. 
Identifying such a reduced parametric space is not a trivial task. Some studies have utilized deep learning algorithms, such as autoencoders, and use the resulting latent space as the reduced parametric space \cite{kontolati2023learning, ZHANG2023116214}. 
Some research has also attempted to increase the resolution of the solution using such algorithms.\cite{koopas2024introducing}. In what follows, we will focus on a more intuitive approach based on Fourier transforms.}

\subsubsection{Fourier-based parameterization}
\begin{figure}[H] 
  \centering
  \includegraphics[width=0.9\linewidth]{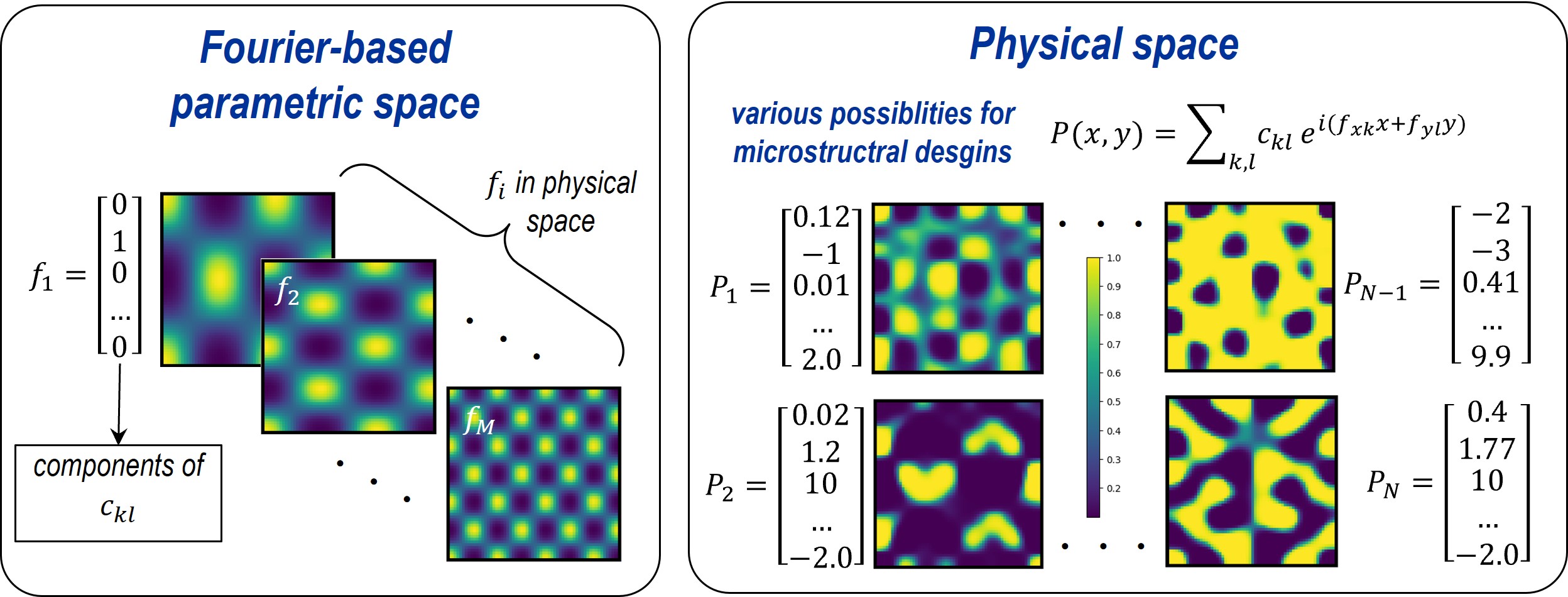}
  \caption{ Fourier-based parameterization helps to decrease the input parametric space dimension.}
  \label{fig:ff}
\end{figure}

Given our interest in periodic design and achieving higher resolution for the solution space without any downsampling, we utilize the Fourier transform method. 
In short, the idea is to combine various frequencies of sinusoidal and cosinusoidal functions to describe spatially varying fields \cite{white2018toplogical}, here the heat conductivity. As a result, and as depicted in Fig.~\ref{fig:ff}, we can reduce the number of inputs to $M$ desired combinations of frequencies, which is usually a much lower number than the number of degrees of freedom in the system, denoted by $N$. 

The Fourier-based function for representation of the conductivity map in a 2D setting has the following form through a combination of sine and cosine functions
\begin{equation}
\label{eq:foruier_eq}
\begin{aligned}
    k_f(x,y) = \sum_i^{n_{sum}} [&A_i \sin{(f_{x,i}~x)} \cos{(f_{y,i}~y)} \\
                              +  &B_i \cos{(f_{x,i}~x)} \sin{(f_{y,i}~y)} \\   
                              +  &C_i \sin{(f_{x,i}~x)} \sin{(f_{y,i}~y)}  \\
                              +  &D_i \cos{(f_{x,i}~x)} \cos({f_{y,i}~y)}].
\end{aligned}
\end{equation}
In Eq.\,\ref{eq:foruier_eq}, and $\bm x = (x, y)^T$, $\{A_i,~B_i,~C_i,~D_i\}$ represents the amplitudes of the corresponding frequency number and $\{f_{x,i},~f_{y,i} \}$ represents the frequency in the $x$- and $y$-direction. 
Finally, we pass the property function $E_f$ through a so-called sigmoidal projection to have more realistic and interesting characteristics for the microstructure
\begin{equation}
\label{eq:sigmoid}
\begin{aligned}
    k(x,y) = (k_{max}-k_{min})\cdot\text{Sigmoid}\left(\beta(k_f-0.5)\right) + k_{min}. 
\end{aligned}
\end{equation}
With the above projection, we ensure that the value of thermal conductivity lies between $k_{min}=0.01$ and $k_{max}=1.0~[W/m^2]$. With the $\beta$ parameter, one can also control the transition between the two phases, which is set to $\beta=5$ for the samples in this study. Some of the main modes, as well as possible samples that this approach can process, are shown in the lower part of Fig.~\ref{fig:ff}.

\textcolor{black}{A known limitation of Fourier-based parameterization is that the number of coefficients must increase rapidly to represent sharply localized features or high-gradient variations, which can make the parametrization inefficient. Consequently, this approach is most effective for smoothly varying fields, and alternative latent representations (e.g., autoencoders) may be preferable for highly irregular or non-periodic designs.}

\subsection{Gradient-based optimization}
\label{sec:grad_opt_description}
\textcolor{black}{After the discretization of the computational domain and governing equations, the design and analysis problem in Eq. \ref{eq:math_problem_def} can be expressed in its discrete form as}
\begin{equation}
\begin{aligned}
& \underset{\boldsymbol{c} }{\text{min}}
& & J(\boldsymbol{k}, \boldsymbol{T}), \\
& \text{subject to} 
& & \boldsymbol{k} = \boldsymbol{A} \,\boldsymbol{c},
\\
& & &\boldsymbol{r} \,=\, \boldsymbol{K}\left(\boldsymbol{k}\right) \boldsymbol{T} -\boldsymbol{f}= \boldsymbol{0}, \\
& & & h_{j} = 0, \; \, \; j = 1, \ldots, m_{h}, \\
& & & g_{j} \leq 0, \; \; \, j = 1, \ldots, m_{g}. 
\end{aligned}
\end{equation}

\textcolor{black}{Here, $\bm{T}$, $\bm{c}$, and $\bm{k}$ are vectors containing the discretized values of the solution field, control parameters, and physical design variables. } To find the optimum solution, we follow first-order optimization with an adapted version of Rosen’s gradient projection technique \citep{rosen1961gradient}. This method iteratively adjusts the true control variables based on the projected gradients, ensuring that each step moves towards a reduction in the objective function while satisfying all imposed constraints. This approach is detailed as

\begin{equation}
\label{eq:update_rule}
^{k+1}\boldsymbol{c} \,=\, ^{k}\boldsymbol{c} - \alpha \underbrace{\left[ \boldsymbol{I} -\boldsymbol{P}\left( \boldsymbol{P}^T\boldsymbol{P}\right)^{-1}  \boldsymbol{P}^T\right] \frac{dJ}{d\boldsymbol{c}}}_{\text{projection}} - \overbrace{ \left[\boldsymbol{P}\left( \boldsymbol{P}^T\boldsymbol{P}\right)^{-1}  \right]\boldsymbol{g}^a}^{\text{correction}},
\end{equation}
where $\alpha$ is the projection step size. The matrix $\boldsymbol{P}\in \mathbb{R}^c \times \mathbb{R}^a$
 contains the gradients $d\boldsymbol{g}^a/d\boldsymbol{c}$
 in a column-wise arrangement, with $\boldsymbol{g}^a \in \mathbb{R}^a$
 being the vector of active constraints. The update rule equation above includes both a projection and a correction term. The projection term projects the steepest descent direction onto a subspace tangent to the active constraints, while the correction term ensures that the update direction remains feasible in the presence of nonlinear constraints.

To update the design variables detailed in Eq. \ref{eq:update_rule}, it is necessary to determine the gradients of both the objective function and the constraints with respect to the true design variables. These calculations must also comply with the underlying physical governing equations that characterize the system. By employing the chain rule for differentiation, the gradients of any state-dependent function, such as the objective function, can be expressed as   
\begin{equation}
\label{eq:chain_rule_diff}
 \frac{dJ}{d\boldsymbol{c}} = \left(\frac{\partial J}{\partial\boldsymbol{T}}\frac{d \boldsymbol{T}}{d\boldsymbol{k}} \,+\, \frac{\partial J}{\partial\boldsymbol{k}}\right)\frac{d \boldsymbol{k}}{d\boldsymbol{c}}.
\end{equation}
While the partial derivatives can often be computed using analytical or automatic differentiation methods, computing the total derivatives involves differentiating both the underlying governing physical equations and the parametrization equations, as per
\begin{align}
\label{eq:total_state_param_derv}
\frac{d \boldsymbol{r}}{d\boldsymbol{k}} &= \frac{\partial \boldsymbol{r}}{d\boldsymbol{T}}\frac{d \boldsymbol{T}}{d\boldsymbol{k}} + \frac{\partial \boldsymbol{r}}{\partial\boldsymbol{k}} = \boldsymbol{0},\,\, \text{then}\,\, \frac{d \boldsymbol{T}}{d\boldsymbol{k}} = - \left(\frac{\partial \boldsymbol{r}}{d\boldsymbol{T}}\right)^{-1}\frac{\partial \boldsymbol{r}}{\partial\boldsymbol{k}},\\
\frac{d \boldsymbol{k}}{d\boldsymbol{c}} &=\boldsymbol{A}.
\end{align} 
Incorporating the earlier equations into the sensitivity Eq.~\ref{eq:chain_rule_diff}, we derive the expression
\begin{equation}
\label{eq:expanded_chain_rule_diff}
 \frac{dJ}{d\boldsymbol{c}} = \left(-\frac{\partial J}{\partial\boldsymbol{T}} \left(\frac{\partial \boldsymbol{r}}{d\boldsymbol{T}}\right)^{-1}\frac{\partial \boldsymbol{r}}{\partial\boldsymbol{k}} \,+\, \frac{\partial J}{\partial\boldsymbol{k}}\right)\boldsymbol{A}.
\end{equation}
As a matter of fact, evaluating sensitivities requires the availability of the inverse of the stiffness matrix. However, this becomes costly for problems involving fine meshes. To mitigate this, the adjoint technique is commonly employed. This method involves initially computing the adjoint variables $\boldsymbol{\lambda}$, followed by the subsequent calculation of sensitivities  
\begin{align}
\label{eq:adjoint_based_SA}
 \boldsymbol{\lambda}&= -\left(\frac{\partial \boldsymbol{r}}{d\boldsymbol{T}}\right)^{-T} \frac{\partial J}{\partial\boldsymbol{T}},\\
 \frac{dJ}{d\boldsymbol{c}} &= \left(\boldsymbol{\lambda} ^T\frac{\partial \boldsymbol{r}}{\partial\boldsymbol{k}} \,+\, \frac{\partial J}{\partial\boldsymbol{k}}\right)\boldsymbol{A}.
\end{align} 
It should be noted that although adjoint-based sensitivity analysis avoids the explicit inversion and storage of the tangent matrix, it necessitates a linear solve for each objective and constraint function. Therefore, this approach may become prohibitively costly when multiple adjoint-based sensitivity analyses are required.

\section{Finite operator learning}
\label{sec:FOL}
\begin{figure}[H] 
  \centering
  \includegraphics[width=0.99\linewidth]{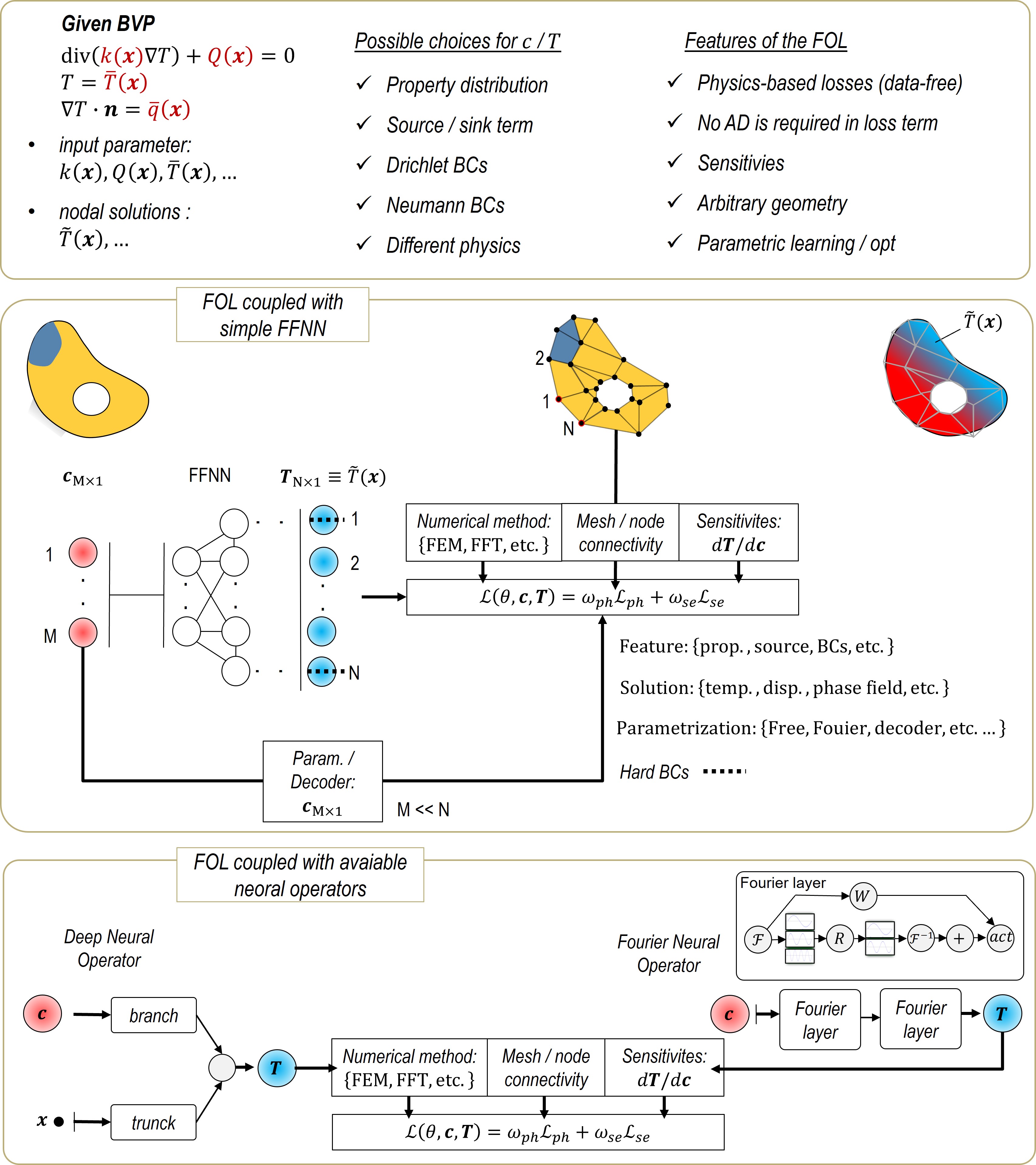}
  \caption{ \textcolor{black}{Work flow and network architecture for finite operator learning.} }
  \label{fig:NN_idea}
\end{figure}

Building upon the concept of physics-informed neural networks \citet{RAISSI2019} and deep energy method \citet{SAMANIEGO2020112790}, we construct a neural network based on the discretized weak form of the governing partial differential equations, as detailed in Section \ref{sec:governing_equation}. The input layer consists of the design variables or parameters, while the output layer represents the unknown solution field, specifically nodal temperatures in this work. Additionally, constraints on the network's inputs and outputs - or any combination thereof - can be effectively enforced through a multi-objective loss function approach. This work is based on feed-forward neural networks. By denoting the information transfer from the $l-1$ layer to $l$ using the vector $\bm{z}^l$, each component of vector $\bm{z}^l$ is computed by
\begin{equation}
\label{eq:NN_1}
    {z}^l_m = {a} (\sum_{n=1}^{N_l} w^l_{mn} {z}_n^{l-1} + b^l_{m} ),~~~l=1,\ldots,L. 
\end{equation}
In Eq.\,(\ref{eq:NN_1}), ${z}^{l-1}_n$ is the $n$-th neuron within the $l-1$-th layer. The component $w_{mn}$ shows the connection weight between the $n$-th neuron of the layer $l-1$ and the $m$-th neuron of the layer $l$. Every neuron in the $l$-th hidden layer owns a bias variable $b_m^l$. The number $N_I$ corresponds to the number of neurons in the $l$-th hidden layer. The total number of hidden layers is $L$. The letter $a$ stands for the activation function in each neuron. As a result, our feed-forward neural operator $\mathcal{N}$ is mathematically expressed as
\begin{align}
\label{eq:in_out}
    \bm{c} &= [c_i],~~\widetilde{\bm{T}} = [\widetilde{T}_j],~~~i= 1 \cdots M,~~~j = 1 \cdots N, \\
    \widetilde{\bm{T}} &= \mathcal{N} (\bm{c}; \bm{\theta}),~~~\bm{\theta} = \{\bm{W},\bm{b}\}.
\end{align}
\color{black}
Again, $\bm{c}$ is the vector of $M$ design parameters,  $\widetilde{\bm{T}}$ is the vector of $N$ predicted nodal temperatures, and the trainable parameters of the network are denoted by $\bm{\theta}$. Fig.~\ref{fig:NN_idea} schematically represents the proposed network for finite operator learning. 

Next, we design the network's loss function to enforce the desired constraints on the inputs and outputs - or any combination thereof - in a multi-objective manner as
\begin{align}
\label{eq:total_loss_sum}
\mathcal{L}_{tot} & = w_{ph}\,\mathcal{L}_{ph} + w_{bc} \,\mathcal{L}_{bc} + w_{se} \,\mathcal{L}_{se}.
\end{align}
Different terms in the above equation read as follows:
\begin{itemize}
    \item $\mathcal{L}_{ph}(\bm{k},\widetilde{\bm{T}})$ ensures that the predicted nodal temperatures comply with the governing physical equations. This is achieved by minimizing either the energy functional or the sum of the nodal residuals 
    \begin{align}
    \mathcal{L}_{en}(\bm{k},\widetilde{\bm{T}}) & = \sum_{e=1}^{n_{el}}\widetilde{\boldsymbol{T}}^T_e~\widetilde{\boldsymbol{r}}_e,~ \text{where}~~\widetilde{\boldsymbol{r}}_e=\boldsymbol \sum_{n=1}^{n_{int}} ~\boldsymbol B_e^T ~(\boldsymbol N_e^T \boldsymbol k_e)~\boldsymbol B_e ~\widetilde{\boldsymbol T}_e - {\boldsymbol f}_e,\\
    \mathcal{L}_{re}(\bm{k},\widetilde{\bm{T}}) & = \sum_{i=1}^{n_{no}} \widetilde{r}_i^2,~ \text{where}~~\widetilde{r}_i = \sum_{e \in \Omega_i} \widetilde{\boldsymbol{r}}_e.
    \end{align}
    Here, $\widetilde{\boldsymbol{r}}$ denotes the nodal residuals computed from the predicted temperatures and $\Omega_i$ is the set of elements sharing node $i$. Note that the subscript $e$ indicates that the corresponding variable is evaluated at the element level.
    \item $\mathcal{L}_{bc}(\widetilde{\bm{T}})$ \textcolor{black}{ensures compliance with any potentially complex boundary conditions, if applicable; otherwise, $w_{bc}$ is set to zero. Note that Dirichlet boundary conditions are automatically and strongly satisfied, since the network predicts only the unknown nodal values. Neumann conditions are naturally incorporated through the weak form. For more involved boundary conditions, where a strong imposition is not feasible, the user can activate this term to enforce them in a soft manner.} 
    \item $\mathcal{L}_{se}(\bm{c},\widetilde{\bm{T}})$ ensures that the network’s output-to-input sensitivity, denoted by ${d\widetilde{\boldsymbol{T}}}/{d\boldsymbol{c}}$, complies with the requirement that the residuals must be stationary with respect to the design variables, as outlined in Eq. \ref{eq:total_state_param_derv}. Specifically, the equation governing this condition is
    \begin{equation}
    \begin{aligned}
        \frac{d \widetilde{\boldsymbol{r}}}{d\boldsymbol{c}} & = \frac{\partial \boldsymbol{\widetilde{r}}}{d\widetilde{\boldsymbol{T}}}\,\frac{d \widetilde{\boldsymbol{T}}}{d\boldsymbol{c}} + \frac{\partial \widetilde{\boldsymbol{r}}}{\partial\boldsymbol{k}} \frac{d {\boldsymbol{k}}}{d\boldsymbol{c}} \\
        & = \boldsymbol{K} \frac{d \widetilde{\boldsymbol{T}}}{d\boldsymbol{c}} + \frac{\partial \widetilde{\boldsymbol{r}}}{\partial\boldsymbol{k}} \boldsymbol{A}
        \\
        & = \boldsymbol{0}.
    \end{aligned}
    \end{equation}
    where $\boldsymbol{K}$ is the tangent stiffness matrix of FEM, $\boldsymbol{A}$ is the tangent matrix of the parameterization, and ${\partial \widetilde{\boldsymbol{r}}}/{\partial\boldsymbol{k}}$ is the tangent matrix of FEM w.r.t the physical design field and it can be analytically calculated. 
    By aggregating the violations of the specified stationary condition across all nodes, the corresponding loss function is defined as
    \begin{equation}
    \mathcal{L}_{se}(\bm{c},\widetilde{\bm{T}})  = \sum_{i=1}^{n_{nodes}} \sum_{j=1}^{n_{c}} \,\,\left(\frac{d \widetilde{r}_i}{dc_j}\right)^2.     
    \end{equation}
    It should be noted that by incorporating these target derivatives into the total loss function, we essentially engage in what is known as Sobolev training \cite{czarnecki2017sobolev}. This approach not only ensures the correct output-to-input sensitivity of the network but also results in surrogate models that deliver higher accuracy in predicted values.  
\end{itemize} 

Finally, the total loss term is minimized concerning the trainable parameters to find the optimal trainable parameters $\bm{\theta}^*$ via a desired optimizer 
\begin{align}
\label{minimize}
\bm{\theta}^* = \arg \min_{\bm{\theta}} \mathcal{L}_{tot}(\bm{\theta},\bm{c},\widetilde{\bm{T}}).
\end{align}
\noindent

\textcolor{black}{\textbf{Remark 1} In FOL, the derivatives are approximated using the shape functions from the finite element method. In simpler terms, the loss function in this work is an algebraic equation.}
\textcolor{black}{Moreover, the choice between energy-based loss and the residual-based loss depends on the underlying PDE characteristics. Specifically, for elliptic problems, minimizing the FE energy functional directly enforces the governing equations through the principle of minimum potential energy, making $L_{en}$ a robust and efficient option. In contrast, $L_{re}$ corresponds to enforcing the weak form residual and becomes preferable when dealing with non-variational problems where a valid energy functional does not exist.}
\textcolor{black}{Finally, the sensitivity–based term is only required for inverse design or optimization tasks when accurate sensitivities are essential. For purely forward PDE problems, or when the design space is sufficiently sampled, this term can be omitted. In such cases, the framework reduces to a single physics–based loss, with no need for additional loss balancing.}


\newpage
\section{Results: Parametric learning and sensitivity analysis}
\label{sec:res_part1}
\begin{figure}[H] 
  \centering
  \includegraphics[width=0.7\linewidth]{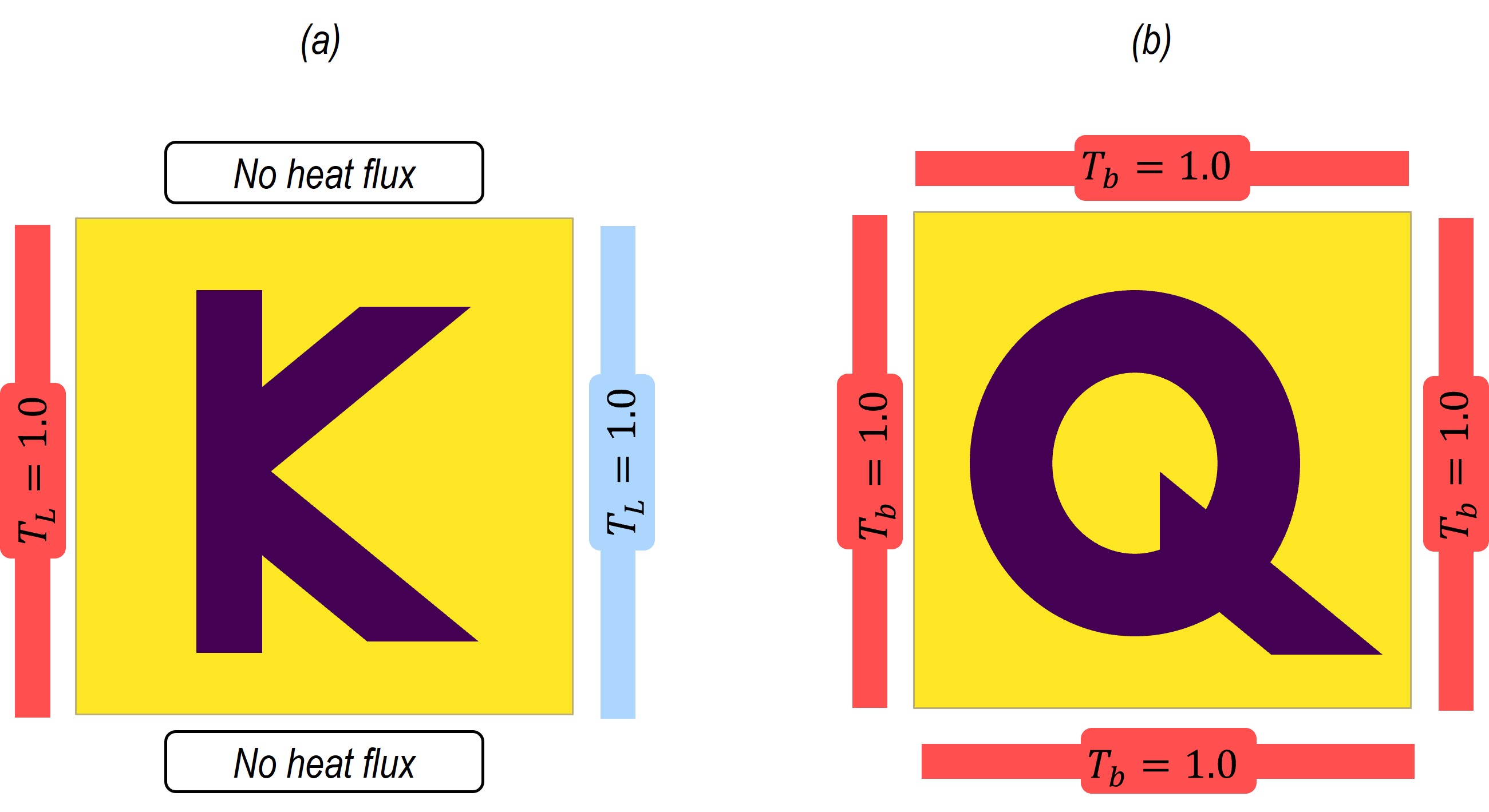}
  \caption{ \textcolor{black}{Given the geometry and boundary conditions for the thermal problem: (a) The property distribution (thermal conductivity, $k$) is under parametric learning. (b) The source term $Q$ is under parametric learning. For both cases, a structured mesh of $51 \times 51$ is utilized.}  }
  \label{fig:examples}
\end{figure}

\begin{figure}[H] 
  \centering
  \includegraphics[width=0.99\linewidth]{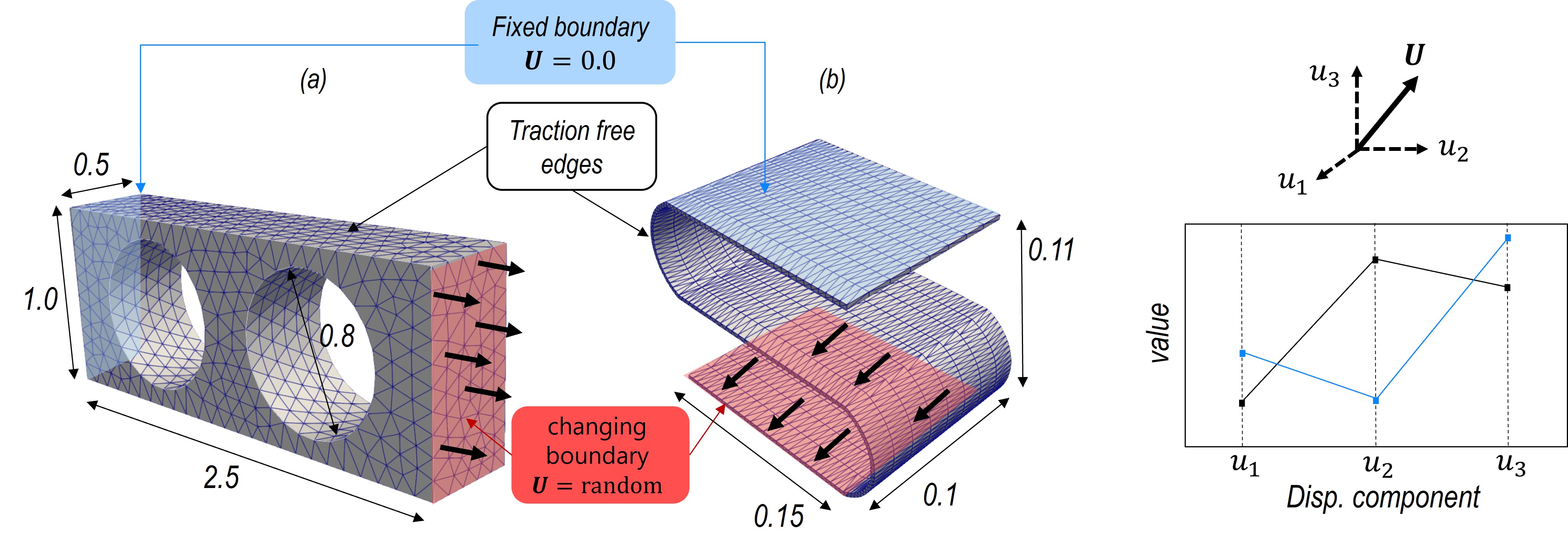}
  \caption{  \textcolor{black}{Given the geometry and boundary conditions for the mechanical problem: (a) The applied displacement on the boundary is under parametric learning using a beam with holes and an unstructured mesh. (b) The applied displacement is under parametric learning using an unstructured mesh and complex geometry. All dimensions are in $mm$.} }
  \label{fig:examples_2}
\end{figure}

This section is divided into three main parts. First, we briefly discuss the generation of collocation fields for the data-free training process. Next, we present the results of parametric learning of the governing equations described in Eq.~\ref{StrongfromThermal}. This includes a comprehensive study on the accuracy of FOL-based predictions for both the primal field and sensitivities. \textcolor{black}{Moreover, for the parametric training we focus on three aspects and two different physics to back up our claim on the flexibility of the FOL in handling different test case scenarios. We shall vary not only the position-dependent property map but also the source terms and boundary conditions separately. These test cases are supposed to cover the main tasks in many engineering applications. See also Table \ref{tab:examples} } Lastly, we demonstrate gradient-based optimization using a pre-train-free FOL approach, where the learning process is nested within each design iteration and the gained knowledge is transferred to the next iteration. This is the so-called NAND optimization using FOL for primal and sensitivity analysis. 
\begin{table}[H]
\centering
\caption{\textcolor{black}{Summary of different studies for parametric learning of the forward problem.}}  
\label{tab:examples}
\begin{footnotesize}
\begin{tabular}{ l l }
\hline
Target for parametric learning  &  Governing physics / comments    \\
\hline
Property distribution (Section~\ref{sec:mat_prop} and Fig.~\ref{fig:examples}a)  &  2D steady-state thermal diffusion \\ 
Position-dependent source term (Section~\ref{sec:source_prop} and Fig.~\ref{fig:examples}b)  &  2D steady-state thermal diffusion \\ 
Boundary terms (Section~\ref{sec:BCs_prop} and Fig.~\ref{fig:examples_2})  &  3D mechanical equilibrium + unstructured meshes \\ 
\hline
\end{tabular}
\end{footnotesize}
\end{table} 
\textcolor{black}{In the 2D results provided in this section, we focus on the geometry and boundary conditions described in Fig.~\ref{fig:examples}, unless mentioned otherwise. The geometries and boundary conditions shown in Fig.~\ref{fig:examples_2} are chosen for the 3D examples, with parametric learning on the boundary terms and mechanical problem using unstructured meshes. With these examples, we aim to demonstrate the performance of the physics-informed FOL framework for different types of PDEs and the parametric learning of various important terms.} 

Furthermore, heat transfer in the y-direction is considered the objective for the sensitivity study and optimization, while heat transfer in the x-direction is treated as the constraint. These functions are mathematically expressed as 
\begin{align}
J &= \int_{\Omega}(-k\,\nabla_y T)^2 \,dV,\label{eq:qy}\\
h &= \int_{\Omega}(-k\,\nabla_x T)^2 \,dV - 0.125 ,\label{eq:qx}
\end{align}
where $J$ and $h$ are the objective and constraint functions, respectively. For the full optimization problem definition, refer to Section \ref{sec:grad_opt_description}.

\textcolor{black}{Finally, we would like to mention that extending the presented frame and studies to unstructured meshes or arbitrarily shaped geometries in 2D and 3D is straightforward. 
All these aspects are discussed in the results section. 
Readers are also encouraged to see \cite{rezaei2024finite, yamazaki2024finite} for comparison against other neural operators as well as extension to transient problems.} 

\textcolor{black}{It is noteworthy that the training can be conducted on personal laptops without GPU acceleration, owing to the efficiency of the proposed framework. However, we primarily utilized a Quadro RTX 6000 GPU unless stated otherwise.}

\textcolor{black}{The algorithms developed in this study are implemented using JAX \cite{jax2018github} software. The codes are also implemented in the SciANN package \cite{SciANN}, and the methodology can be adapted to other programming platforms as well.}

\subsection{Sample preparation} 
\label{sec:sample_prep}
\textcolor{black}{We have discussed two sets of parametrization approaches in this work. Regardless of the user's choice, each parametrization approach requires exploring the possible options in the parametric space. By providing the network with a sufficiently diverse set of inputs, the performance of the deep learning model will improve. 
In \ref{sec:app_sample}, we discuss all the details regarding our strategies to explore each design space. 
}

\newpage   
\subsection{Parametric learning on property distribution} 
\label{sec:mat_prop}
We start by learning the solution based on the given property distribution. This section is of particular interest, as it opens up numerous opportunities in the design of a new generation of materials through a multiscale approach.
In the first set of studies, the Dirichlet boundary conditions are satisfied in a soft manner via a penalty term to demonstrate the framework's flexibility in handling such cases. As will be explained later, these terms can easily be imposed in a hard manner as well. Furthermore, the loss term related to sensitivities is switched off at first to evaluate the performance of the network solely based on the discretized energy form of the problem. For the loss term related to Dirichlet boundary conditions, we assigned higher weightings to expedite the satisfaction of the boundary terms. Following a systematic study that began with equal weightings for both loss terms, we selected $w_{db}=10$. 
A summary of the network's (hyper)parameters is presented in Table~\ref{tab:NN_para}. Note that whenever multiple options for a parameter are introduced, we examine the influence of each parameter on the results.
\begin{table}[H]
\centering
\caption{Summary of the network parameters for learning based on property distribution.}  
\label{tab:NN_para}
\begin{footnotesize}
\begin{tabular}{ l l }
\hline
Parameter                          &  Value    \\
\hline
Inputs, Outputs                  &  $\{k_i\}$, $\{T_i\}$ \\ 
Activation function                &  tanh, swish, sigmoid, linear \\ 
Number of layers and neurons for each sub-network ($L$, $N_l$)  &  (2, 10) \\
Optimizer                         &  Adam \\ 
Number of initial samples                         &  $2000$, $4000$ \\ 
Batch size                         &  $50$, $100$ \\ 
(Learning rate $\alpha$, number of epochs)  &  $(10^{-3},5000)$ \\ 
\hline
\end{tabular}
\end{footnotesize}
\end{table} 

\begin{figure}[H] 
  \centering
  \includegraphics[width=1.0\linewidth]{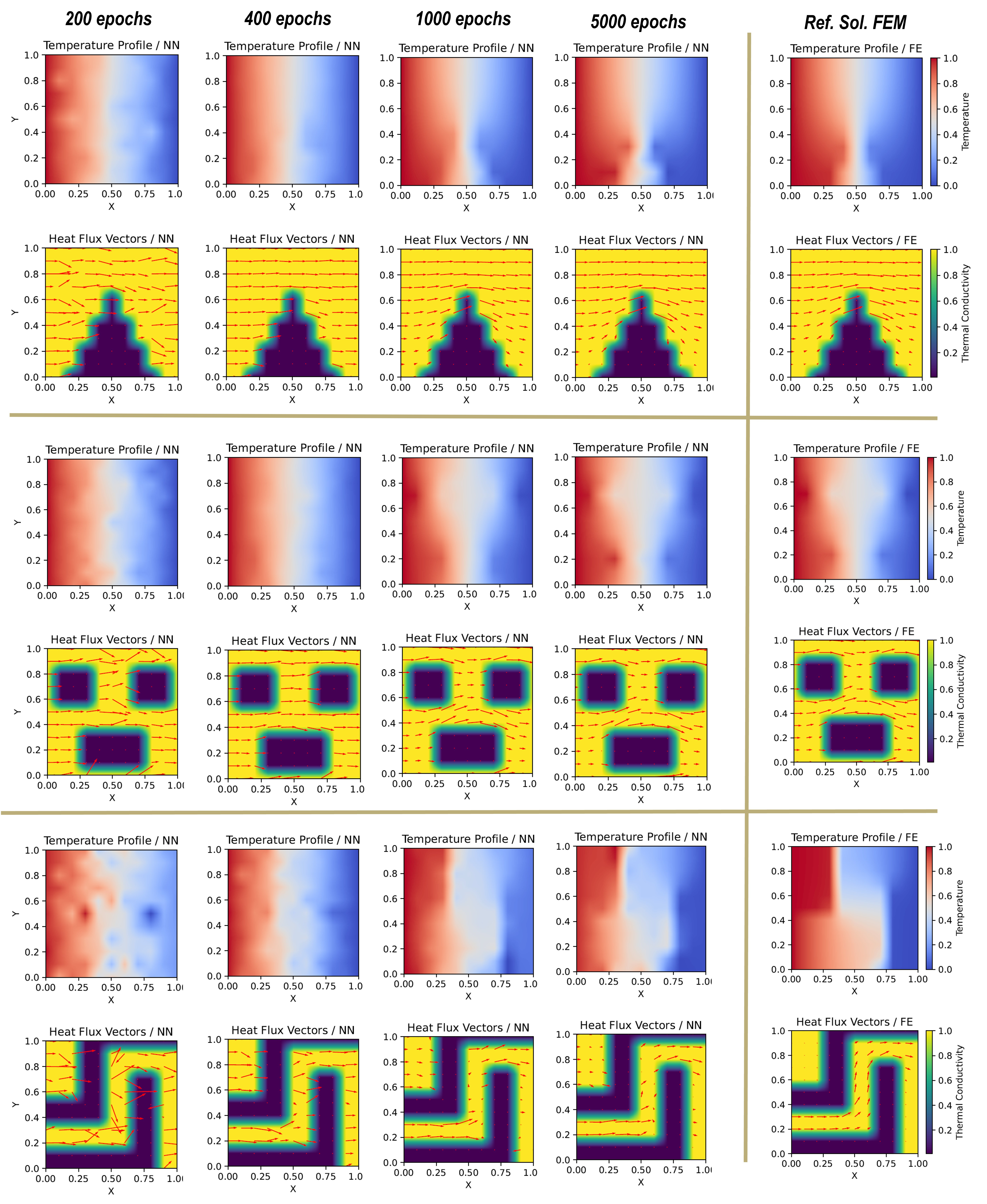}
  \caption{Predictions of the trained FOL for different numbers of epochs.}
  \label{fig:evol_2}
\end{figure}

In Fig.~\ref{fig:evol_2}, the predictions of the same neural network for various sample test morphologies (i.e., conductivity maps) are depicted across different numbers of epochs, presented in separate columns. The last column displays the reference solutions obtained by the finite element method, utilizing the same number of elements (10 by 10 elements or 11 by 11 nodes). It is important to note that post-training, the results are assessed on finer grid points ($165 \times 165$) using simple shape function interpolations, which lend a smoother appearance to the results. For conciseness, all temperature values are denoted in Kelvin $[K]$, heat flux is represented in $[W/m^2]$, and thermal conductivity is expressed in $[W/mK]$.

As depicted in Fig.~\ref{fig:evol_2}, the NN's outcomes are unreliable before 1000 epochs. Surprisingly, only after 1000 epochs, the NN demonstrates an acceptable level of training. Training further resulted in some improvements, which are particularly noticeable in sharp discontinuities near the phase boundaries. It is noteworthy that the results showcased in Fig.~\ref{fig:evol_2} correspond to test cases, where these morphologies are unseen by the NN.  More comprehensive comparisons and quantitative error analyses are provided in subsequent sections.
\\ \\
\textbf{Remark 2} 
In the current framework, we expect three different convergence studies. First, we need to train with a sufficient number of epochs. Second, we need to ensure the mesh resolution is adequate for the output layer. Third, for parametric learning, we need to ensure that a sufficient number of (random) samples is provided for the FOL.
\\ \\
\textbf{Remark 3} We exclusively predict temperature profiles in the current approach. Consequently, the flux vector, as presented in subsequent studies, is derived via local numerical differentiation of the temperature. However, an extension of this approach could involve a mixed formulation, predicting the flux values as additional outputs for the network to further enhance predictions, as explored in works such as \cite{Faroughi22, REZAEI2022PINN, Harandi2023}.
\color{black}

\newpage

\subsubsection{Comparison of physics-driven and data-driven deep learning models}
\begin{figure}[H] 
  \centering
  \includegraphics[width=0.75\linewidth]{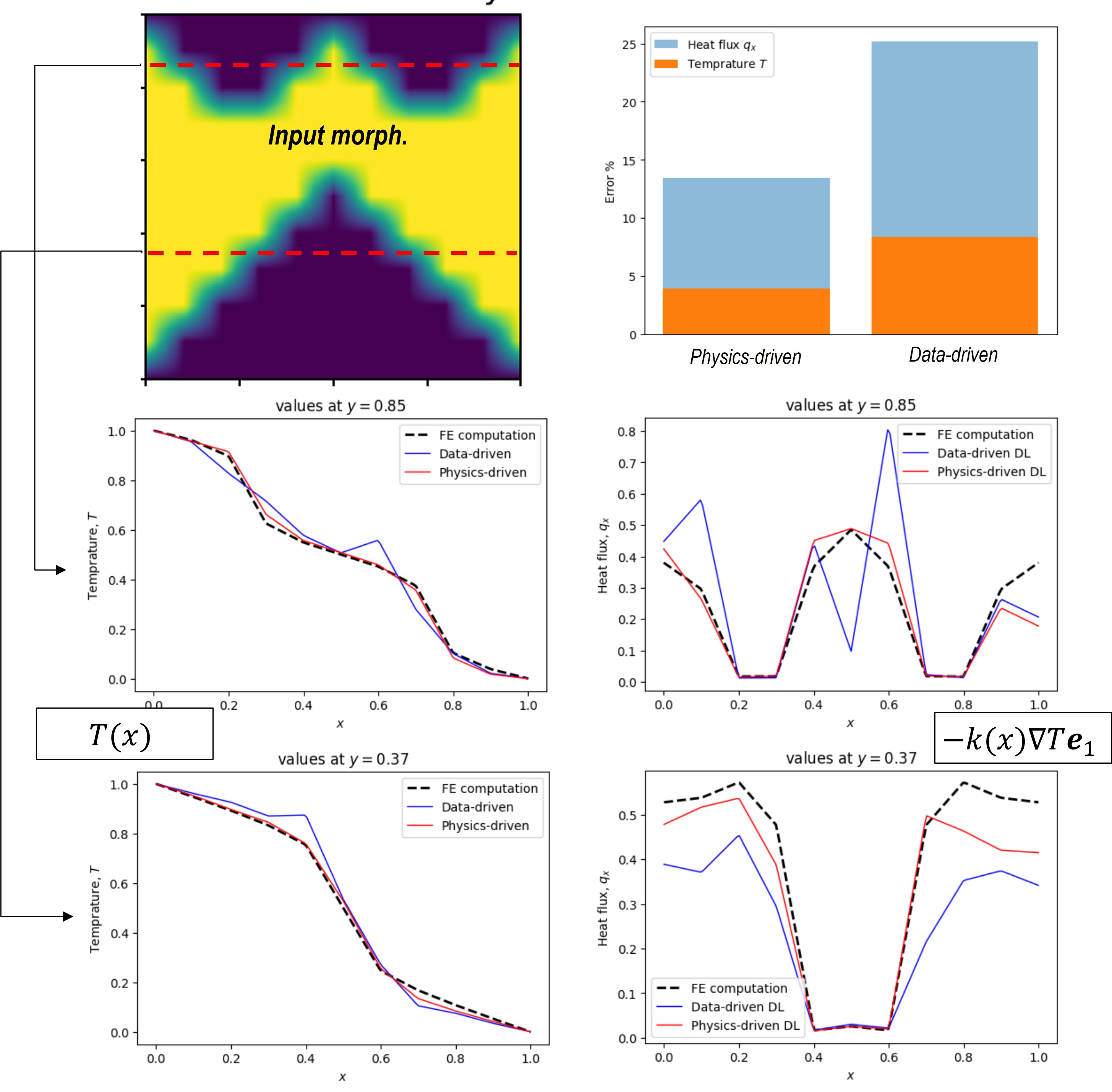}
  \caption{Comparison of data-driven and physics-driven approaches regarding the temperature and flux profiles.}
  \label{fig:err_dd}
\end{figure}

In this part, we seek to demonstrate the implications of using traditional numerical models, gathering their data, and training a neural network in a supervised manner. This latter approach, termed data-driven, is widely pursued by many researchers. To conduct this comparison, we utilize the same set of 4000 sample data, perform finite element calculations for each, and store the associated solutions. We employ an identical network architecture to train the data-driven network. The comparison results are illustrated in Fig.~\ref{fig:err_dd} for a specific test case. For an unseen test case, the physics-driven method provides more accurate predictions for temperature and flux distribution. The error in this study is based on $Err = \frac{\sqrt{\sum_i (T_{NN}(i)-T_{FE}(i))^2}}{\sqrt{\sum_i (T_{FE}(i))^2}}\times 100$.  
The observations and conclusions mentioned above are not confined to the test case. This significantly underscores the appeal of the physics-driven model for two primary reasons: 1) It does not rely on labeled data for training, and 2) it delivers more accurate predictions for unseen cases. However, we should report that the supervised training requires a lower training cost (approximately 10 times less, on average) than the physics-based approach. Moreover, it is challenging, if not impossible, to assert that there is no potentially better NN for improving the data-driven approach. Therefore, further investigations using other types of architectures for physics-driven and data-driven neural networks are important for future studies.


\color{black}
\subsubsection{Inflaucne and importance of the sensitivity term}
\label{sec:res_sen}
\begin{figure}[H] 
  \centering
  \includegraphics[width=0.86\linewidth]{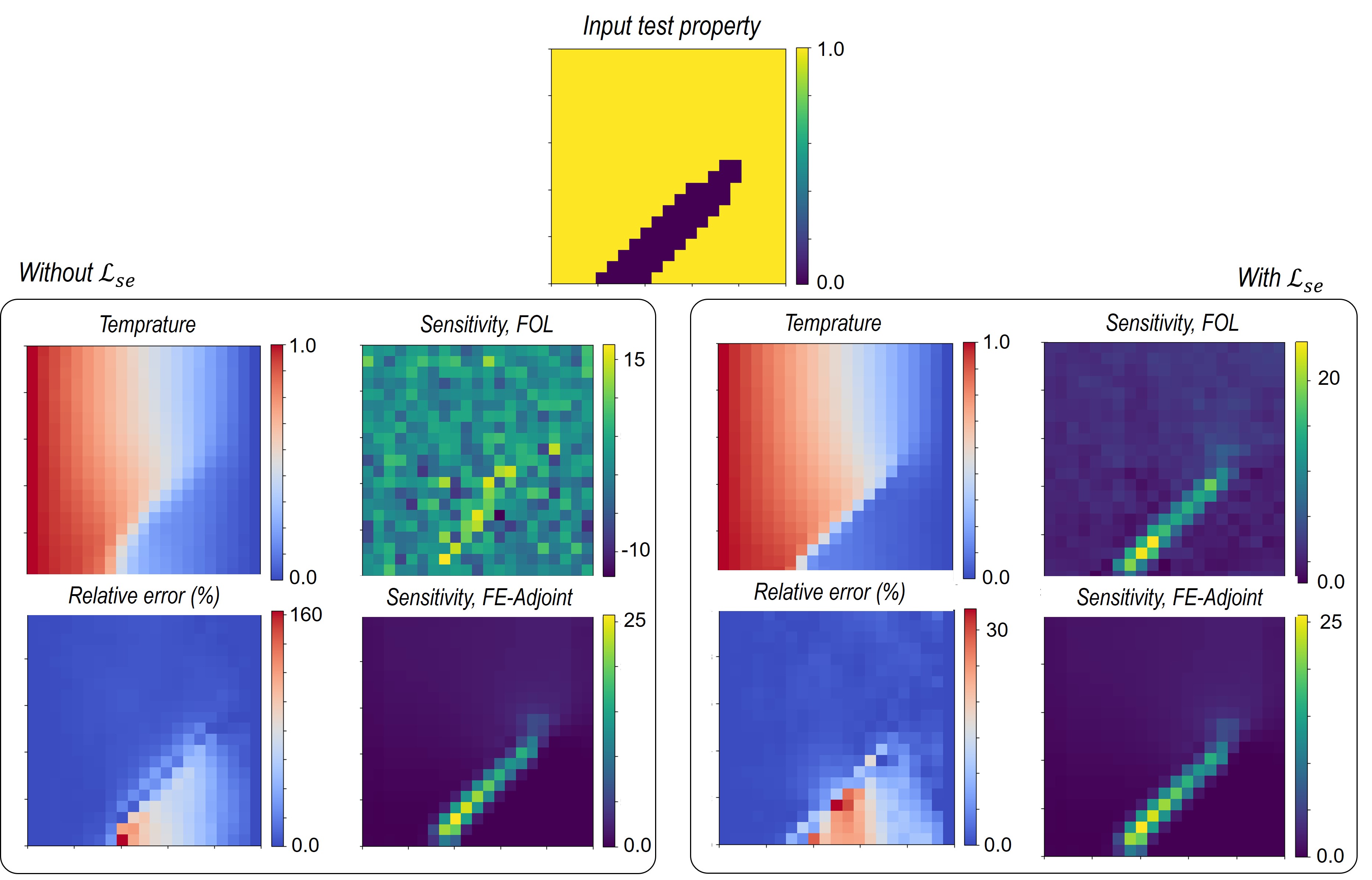}
  \caption{ \textcolor{black}{Influence of adding the loss term for sensitivities. Left: Sensitivity values are still not accurate enough after simple parametric learning. Right: Adding the additional term for sensitivities significantly improves the sensitivity values. Here, the input property distribution includes a sharp transition between the two phases.}  }
  \label{fig:sen_1}
\end{figure}

According to Eq.~\ref{eq:total_loss_sum}, in addition to the specific physical loss term introduced in this work, we also propose adding an additional loss term for the sensitivities or the derivative of the output (solution space) with respect to the inputs (desired parametric space). Based on the description so far, even without this additional term (i.e., by setting $w_{se}=0$), parametric learning is possible, and acceptable solutions can be achieved.
However, we argue for the first time that even though the network is trained based on the governing equations and physics of the problem, the sensitivities are still not acceptable. This finding is illustrated in Figs.~\ref{fig:sen_1} and \ref{fig:sen_2} for the free-parameterization case. 
These results are shown for two different test cases with sufficient variety, but the issue persists in other cases as well. 
Here, the response of sensitivities are calculated for heat flux in the $x$-direction based on Eq.~\ref{eq:qx}.
Moreover, the problem setup, including the boundary conditions and material properties, is the same as in the previous section.

Interestingly, adding the additional loss term not only significantly improves the results for the sensitivities but also enhances the quality of the results for the forward problem (i.e., temperature predictions).
See also the litriture on Sobolev training \cite{VLASSIS2021113695,son2021sobolev,avrutskiy2020enhancing,czarnecki2017sobolev}.
For the results reported in this section, we used a mesh of $21 \times 21$, and the samples for the training, as well as the network hyperparameters, are consistent with the previous section as described in \ref{sec:app_sample}.
\textcolor{black}{It should be noted that providing more initial samples to cover a sufficient variety of the input space will improve the accuracy of sensitivities even without the additional term. 
However, this approach is challenging for practical cases due to a large variety of possible choices. 
A second strategy would be to reduce the input parametric space. 
This approach is further explored in the following sections with Fourier-based parameterization.
}

\subsubsection{Fourier-based parameterization for FOL: Mesh and sample convergence}
\label{sec:res_fourier_parametric}
We now switch to the second formulation for the FOL framework, which is primarily based on a reduced set of input parameters using Fourier analysis (see also Fig.~\ref{fig:NN_idea}). 
Moreover, we take advantage of applying hard boundary constraints. We use $T_R = 0.1~$K and $T_L = 1.0~$K to show the flexibility of the method to handle other BCs. 
Note that for the rest of the paper, all the loss terms defined in Eq.~\ref{eq:total_loss_sum} are now active unless it is mentioned otherwise for further studies.

Let's turn into Eq.~\ref{eq:foruier_eq}. For the rest of the study, and for simplification purposes, only the constant term as well as the last term involving the multiplication of two cosine functions are considered (i.e., $A_i=0,~B_i=0,~C_i=0$). However, even by limiting ourselves to these terms, one can produce many designs for the property function $k_f$.
We select three frequencies for each direction. Considering the constant term, this choice gives us $M = 3 \times 3 + 1 = 10$ different terms, which can be added up to construct $k(x,y)$ according to Eq.~\ref{eq:sigmoid} and Eq.~\ref{eq:foruier_eq}. See also Table \ref{Table:Fourier_param} for a summary of the Fourier-based parameterization for the FOL model.
\begin{table}[H]
\centering
\caption{Summary of the network parameters for learning based on property distribution using the Fourier-based FOL.}
 \begin{tabular}{ll} 
    \hline
        Parameter &   Value\\
    \hline
     Inputs, Outputs                  &  $\{\text{freq}_i\}$, $\{T_i\}$ \\ 
     Frequency in the $x$- direction ${f_{x, i}}$ & $\left\{3,~5,~7\right\}$\\
     Frequency in the $y$- direction ${f_{y, i}}$ & $\left\{2,~4,~7\right\}$\\
     Number of random samples & $8000$ \\
     Neurons in the input layer (input features, $\text{freq}_i$) & $10$\\
     Mesh (Neurons in the output layer without DBCs) &  $51 \times 51~(5202)$\\
     Act. func., hidden layers, learning rate, Epoch num. & Swish, [300, 300], 0.001, 1000 \\
    \hline\\    
    \end{tabular}
    \label{Table:Fourier_param}
\end{table}

\begin{figure}[H]
  \centering
  \begin{subfigure}[b]{0.9\linewidth}
    \centering
    \includegraphics[width=\linewidth]{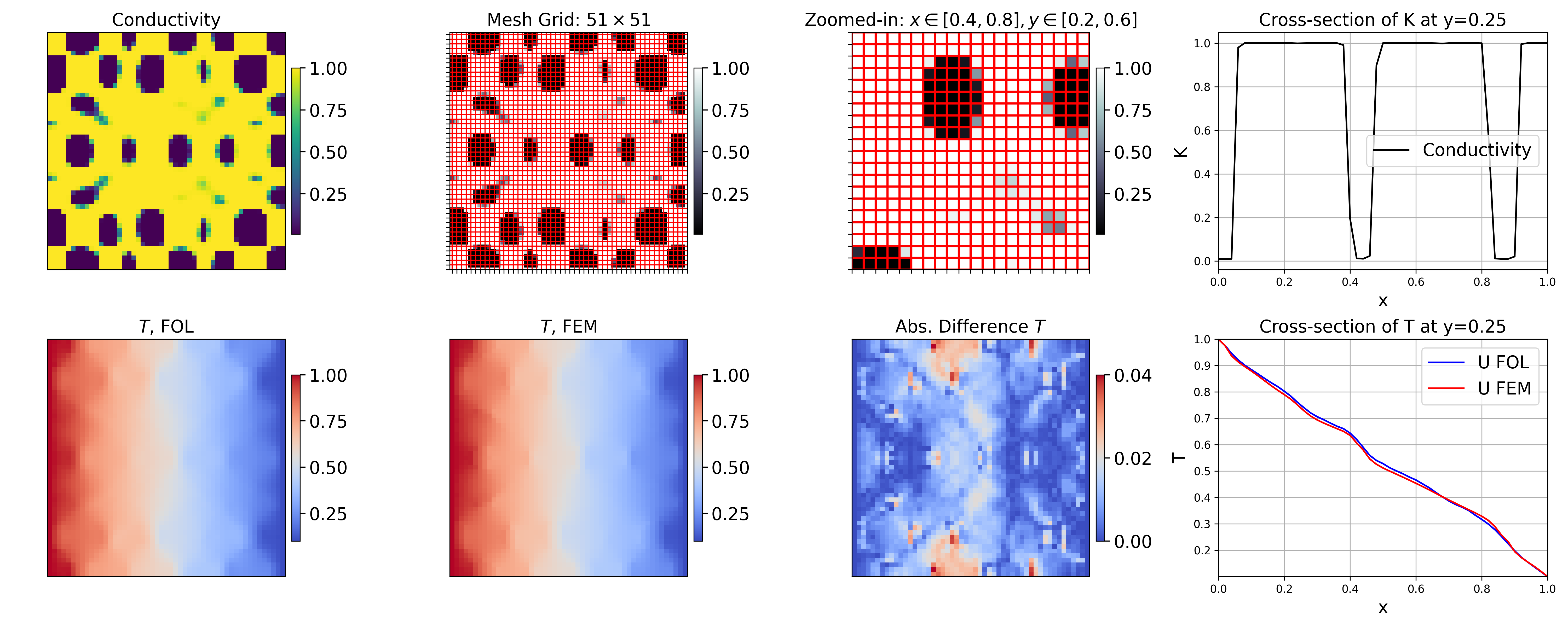} 
    \caption{Comparison of the temperature profile between FOL and FEM.}
    \label{fig:66_plot_mesh_vec_data}
  \end{subfigure}
  \hfill
  \begin{subfigure}[b]{0.9\linewidth}
    \centering
    \includegraphics[width=\linewidth]{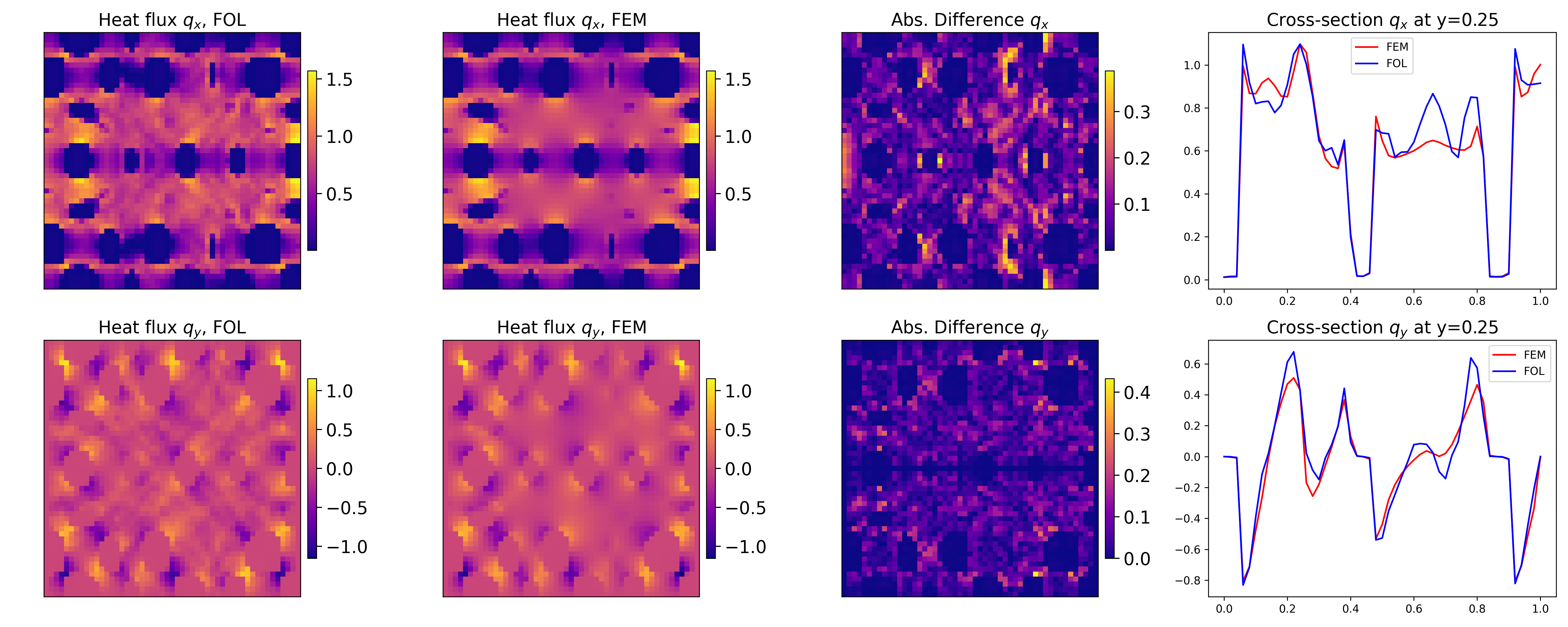}
    \caption{Comparison of heat flux profiles between FOL and FEM.}
    \label{fig:66_plot_mesh_vec_grad_data}
  \end{subfigure}
  \hfill
  \begin{subfigure}[b]{0.7\linewidth}
    \centering
    \includegraphics[width=\linewidth]{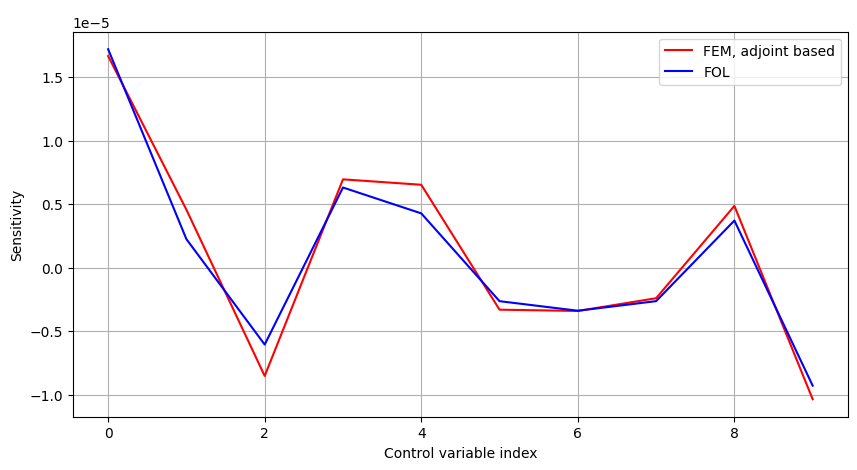}
    \caption{Comparison of FOL-based and FEM-based sensitivities for the given microstructure.}
    \label{fig:66_Qx_i2_sens_comparisons}
  \end{subfigure}
  \caption{Results of pre-trained FOL for a random unseen microstructure using $\left[5.3, 6.0, 7.7, 5.1, 5.1, 6.8, 5.5, 8.3, 8.1, 7.5 \right]$.}
  \label{fig:forward_results}
\end{figure}

The results reported in Figs.~\ref{fig:66_plot_mesh_vec_data}, \ref{fig:66_plot_mesh_vec_grad_data}, and \ref{fig:66_Qx_i2_sens_comparisons}, show an acceptable agreement between the two methods for forward temperature profile, obtained heat fluxes and sensitivities, respectively. All these results are obtained for the $51 \times 51$ and no interpolation or any other additional post-processing step is done.
Again, we noticed that the results for the spatial derivatives of the temperature field show more fluctuations as the network is solely trained to obtain the temperature field. Nevertheless, the errors for the heat flux components $q_x$ and $q_y$ seem to be acceptable, especially when considering the averaged values (see also the results in Fig.~\ref{fig:err_dd} and relative quantitative comparisons).

\begin{figure}[H] 
  \centering
  \includegraphics[width=0.90\linewidth]{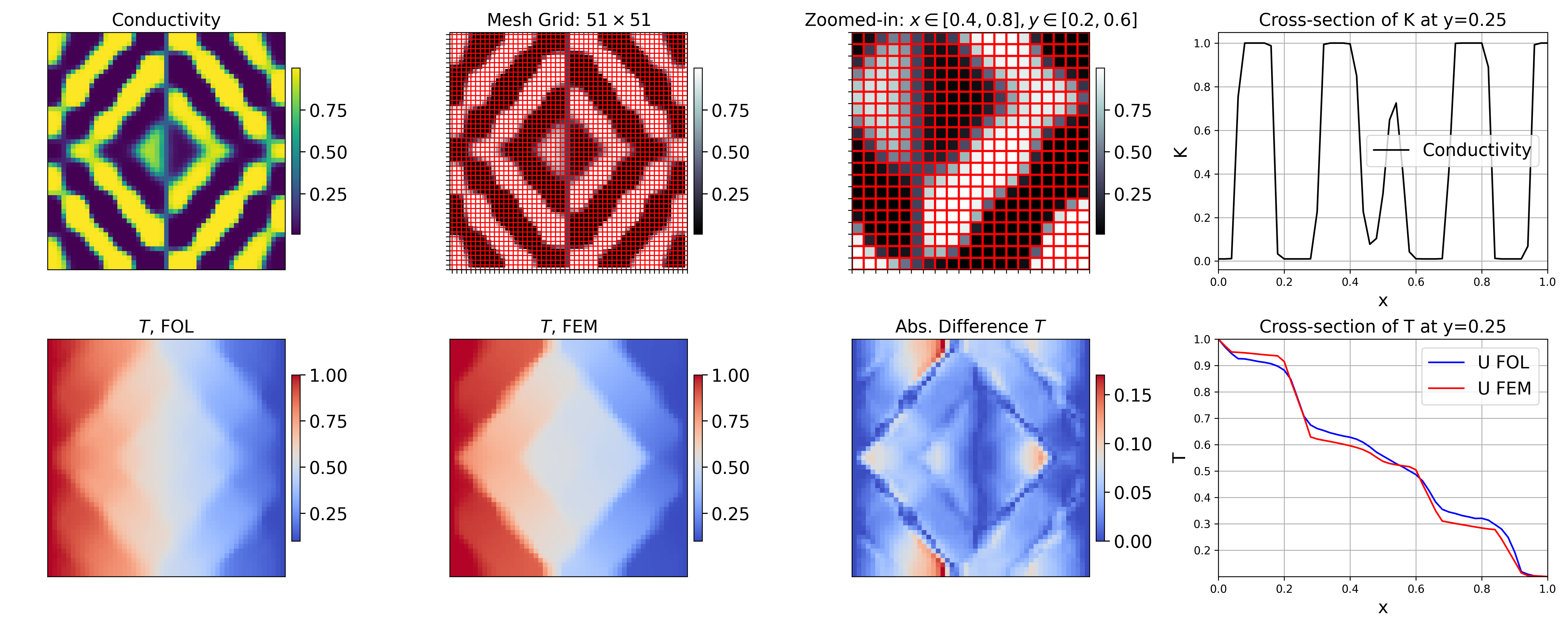}
  \caption{ Comparison of the deformation profile between FOL and FEM for a random microstructure using $\left[0.7, -0.5, -0.0, 0.3, 0.9, 1.6, -0.2, 0.9, -0.3, -1.3 \right]$. }
  \label{fig:73_plot_mesh_vec_data}
\end{figure}

\begin{figure}[H] 
  \centering
  \includegraphics[width=0.90\linewidth]{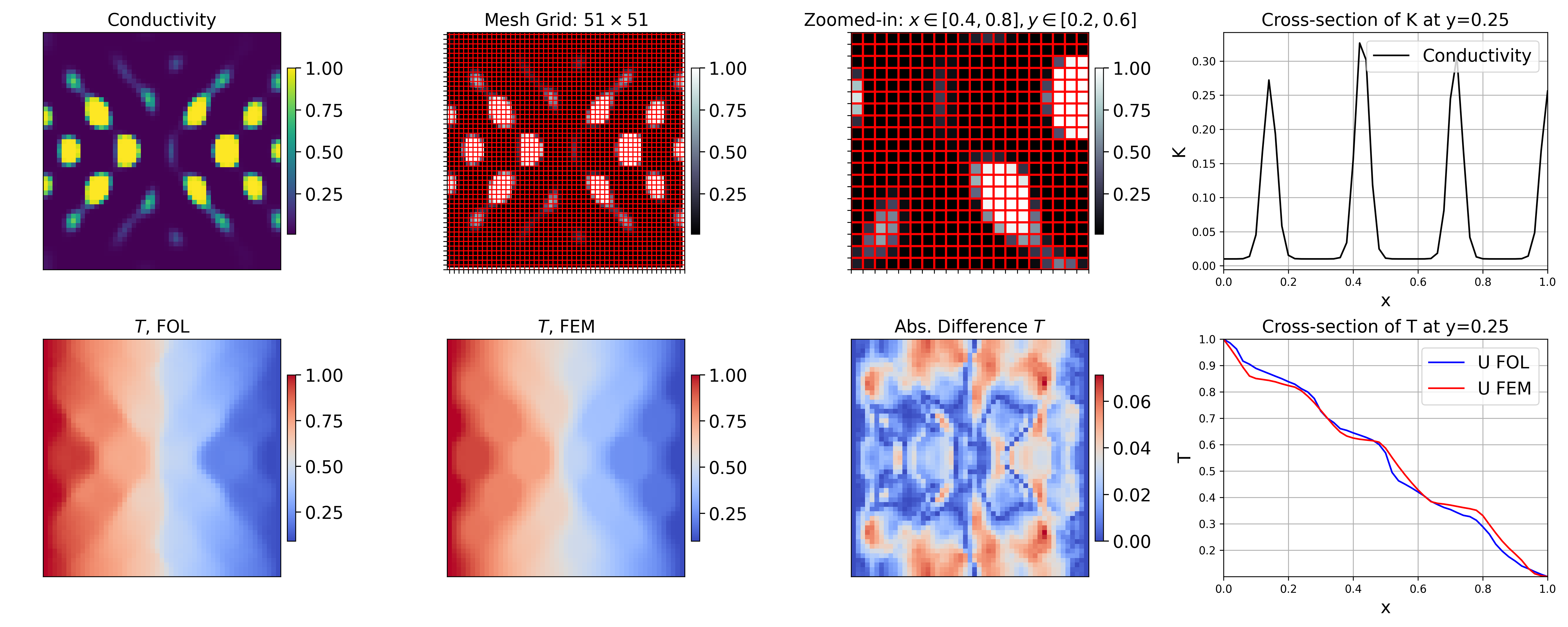}
  \caption{ Comparison of the deformation profile between FOL and FEM for a random microstructure using $\left[-1.7, 0.7, -0.8, 0.6, 0.3, 0.5, -0.8, -0.9, 1.8, -0.6 \right]$. }
  \label{fig:28_plot_mesh_vec_data}
\end{figure}

In Figs.~\ref{fig:73_plot_mesh_vec_data} and \ref{fig:28_plot_mesh_vec_data} further unseen test cases are shown for the temperature predictions. Note that in these selected samples, we have extreme conditions of isolation within the microstructure, and therefore, a very sharp solution is expected and observed for the temperature field. Despite the challenging nature of these two test cases, the predictions of the FOL after just one training session seem to agree well with FEM.


\begin{figure}[H] 
  \centering
  \includegraphics[width=0.99\linewidth]{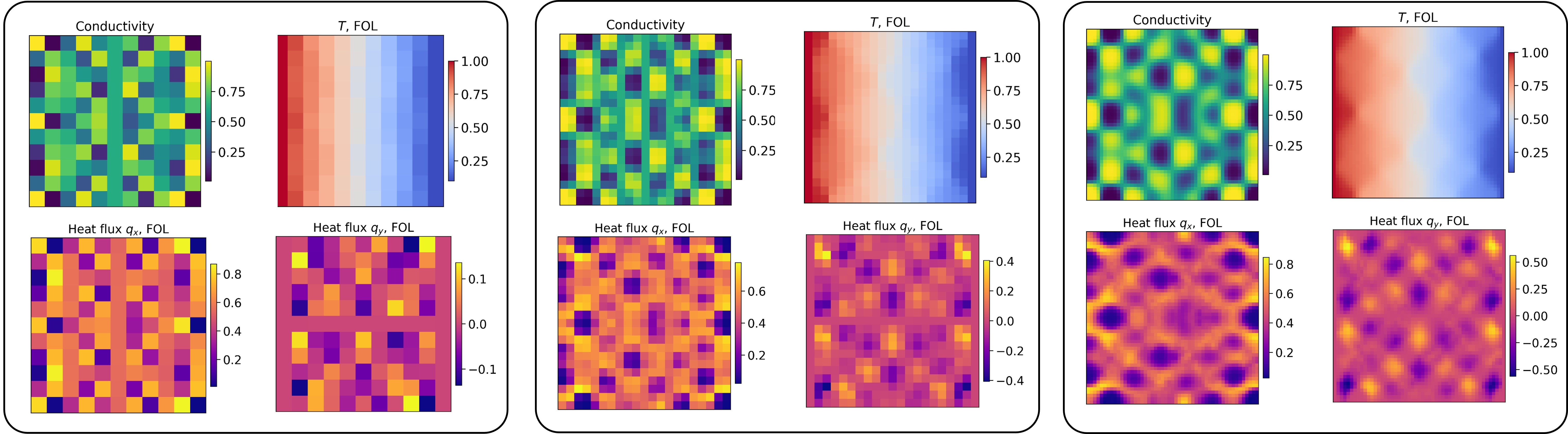}
  \caption{ Mesh convergence analysis for a random microstructure using $\left[-3.6, 0.8, 0.5, 2.0, 3.8, 0.0, -0.8, 2.6, 0.3, -0.3 \right]$. }
  \label{fig:mesh}
\end{figure}

\begin{figure}[ht]
    \centering
    
    \begin{subfigure}[b]{0.41\textwidth}
        \centering
        \includegraphics[width=\textwidth]{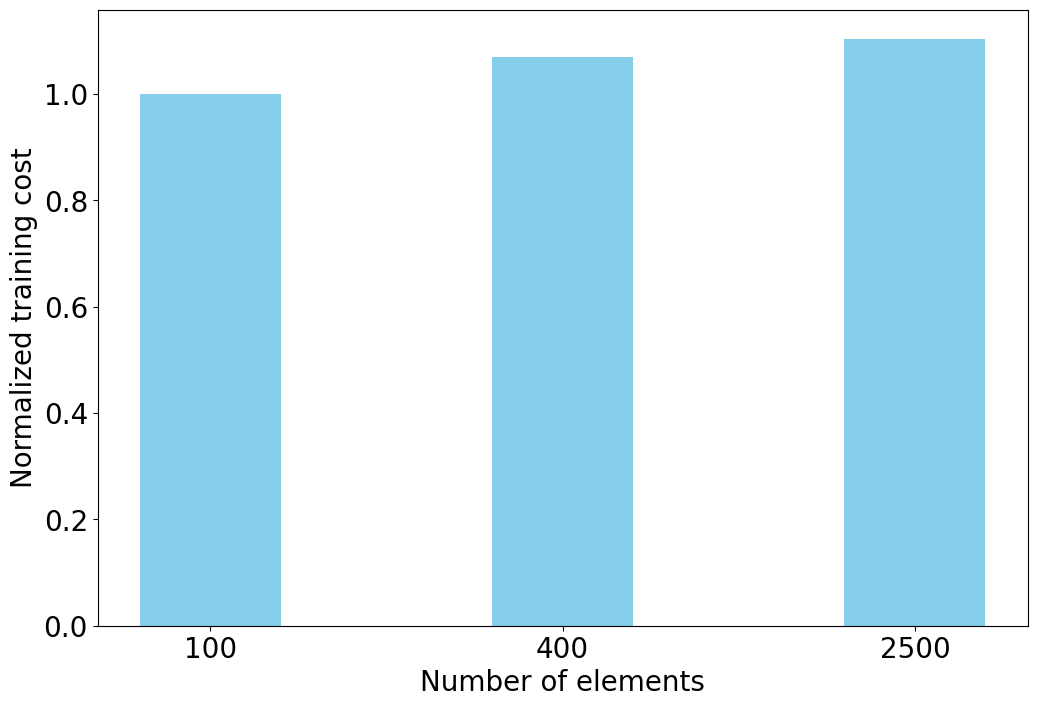}
        \caption{Training cost by increasing the element numbers.}
        \label{fig:cost_comp_1}
    \end{subfigure}
    \hfill
    \begin{subfigure}[b]{0.45\textwidth}
        \centering
        \includegraphics[width=\textwidth]{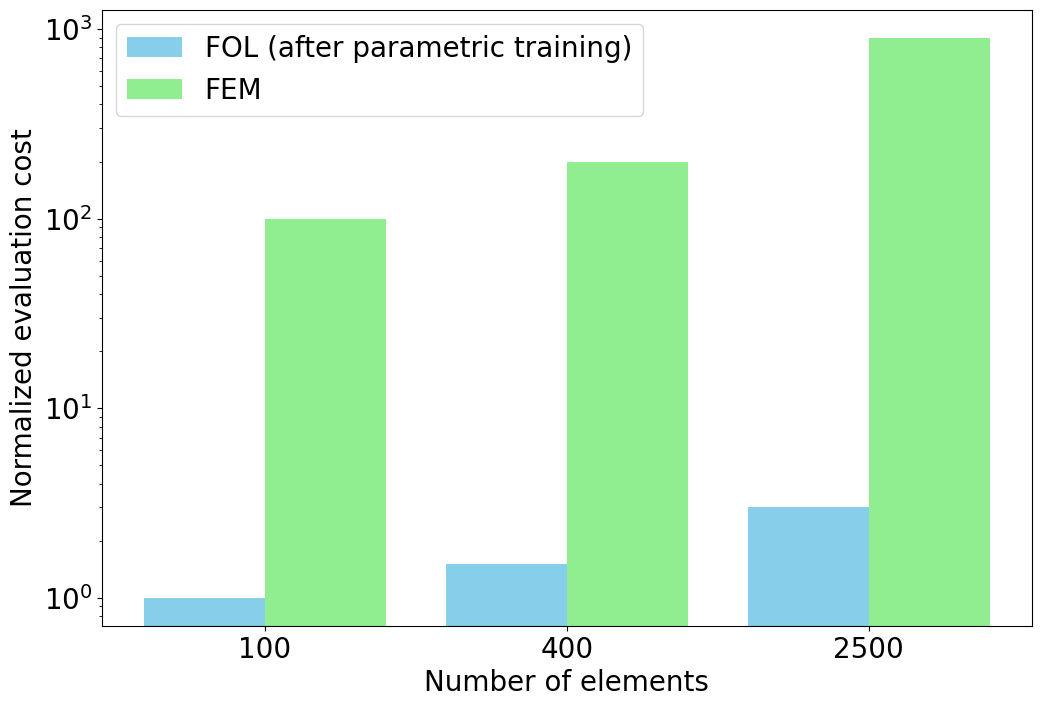}
        \caption{Evaluation cost for different element numbers.}
        \label{fig:cost_comp_2}
    \end{subfigure}
    \caption{Comparison of the training and evaluation costs for FOL and FEM. }
    \label{fig:cost_comp}
\end{figure}

Next, we check for mesh convergence with different choices of $N$ as shown in Fig.~\ref{fig:mesh}. To ensure a fair comparison, we train the same network described in Table \ref{Table:Fourier_param} with just a different number of neurons in the output layer, which is directly determined based on the mesh resolution.
Interestingly, the training and evaluation times (or costs) remain almost unchanged. Moreover, we observe a significant reduction in evaluation cost compared to the FEM as we go to higher mesh densities. See also the results provided in Fig.~\ref{fig:cost_comp} for the normalized training and evaluation costs. We confirm a clear convergence for the temperature profile using denser meshes. For brevity, we recorded the following temperature values for the mesh densities shown in Fig.~\ref{fig:mesh} at point $x_P=0.6$ and $y_P=0.25$, we have $T^{11\times 11}_P=0.455$, $T^{21\times 21}_P=0.460$, and $T^{51\times 51}_P=0.461$.
\textcolor{black}{Generally, the higher the frequency and phase contrast, the greater the number of elements needed. For the current work and chosen parameters, a mesh of $50 \times 50$ was sufficient according to our studies.}

\begin{figure}[ht]
    \centering
    
    \begin{subfigure}[b]{0.41\textwidth}
        \centering
        \includegraphics[width=\textwidth]{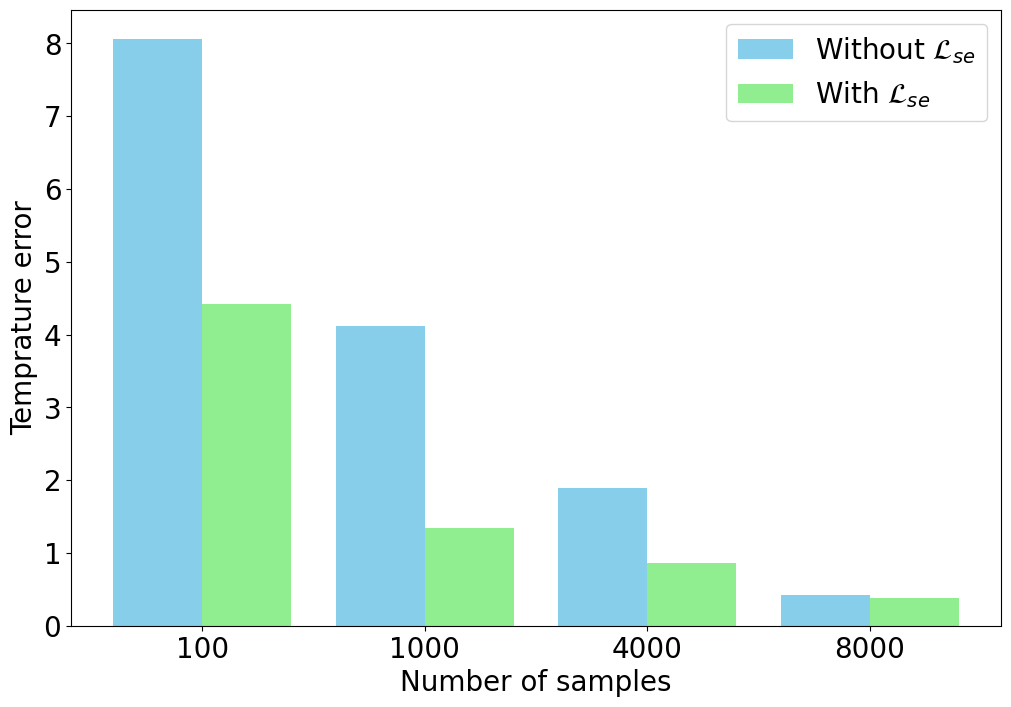}
        \caption{Convergence of the temperature values with respect to the number of samples.}
        \label{fig:sub1}
    \end{subfigure}
    \hfill
    \begin{subfigure}[b]{0.45\textwidth}
        \centering
        \includegraphics[width=\textwidth]{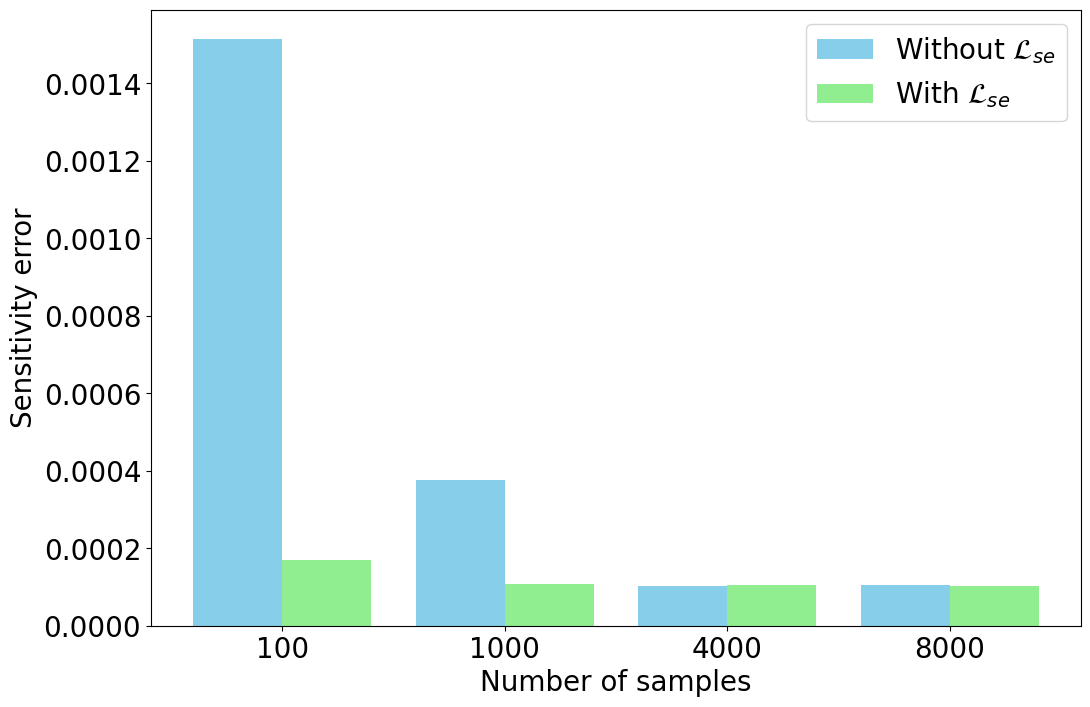}
        \caption{Convergence of the sensitivity values with respect to the number of samples.}
        \label{fig:sub2}
    \end{subfigure}
    \caption{Error analysis for the physics-informed parametric learning. The Sobolev training, or adding an additional loss for sensitivities, significantly improved the \textcolor{black}{mean absolute errors} in the test cases for both the forward prediction of the temperature field and the sensitivity values. }
    \label{fig:conv_sam}
\end{figure}

Finally, we investigate the influence of the additional sensitivity loss or the Sobolev training on the prediction accuracy of the trained network for a set of unseen test cases. To gain better insight, we also varied the number of initial sample fields for each comparison.
In Fig.~\ref{fig:conv_sam} we show the error reduction by increasing the number of samples for training. Furthermore, by adding the sensitivity loss term into the training, we managed to further reduce the average error observed in the test cases.

\color{black}
\subsection{Influence of the network architecture}
\label{sec:vs}

We should emphasize that the central goal of this work is to motivate training neural operators—regardless of their specific architecture—using established numerical methods. 
In this study, we mainly focus on a simple feed-forward network for clarity. However, the proposed framework is fully general: any high-performing neural architecture can, in principle, be trained using finite-element (or other numerical methods) residuals, and we expect this strategy to be more stable, more efficient, and more accurate than classical approaches based solely on automatic differentiation. 
See also recent studies on integrating FNO with FFT-based solvers for micromechanics \cite{HARANDI2025106219}.

\begin{figure}[H]
  \centering
  \includegraphics[width=0.99\linewidth]{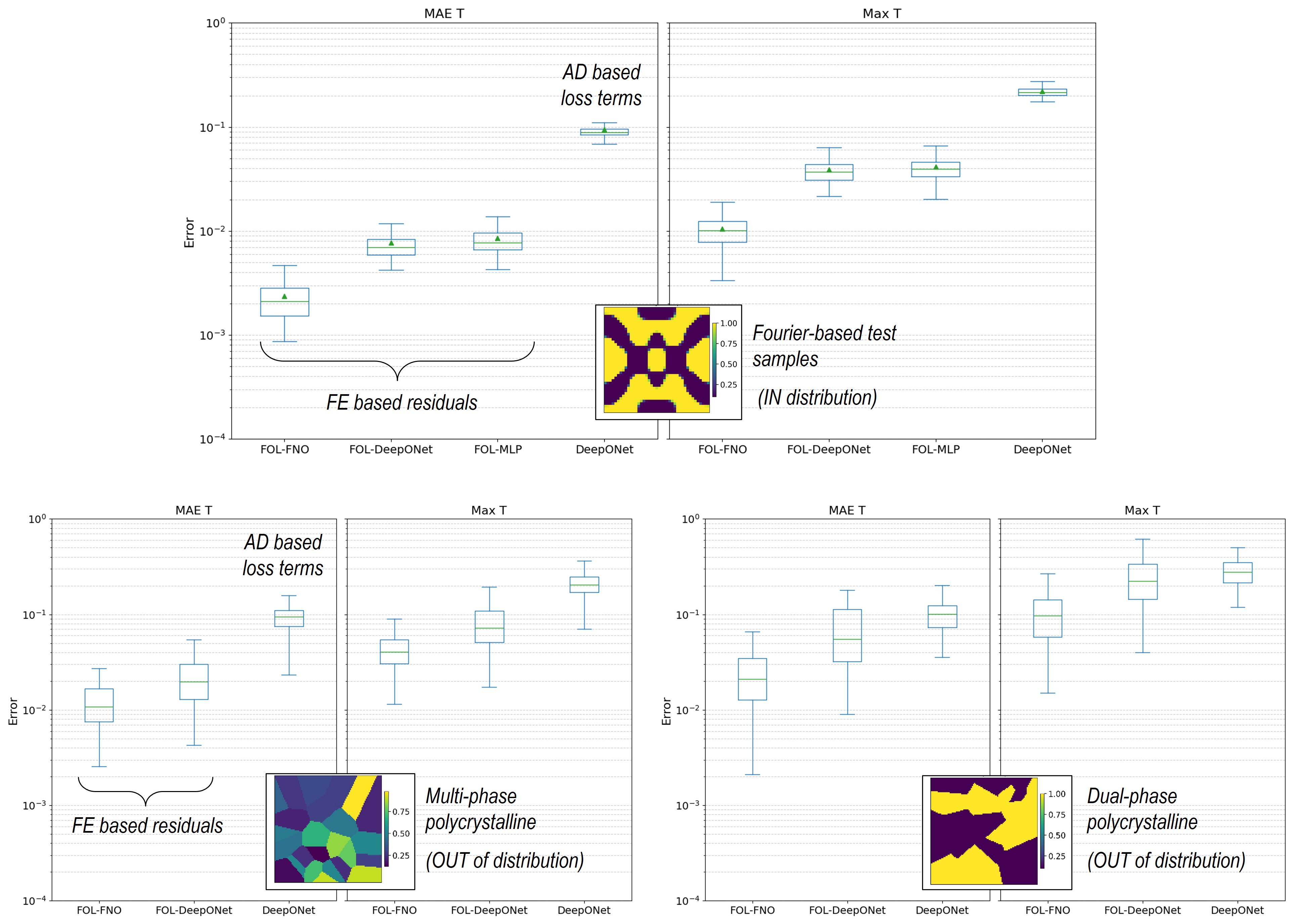}
  \caption{ Comparison of FE based and AD based loss terms for different architectures for IN and OUT of distribution topologies.}
  \label{fig:vs}
\end{figure}

\begin{figure}[H]
  \centering
  \includegraphics[width=0.9\linewidth]{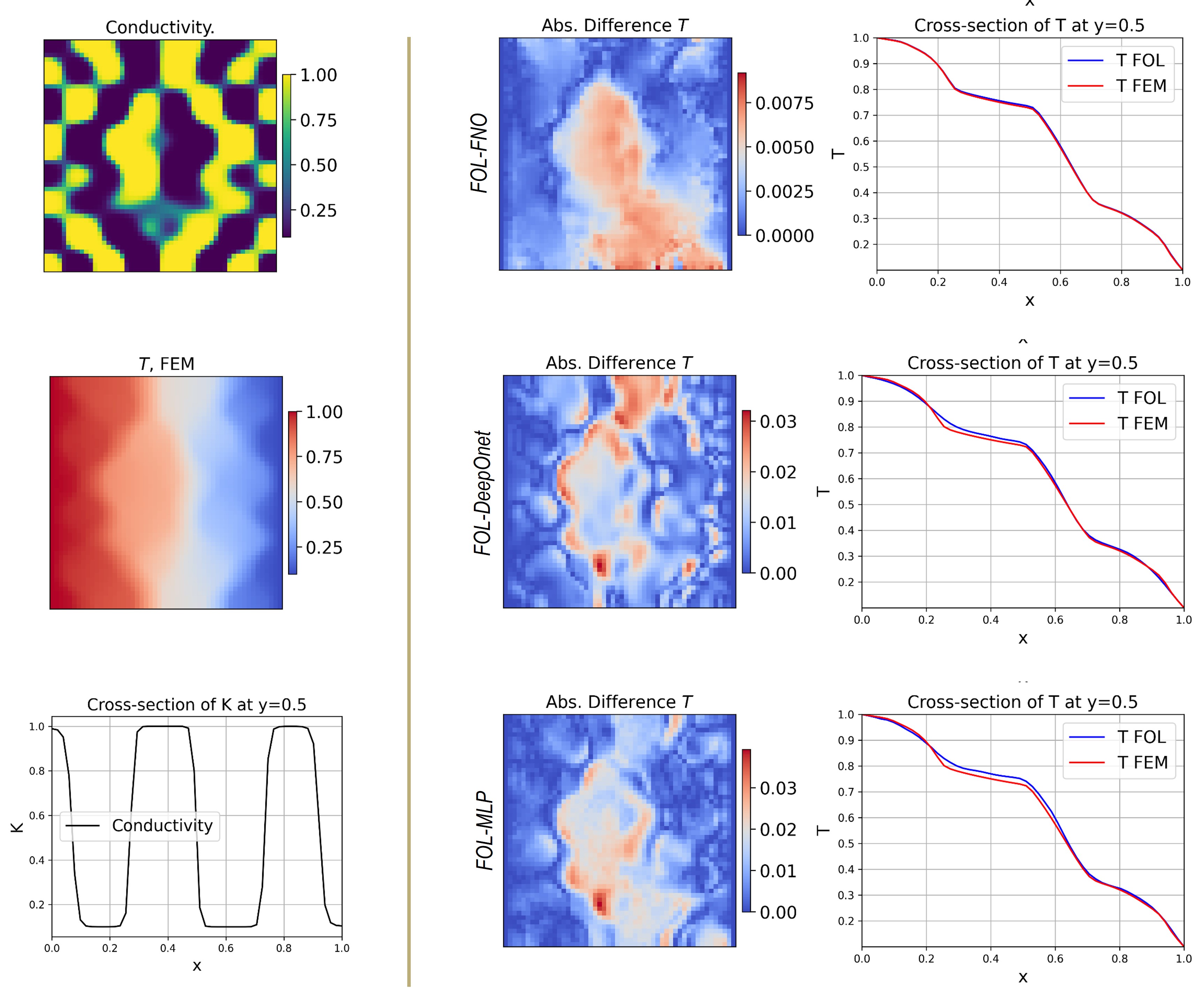}
  \caption{ Comparison of FE-based loss performance across different architectures for both in-distribution and out-of-distribution topologies.}
  \label{fig:vs2}
\end{figure}

To examine this claim, we carried out a direct comparison between the FE-residual training strategy and the standard AD-based physics-informed formulation. 
The results of this comparison are summarized in the Figures \ref{fig:vs} and \ref{fig:vs2}.

Our preliminary observations showed that the baseline physics-informed AD-based DeepONet performs poorly. 
The main issue is constructing physics-based losses entirely via automatic differentiation, which often leads to lower accuracy and instability for problems with sharp or high-frequency features. 
AD-based approach typically leads to trivial solutions dominated by boundary conditions, reflecting only the lowest-frequency modes of the field. Therefore, it seems that additional modifications are always necessary, as discussed in \cite{REZAEI2022PINN}.
For the PI-AD comparison, we focus primarily on DeepONet, since it can more naturally incorporate spatial derivatives. 
The errors (in terms of MAE and maximum pointwise error) are reported statistically over 200 test samples for each case. 
The test set includes both in-distribution samples—generated using similar but unseen Fourier-based parameterizations—and out-of-distribution cases with completely different topologies, including dual-phase and multiphase polycrystalline microstructures. 
Overall, we confirm that the FE-based residual formulation consistently outperforms the AD-based version, independent of the network architecture. 
Interestingly, for the selected square geometry, the FE-based FNO (FOL-FNO) performs best, while DeepONet and a simple MLP exhibit similar performance.

Note that, due to the inherent limitations of MLP-based models, the inference stage cannot be generalized to unseen parameterization strategies or resolutions beyond those used during training. This restriction does not apply to the FNO- and DeepONet-based formulations within the FOL framework, both of which naturally support flexible parameterization and resolution changes.

Although this may appear to be a drawback of the MLP approach, our results indicate that—depending on the application—such a simple architecture can still be sufficiently accurate for a specific class of problems, while avoiding unnecessarily complex architectures and excessive parameterization of the network.

\color{black}

\newpage
\subsection{Parametric learning on source term}
\label{sec:source_prop}
We now eximine the second chosen boundary value problem shown in the right hand side of Fig.~\ref{fig:examples}. 
Here, we intend to learn the temperature distribution as a function of a given position-dependent heat sink introduced into the domain. 
In this example, the thermal conductivity remains constant and equal to $1$ throughout the training. 
Moreover, we have changed the boundary conditions to equal temperature on all the edges. The boundary conditions are strictly enforced, thanks to the FOL framework, where we use $T_b = 1.0~$K. 
The results are based on the second formulation of the FOL framework using Fourier parameterization. 
The summary of the samples generated for this case, as well as the hyperparameters of the network, are reported in \ref{sec:app_sample} and Table \ref{Table:Fourier_param_Q}, respectively.
For these results, only the first loss term in Eq.~\ref{eq:total_loss_sum} remains active, and for simplicity, the sensitivity loss is turned off.

\begin{table}[H]
\centering
\caption{ \textcolor{black}{Summary of the network parameters for learning based on source term distribution using the Fourier-based FOL.} }
 \begin{tabular}{ll} 
    \hline
        Parameter &   Value\\
    \hline
     Inputs, Outputs                  &  $\{\text{freq}_i\}$, $\{T_i\}$ \\ 
     Frequency in the $x$- direction ${f_{x, i}}$ & $\left\{1,~2,~3\right\}$\\
     Frequency in the $y$- direction ${f_{y, i}}$ & $\left\{1,~2,~3\right\}$\\
     Number of random samples & $1000$ \\
     Neurons in the input layer (input features, $\text{freq}_i$) & $10$\\
     Mesh (Neurons in the output layer without DBCs) &  $51 \times 51~(5202)$\\
     Act. func., hidden layers, learning rate, Epoch num. & Swish, [100, 100], 0.001, 1000 \\
    \hline\\    
    \end{tabular}
    \label{Table:Fourier_param_Q}
\end{table}

\begin{figure}[H] 
  \centering
  \includegraphics[width=0.90\linewidth]{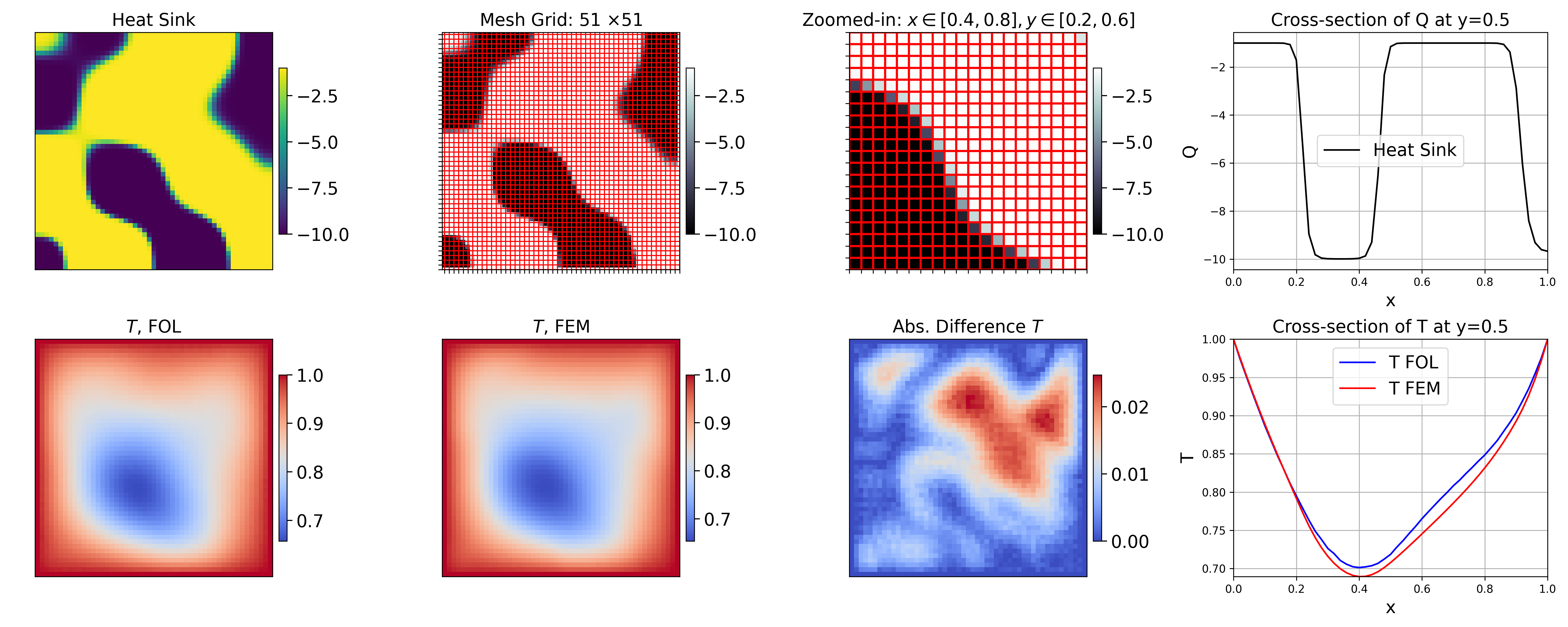}
  \caption{ \textcolor{black}{Comparison of the temperature profile between FOL and FEM for a random unseen heat sink.} }
  \label{fig:sample_Q_1}
\end{figure}

\begin{figure}[H] 
  \centering
  \includegraphics[width=0.90\linewidth]{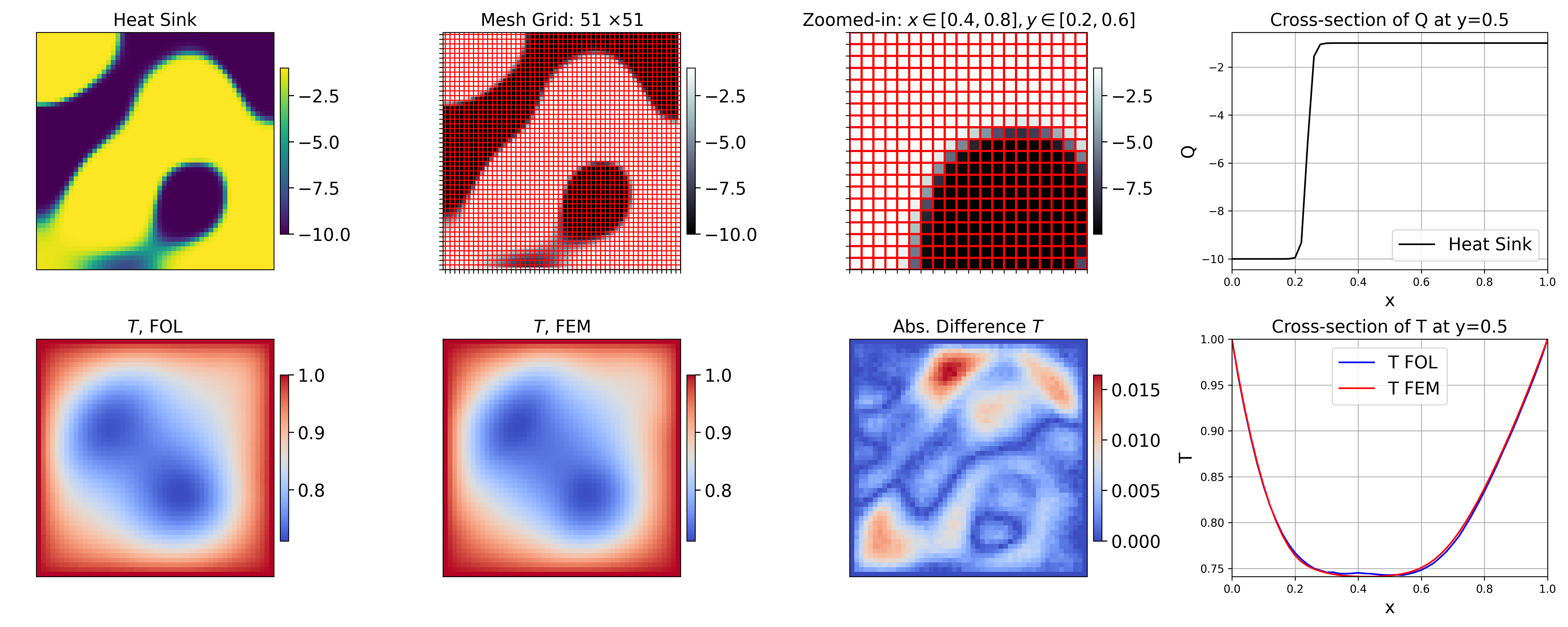}
  \caption{ \textcolor{black}{Comparison of the temperature profile between FOL and FEM for a random unseen heat sink.} }
  \label{fig:sample_Q_2}
\end{figure}

\begin{figure}[H] 
  \centering
  \includegraphics[width=0.90\linewidth]{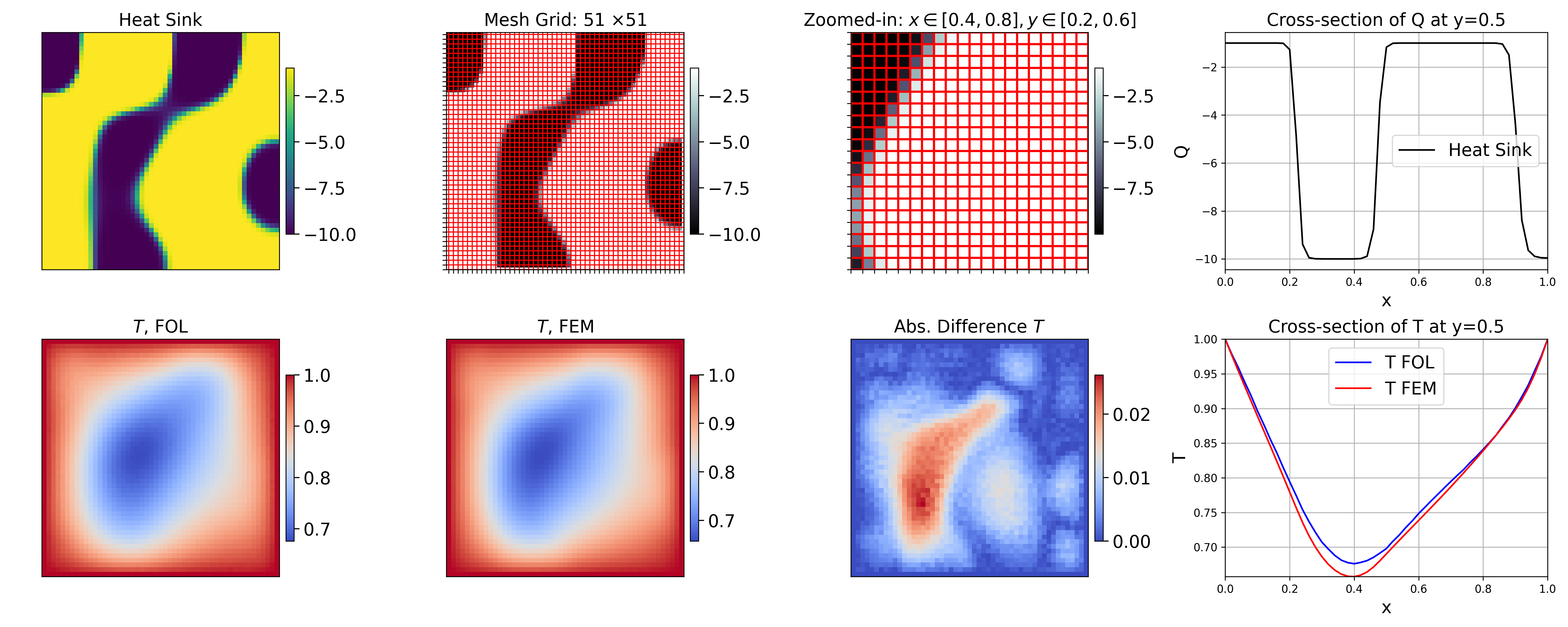}
  \caption{ \textcolor{black}{Comparison of the temperature profile between FOL and FEM for a random unseen heat sink.} }
  \label{fig:sample_Q_3}
\end{figure}

The results reported in Figs.~\ref{fig:sample_Q_1}, \ref{fig:sample_Q_2}, and \ref{fig:sample_Q_3}, show an acceptable agreement between the FOL and FEM for temperature profile. All these results are obtained for the $51 \times 51$ and no interpolation or any other additional post-processing step is done.
We confirm very similar conclusions as before regarding the case of position-dependent thermal conductivity which are not repeated here.

\subsection{Parametric learning on boundary conditions and mechanical problem}
\label{sec:BCs_prop}
Finally, we applied the methodology to different physics, namely mechanical equilibrium, as described in \cite{REZAEI2022PINN}.
The boundary value problems are described in Fig.~\ref{fig:examples_2}. 
Here, we intend to learn the components of displacement field as a function of the applied Dirichlet boundary conditions on a plane. 
In this example, the elasticity and possion ratio remains constant and equal to $1$ and $0.2$ throughout the training.   
The summary of the samples generated for this case, as well as the hyperparameters of the network, are reported in \ref{sec:app_sample} and Table \ref{Table:param_bcs}, respectively.
For these results, only the first loss term in Eq.~\ref{eq:total_loss_sum} remains active, and for simplicity, the sensitivity loss is turned off.

\begin{table}[H]
  \centering
  \caption{ \textcolor{black}{Summary of the network parameters for learning based on boundary conditions.} }
   \begin{tabular}{ll} 
      \hline
          Parameter &   Value\\
      \hline
       Inputs, Outputs                  &  $\{\text{u}_{1,i},\text{u}_{2,i},\text{u}_{3,i}\}$, $\{U_{x},U_{y},U_{z}\}$ \\ 
       Number of random samples & $1000$ \\
       Mesh (Neurons in the output layer) &  $1059\times3$ (beam), $2524\times3$ (S-shape)\\
       Act. func., hidden layers, learning rate, Epoch num. & tanh, [10, 10], 0.001, 1000 \\
      \hline\\    
      \end{tabular}
      \label{Table:param_bcs}
\end{table}

\begin{figure}[H] 
  \centering
  \includegraphics[width=0.99\linewidth]{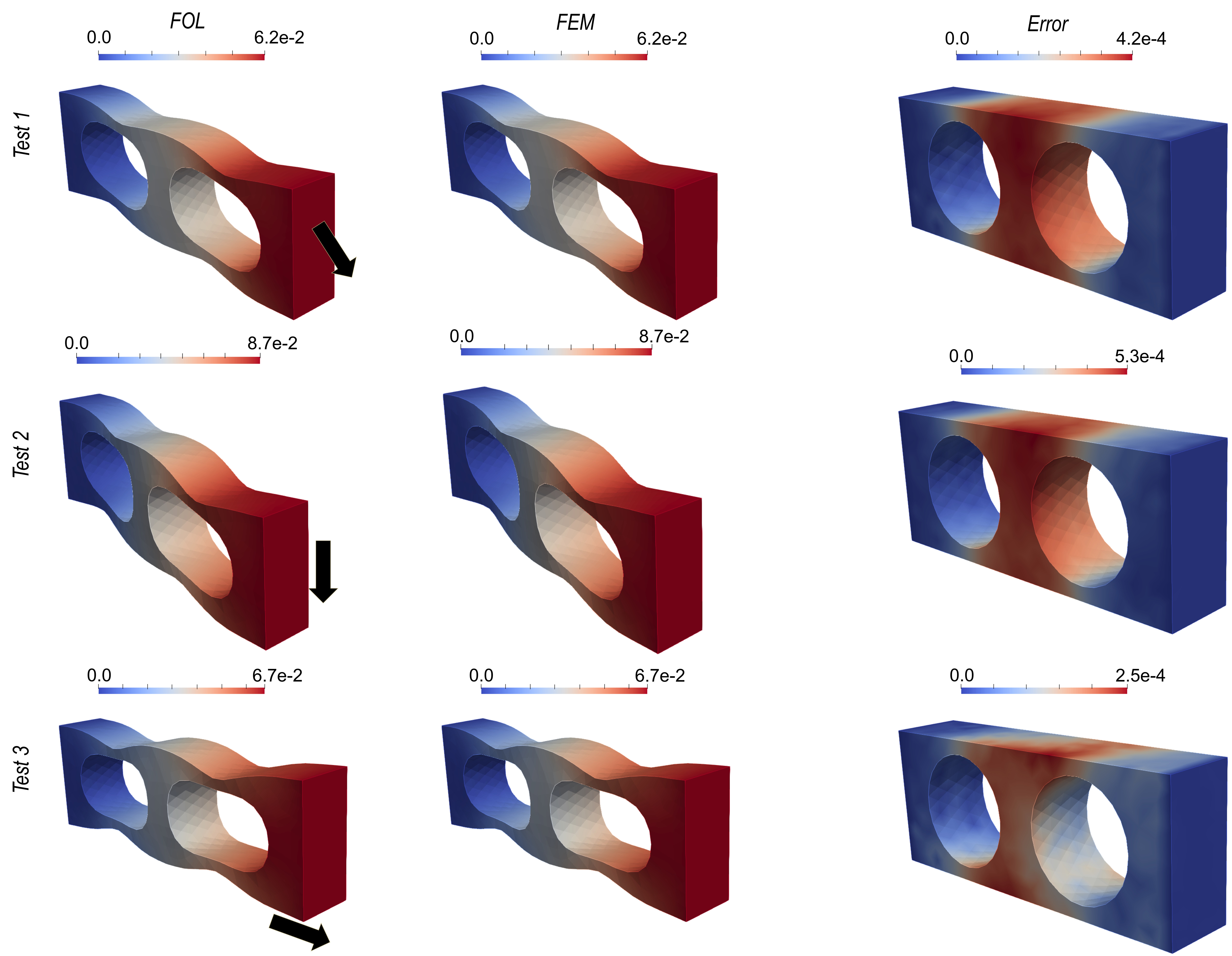}
  \caption{ \textcolor{black}{Comparison of the magnitude of the obtained displacement vector between FOL and FEM for a randomly applied unseen boundary condition.} }
  \label{fig:cube_mech}
\end{figure}

The results of the beam example with holes for three different unseen test cases are shown in Fig.~\ref{fig:cube_mech}. 
The network successfully learned deformation patterns under mixed mode of loadings. 
Thanks to the FOL architecture, the boundary conditions are satisfied in a hard way, which explains the almost zero error in the applied boundary regions. 
This example illustrates the effectiveness of the FOL framework in managing complex geometries with unstructured meshes and parametric learning tailored to the specified boundary conditions.

\begin{figure}[H] 
  \centering
  \includegraphics[width=0.99\linewidth]{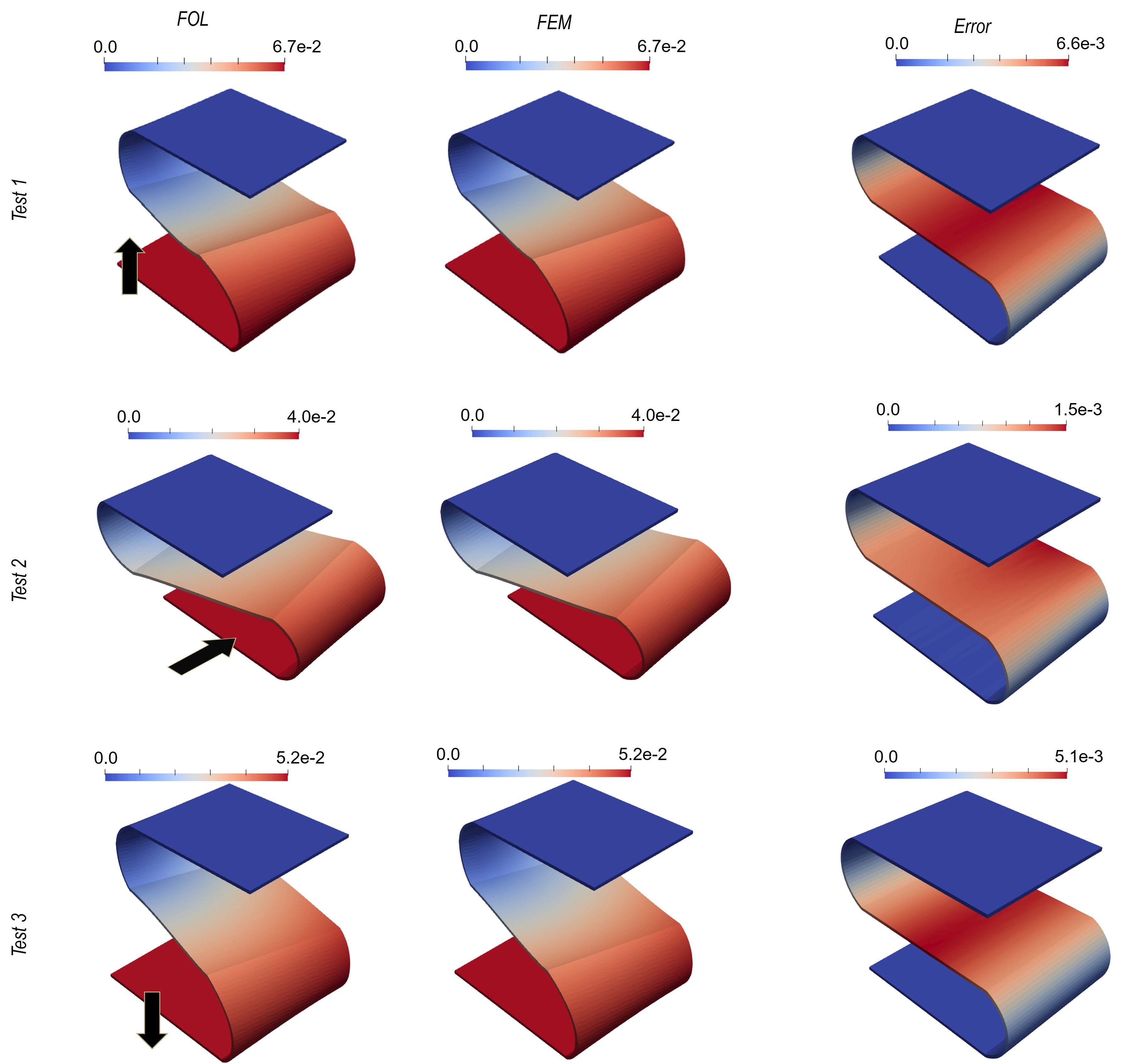}
  \caption{ \textcolor{black}{Comparison of the magnitude of the obtained displacement vector between FOL and FEM for a randomly applied unseen boundary condition.} }
  \label{fig:S_mech}
\end{figure}

\newpage
The results of the so-called 'S-shape' sample for three different unseen test cases are shown in Fig.~\ref{fig:S_mech}. 
As we can see, the network successfully learned different deformation patterns under pressure, tension, shear deformation, and even combinations of them. 
Note that in all cases, the boundary conditions are satisfied in a hard way, which explains the almost zero error in the applied boundary regions. The maximum error occurs in the areas with maximum deflection. 
This example demonstrates the capability of the FOL framework to handle complex geometries using unstructured mesh and parametric learning based on the given boundary conditions.

\color{black}

\newpage

\section{Results: Optimization and matrix-free solver}
\label{sec:opt_res}
\subsection{Optimization}
Encouraged by the strong performance of FOL in forward and sensitivity analysis, we test the methodology to perform gradient-based optimization. As previously discussed in the introduction of this section, our objective is to enhance heat transfer capabilities in the y-direction while constraining it in the x-direction. For this purpose, based on the mesh study results, a mesh resolution of $51\times51$ is chosen, see Fig.~\ref{fig:mesh}. Furthermore, we employ Fourier parametrization with frequencies in $x$- direction $\left\{5,7,9\right\}$ and in $y$- direction $\left\{4,6,8\right\}$. Starting from a uniform conductivity field of 0.5, with Fourier's weights set to zero except for the first, gradient-based optimization is performed iteratively until there is no significant change in the objective value. A hyperparameter study was conducted on the network's structure to determine the optimal number of layers and neurons. Relative errors in constraint values and gradients for different network architectures, both with and without considering the sensitivity loss, are reported in Table \ref{table:FOL_opt_NN_param_study}. These results underscore the importance of the sensitivity loss function in ensuring accurate solution-to-design sensitivities. The selected architecture for the FOL-based results in this Section is highlighted in Table \ref{table:FOL_opt_NN_param_study}. This architecture strikes a good balance between the network's density and its accuracy in both values and sensitivities.
\begin{table}[H]
\label{table:FOL_opt_NN_param_study}
\fontsize{7}{10}\selectfont
\centering
\caption{Errors (\%) in constraint values and gradients for different network architectures, both with and without considering the sensitivity loss. FEM-based values and adjoint-FEM-based gradients are used as references for error calculation. $N=51$. The case used for optimization is highlighted in red.}
\begin{adjustbox}{center}
\begin{tabular}{ l l l | l l l | l l l}
\hline
Layers  &  $w_{se}=1$  &  $w_{se}=0$ & Layers &  $w_{se}=1$  &  $w_{se}=0$ & Layers &  $w_{se}=1$  &  $w_{se}=0$  \\
\hline
$[N/3]$  & (0.96,12.39) & (1.20,218.75) & [$N/3$,$N/3$]  & (2.06,20.35) & (0.19,217.73)& [$N/3$,$N/3$,$N/3$]  & (2.11,24.43) & (0.84,229.57) \\ 
{\boldmath\underline{ $[N]$}} & \textbf{\underline{(1.61,8.39)}} & \textbf{\underline{(1.82,239.37)}} & [$N$,$N$]  & (2.59,8.84) & (0.99,180.90)& [$N$,$N$,$N$]  & (1.60,16.08) & (0.56,218.10) \\ 
$[10N]$  & (1.42,9.65) & (0.56,358.86) & [$10N$,$10N$]  & (3.64,6.77) & (1.41,590.13)& [$10N$,$10N$,$10N$]  & (0.91,6.76) & (1.44,167.49) \\
$[25N]$  & (3.40,10.58) & (2.47,1019.49) & [$25N$,$25N$]  & (3.63,7.08) & (1.59,252.81)& [$25N$,$25N$,$25N$]  & (4.47,5.69) & (1.96,136.96) \\
\hline
\label{table:FOL_opt_NN_param_study}
\end{tabular}
\end{adjustbox}{}
\end{table}

\begin{figure}[H]
  \centering
  \begin{subfigure}[b]{\linewidth}
    \centering
    \includegraphics[width=\linewidth]{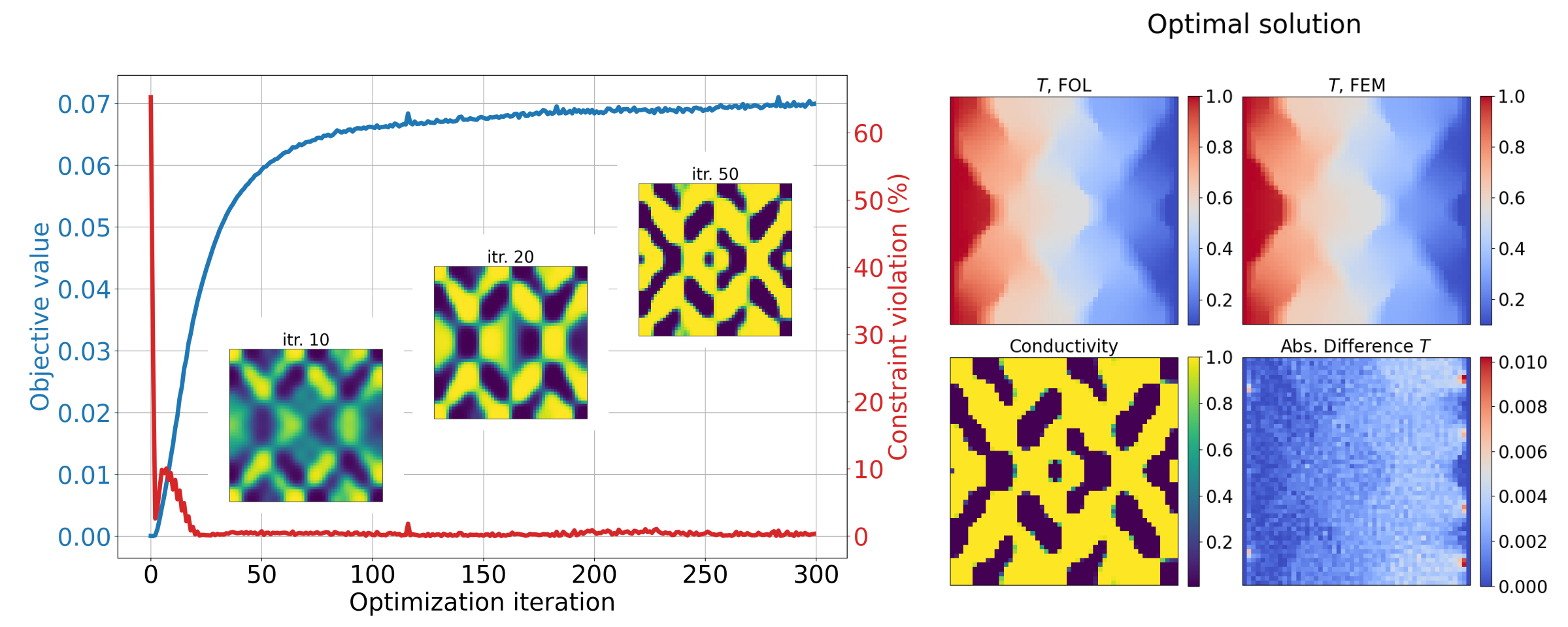} 
    \caption{FOL-driven optimization: temperature and sensitivity fields are obtained by training FOL's neural network at each optimization iteration, without any pertaining.}
    \label{fig:FOL_opt}
  \end{subfigure}
  \hfill
  \begin{subfigure}[b]{\linewidth}
    \centering
    \includegraphics[width=\linewidth]{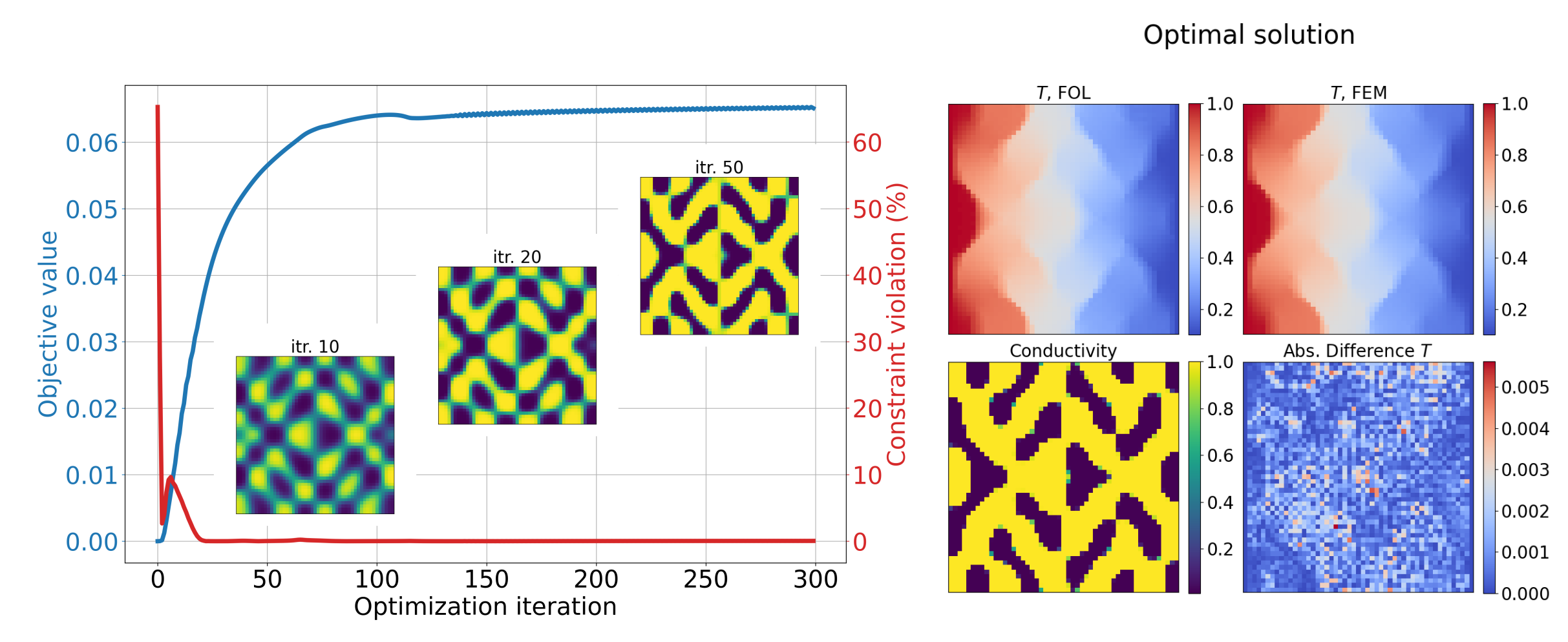}
    \caption{FEM-driven optimization: temperature and sensitivity fields are calculated using FEM and adjoint-FEM, respectively.}
    \label{fig:FEM_opt}
  \end{subfigure}
  \hfill
  \begin{subfigure}[b]{\linewidth}
    \centering
    \includegraphics[width=\linewidth]{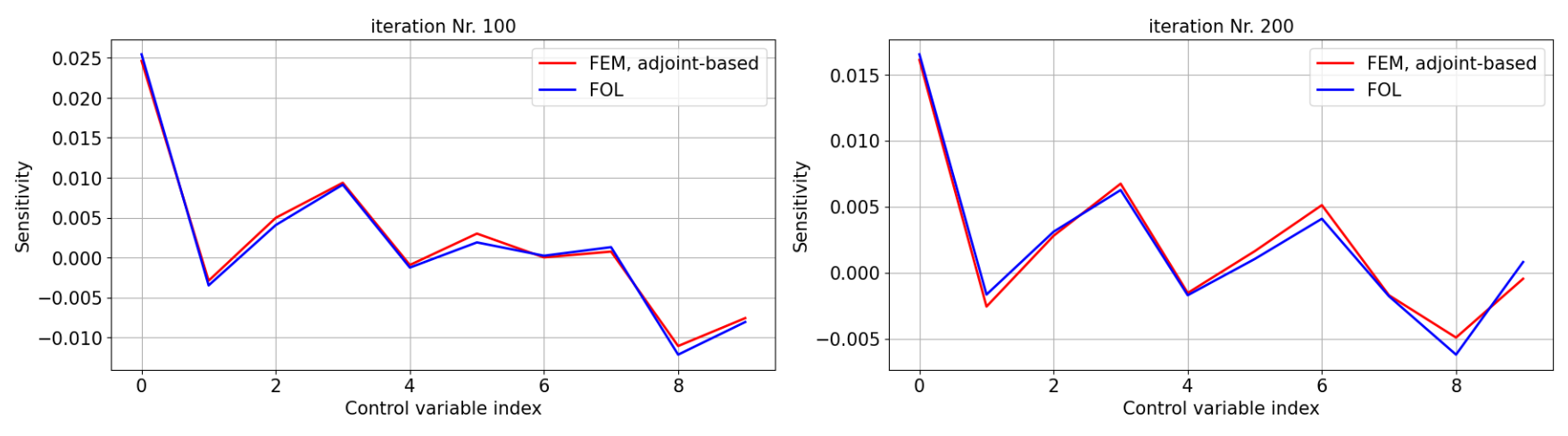}
    \caption{Comparison of FOL-based and FEM-based sensitivities at selected iterations of the FOL-driven optimization.}
    \label{fig:FOL_vs_FE_Sens}
  \end{subfigure}
  \caption{Results of gradient-based optimization using Fourier parametrization with $f_x=\left\{5,7,9\right\},f_y=\left\{4,6,8\right\}$.}
  \label{fig:opt_results}
\end{figure}

The results of maximizing $J$ (Eq. \ref{eq:qy}) while constraining $h$ (Eq. \ref{eq:qx}) to 0.0 are shown in Fig.~\ref{fig:opt_results}. Two cases are considered: a) FOL-driven optimization, where the primal and sensitivity analyses are conducted simultaneously by minimizing the FOL's total loss function (Eq. \ref{eq:total_loss_sum}) at each optimization iteration, and b) FEM-driven optimization, where the primal and sensitivity analyses are performed sequentially using FEM and adjoint FEM, respectively. While both optimal solutions satisfy the constraint, the FOL-driven optimization achieves approximately 7.5\% more improvement in the objective function. This additional improvement accounts for the differences observed in the optimal conductivity and temperature fields. 

When comparing the convergence histories in Figs.~\ref{fig:FOL_opt},~\ref{fig:FEM_opt}, fluctuations are observed in the FOL-driven optimization, which may be attributed to possible errors in the temperature field, sensitivities, or both. Motivated by this observation, in Fig.~\ref{fig:FOL_vs_FE_Sens}, we compare the FOL-based and adjoint-based sensitivities of the constraint function $h$ at certain optimization iterations. Both quantitatively and qualitatively, the results are in very close agreement. The minor deviations can be attributed to local errors in the FOL-based temperature field compared to the reference. These local errors directly lead to discrepancies in the sensitivities. Furthermore, the observed fluctuations can be attributed to the fixed number of training epochs and may also indicate the challenge of balancing loss functions and tuning the network's hyperparameters.

\begin{figure}[H] 
  \centering
  \includegraphics[width=0.7\linewidth]{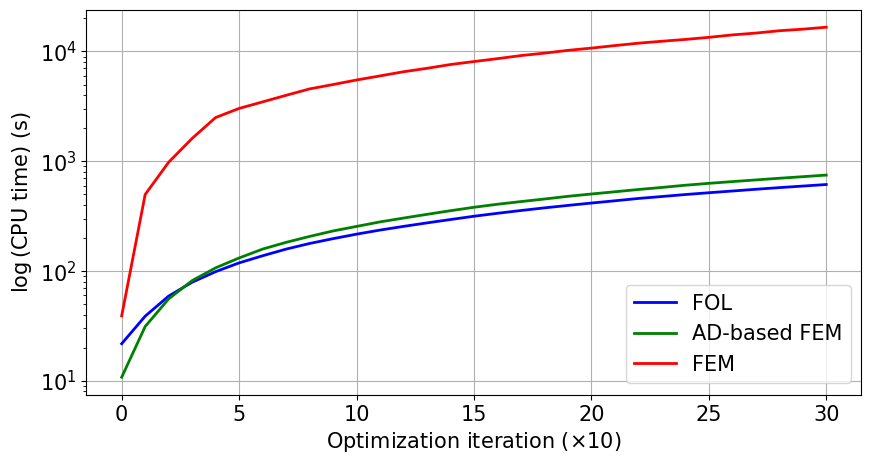}
  \caption{History of the total computational time for FOL-driven and FEM-driven optimizations. AD-based FEM computes residual and stiffness matrices using the automatic differentiation capabilities of JAX.}
  \label{fig:FOL_vs_FE_OPT_time}
\end{figure}

Finally, we compare the computational performance of the FOL-driven and FEM-driven optimizations, as shown in Fig.~\ref{fig:FOL_vs_FE_OPT_time}. 
\textcolor{black}{It should be noted that both FOL and FEM implementations utilize the JAX framework in Python. While the traditional FEM is implemented by computing element stiffness matrices and assembling them into the global matrix, AD-based FEM leverages automatic differentiation to derive the residuals and stiffness matrix.} 
The first observation is that FOL-based optimization outperforms traditional FEM and AD-FEM by factors of 27.09 and 1.2, respectively. However, it should also be mentioned that the linear solves in FEM computations are performed using the solve function from the linear algebra library of JAX. Another important point to consider is that the computational cost of FOL-based sensitivity analysis is completely independent of the number of response functions, whereas the cost of FEM-based sensitivity analysis scales directly with it.

\color{black}
\subsection{FOL as matrix-free PDE solver}
\label{sec:3D}
As discussed in the previous section, when the batch size is set to one and only one sample is provided for training, the proposed methodology reproduces FEM-based results. However, unlike Newton-based solvers, FOL eliminates the need to construct and invert the tangent matrix of the underlying discretized PDE. Instead, the tangent matrix of the neural network is used to iteratively update the predicted solution during the backward propagation process. This occurs without explicitly constructing the matrix, which has two main advantages. First, the required memory is significantly reduced, especially when we are dealing with many nodes (or degrees of freedom). Second, we accelerate the solving process compared to similar FEM methods based on automatic differentiation.

The question arises about how accurate and fast the new solver is. Answering all the details and making a fair comparison between FOL and other numerical methods is a tedious task since many factors, from hardware to software platforms, play a role. Nevertheless, we intend to open up the discussion on this matter with the following example. To demonstrate the performance of the method in this context, we examine a nonlinear version of the heat equation where the heat conductivity coefficient is assumed to be temperature-dependent according to
\begin{align}
k(T) = m_1 + \beta~T^{m_2}. 
\label{eq:Q_T}
\end{align}
Here, $m_1=2$ and $m_2=4$ are selected and the above term is inserted in Eq.~\ref{StrongfromThermal}. Our motivation is mainly for academic study, and we are not investigating any particular physical system and more investigations with greater nonlinearity should be conducted in future studies.

Utilizing unstructured meshes is suitable for discretizing complex geometries. To demonstrate capability of FOL in such cases, in the following example, we perform the analysis on 3D complex geometries. 

\begin{figure}[H] 
  \centering
  \includegraphics[width=0.9\linewidth]{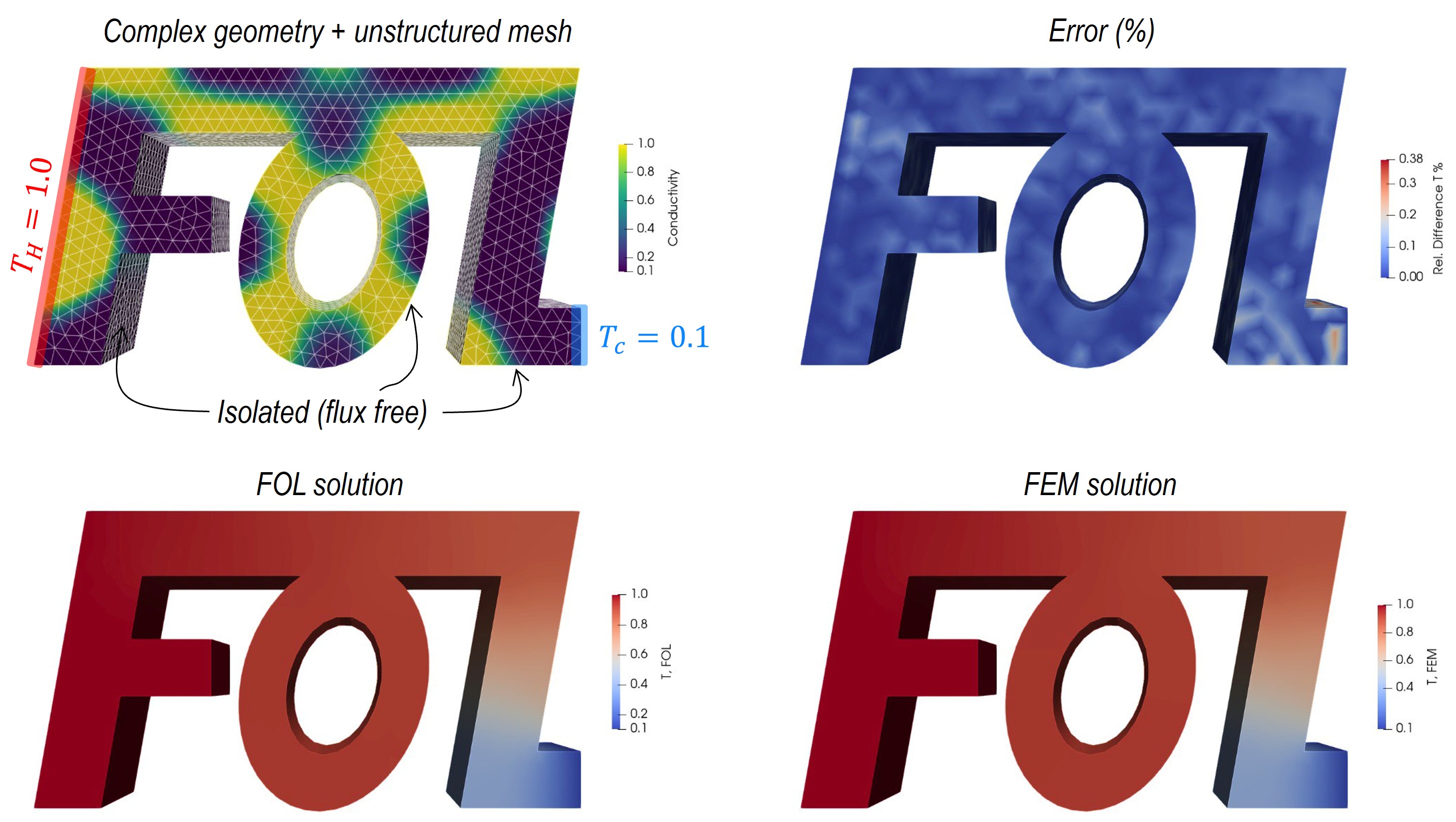}
  \caption{\textcolor{black}{Comparison of FOL and FEM for a nonlinear PDE (thermal diffusion) in a complex geometry}.}
  \label{fig:3D}
\end{figure}

In the first example, the far left and right surfaces are subjected to Dirichlet boundary conditions with fixed temperatures of $1.0$ and $0.1$, respectively. All other surfaces, including the curved and straight ones, are flux-free or isolated. The domain of interest is also heterogeneous, with the initial conductivity field varying spatially as shown in Fig.~\ref{fig:3D}. The performance of the matrix-free solver FOL is extraordinary, showing a maximum pointwise error of $0.4$ \%.

For the chosen mesh, we have 10 inputs representing the contributions of different frequency numbers, exactly as before. However, we now have 5485 neurons in the output layer, which represent all the degrees of freedom (i.e., discretized solution space). For this training, we used only one layer with one neuron since we are not interested in the parametric solution or sensitivities. This choice is sufficient to solve all the unknowns. Furthermore, the system is trained for 1000 epochs with a learning rate of $0.001$ and swish function. 

\begin{figure}[H] 
  \centering
  \includegraphics[width=0.9\linewidth]{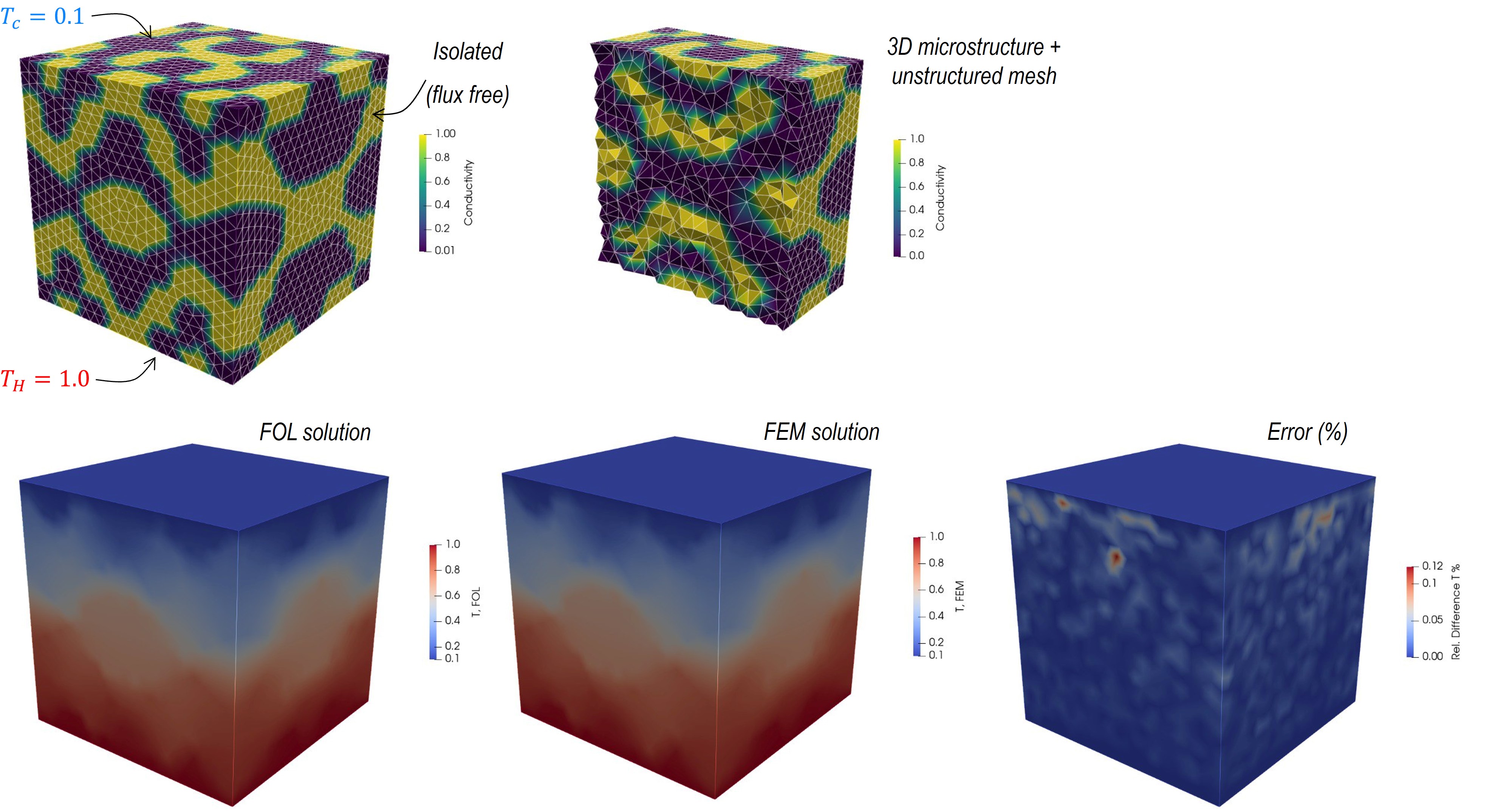}
  \caption{ \textcolor{black}{Comparison of FOL and FEM for a nonlinear PDE (thermal diffusion) in a heterogeneous domain}.}
  \label{fig:3D_RVE}
\end{figure}

In the second example which is shown in Fig.~\ref{fig:3D_RVE}, the top and bottom surfaces are subjected to Dirichlet boundary conditions with fixed temperatures of $0.1$ and $1.0$, respectively. All other surfaces are flux-free. 
Again, the performance of the matrix-free solver FOL is very acceptable.
For the second 3D example, we have 10 inputs as before and 6807 neurons in the output layer. The training hyperparameters remain as before. 

Finally, we examine the mechanical problem in 3D with a rather complex geometry and unstructured mesh. The geometry, boundary conditions, and mesh types are shown in Fig.~\ref{fig:3D_RVE}. 
\textcolor{black}{The same network setup is used, and the results for the deformation vector magnitude and two strain components from the two methods agree well enough. 
We observe that the maximum deformation error coincides with the regions of largest displacement, even though these areas are not subjected to external loading due to structural effects. A similar trend is observed for the strain components, where higher errors appear in critical regions of the structure. These observations suggest that future work should explore alternative optimizers and network architectures to further reduce error levels. }

\begin{figure}[H] 
  \centering
  \includegraphics[width=0.99\linewidth]{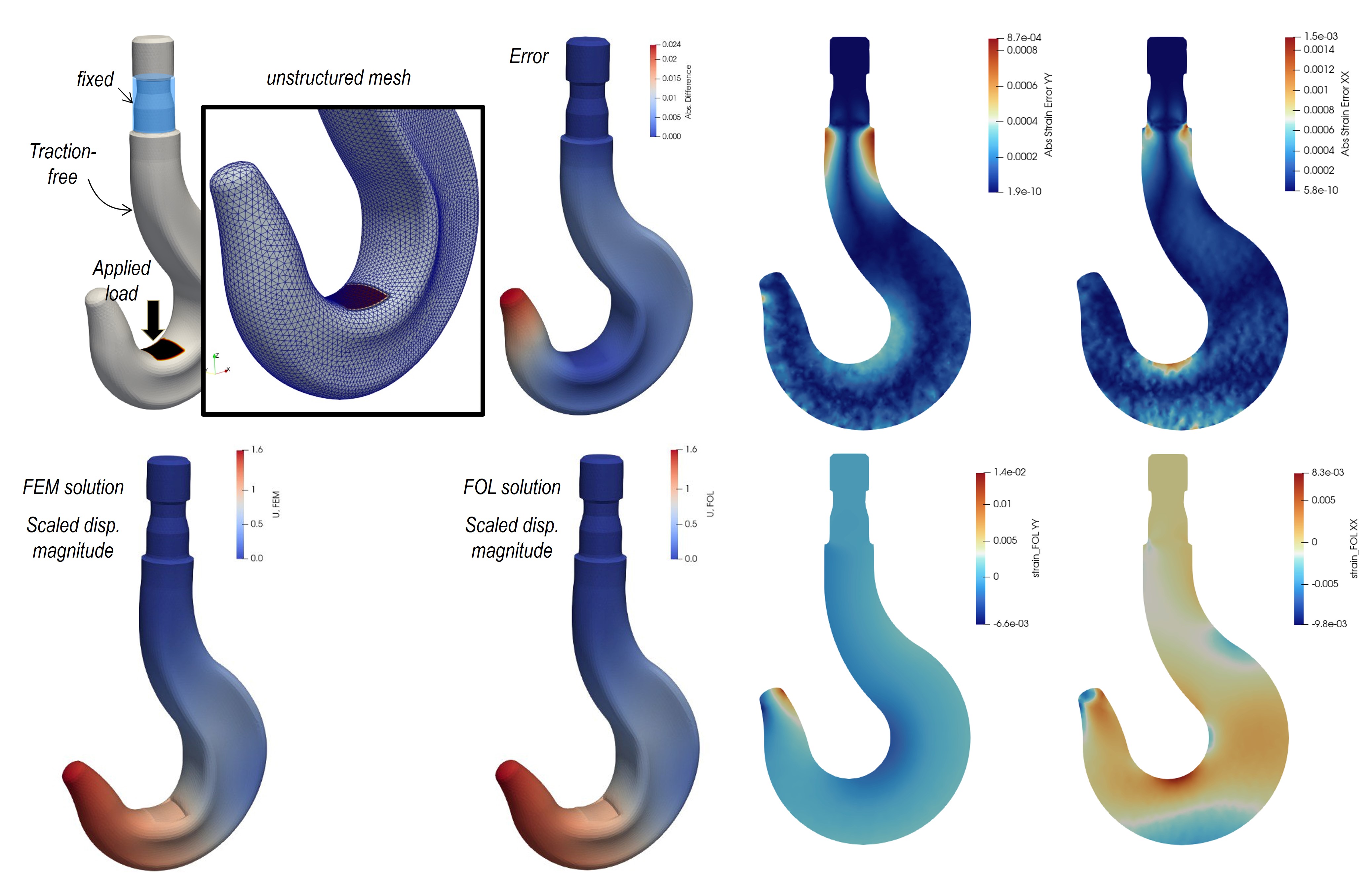}
  \caption{ \textcolor{black}{Comparison of FOL and FEM for a complex geometry and mechanical problem in 3D.} }
  \label{fig:hook}
\end{figure}

\color{black}
The above mentioned examples for optimization and matrix-free solver opens up more opportunities for future reserach. In other words, although the provided examples are intentionaly selected to cover some important aspects such as complex unstructured 3D mesh and geometry, but they are by no means complete and much further detailed studies are needed to examine different aspect of FOL as a new DL-based nonlinear solver in computational mechanics. 
Further examples and explanations, including non-elliptic PDEs and multiphysics problems, will be the subject of future studies. 
\color{black}

\section{Conclusion and outlooks}
This work showcases a simple-to-use, data-free, physics-based neural operator technique for parametric learning of equations across arbitrarily shaped domains. The idea is to take any discretized parametric design space as an input layer and set the desired solution at a finite number of sensor points as the output layer. Next, the output and input will be used in the given set of governing physical equations while the boundary terms can be satisfied in a hard manner. Based on the proposed architecture, the current method can be coupled with any available numerical schemes.
For the current work, we take advantage of the finite element method and use the discretized version of the weak form or energy form of the problem. By using shape functions and their spatial derivatives, we avoid the need for automatic differentiation in constructing the physical loss functions, which enhances training efficiency.

The proposed methodology also ensures accurate solution-to-design sensitivities (output-to-input) by incorporating the stationarity of the residuals into the loss function (also known as Sobolev training). As a result, one can directly use the method to perform tasks in gradient-based optimization without any additional cost for adjoint-based sensitivity analysis.
In particular cases, when optimization and parametric learning are not of interest, the method reduces to a matrix-free PDE solver, which requires much less memory space for calculations.

We applied our methodology to a steady-state thermal diffusion problem within a strongly heterogeneous domain (i.e., two-phase microstructure). Our findings demonstrate that even when considering a relatively high phase contrast, the method accurately predicts the temperature and flux profile. Moreover, thanks to the Sobolev training and the additional physical loss term for sensitivities, we performed a gradient-based topology optimization where an optimal design for the microstructure topology is discovered under given objective and constraint functions. By comparing calculation times and optimal solutions, we concluded that the FOL method can outperform classical adjoint-based techniques for gradient-based optimization.

A comprehensive comparison among various available numerical methods is not straightforward due to several reasons. Firstly, some of the newer methods, particularly those based on deep learning, are still under development. Secondly, depending on the application, dimensionality, computational resources, and implementation complexities, certain methods may prove more beneficial than others. Setting these aspects aside, a basic comparison between FEM, PINN, and FOL is presented in Table \ref{tab:compare}.

The main advantages of the current work are as follows: First, training the network in a data-free and parametric manner. Second, sensitivities are correctly captured based on Sobolev training. Third, the method can be utilized for fast forward prediction and inverse design, even when considering significant jumps in the solution field.

\begin{table}[H]
\fontsize{7}{10}\selectfont
\centering
\caption{ \textcolor{black}{Comparison of FOL against PINN (and its other extensions such as DEM) as well as FEM.} }
\label{tab:compare} \begin{adjustbox}{center}
\begin{tabular}{ l l l l }
\hline
    &  FEM  & \textcolor{black}{PINN (and its other extensions such as DEM)} &  FOL  \\
\hline
Implementing/Transferring codes  &  Hard (software specific)  & Easy (only network evaluation) & Easy (only network evaluation)  \\
Training time  &   - (does not apply)  & Moderate-high (based on AD) & Moderate (based on shape functions) \\
Parametric learning  &  No  & Yes (upon using extra inputs or branch net) & Yes \\
Order of derivations  &  Low (weak form)  & High (unless using DEM or Mixed-PINN) & Low (weak form)  \\
Handling complex geometries  &  Yes  & Yes & Yes \\
Handling strong heterogeneity  &  Yes  & Yes, (see Mixed PINN, XPINN, cPINN, hp-VPINNs) & Yes \\
Data requirement  &  Only model param.  & Only model param. & Only model param.  \\ 
&  & no additional solution is required & no additional solution is required \\
\hline
Main limitation &  Comput. cost + discretization error  & Training cost + \textcolor{black}{problem with sharp sol.} &  discretization error  \\
Main advantage  &   Accurate solutions for various PDEs  & Fast eval. + no data is required & \textcolor{black}{Fast and accurate eval. + data-free}  \\
 & - & - & \textcolor{black}{parametric learning + complex geom.} \\
 & - & - & \textcolor{black}{handles sharp solution} \\
\hline
\end{tabular}
\end{adjustbox}{}
\end{table}

\textcolor{black}{In this work, through a simple example, we compared the performance of several available neural operators when combined with the proposed FOL framework. Although FNO has known limitations in handling complex or highly irregular geometries, its FOL-based formulation performs very well on the simple 2D square domain considered here. Likewise, both the MLP-based model and DeepONet demonstrate strong performance under the same setting. Nevertheless, further investigations on more complex geometries and higher-dimensional problems are required to fully assess their capabilities and limitations.}
The FOL framework also opens up new horizons for computational physics in various fields (e.g. multiphysics problems \cite{Harandi2023}). 
\textcolor{black}{We emphasize that throughout this work we extensively employed Fourier-based parametrization due to its strong capability to reduce the design input space while still representing complex topologies. 
While the Fourier-based parameterization used here allows the material field to be evaluated on arbitrary geometries and unstructured meshes through FEM-based coordinate sampling, its expressiveness may become limited for highly non-periodic or localized topological features. Future work will therefore explore replacing or augmenting the Fourier representation with learned latent spaces obtained from autoencoders or similar architectures. Such parameterizations can provide more compact, geometry-aware, and application-specific representations, improving flexibility while retaining compatibility with the proposed operator-learning framework.}
\textcolor{black}{Even when free parametrization is combined with downsampling through a U-Net–based autoencoder, the FOL framework can be applied to obtain low-resolution evaluations, which can subsequently be translated back into high-resolution solutions \cite{NajafiKoopas2025}.} 

It is worth mentioning that one can easily represent the input layer based on other desired parameter spaces (e.g., fiber-reinforced composites or Voronoi tessellation based parameterization for polycrystalline materials \cite{Rezaei2025npj}). The proposed approach can also be applied to transient problems by employing FOL in an autoregressive manner \cite{yamazaki2024}.
Moreover, one is not limited to the FEM method to construct the loss terms in the proposed method and any other numerical schemes, especially spectral solvers, can also be utilized for faster speedup in multiscale computational systems \cite{HARANDI2025106219}.

\textcolor{black}{Finally, we should note that the potential benefits of the proposed framework become even more significant in nonlinear settings. In general, the computational burden of solving PDEs grows rapidly with increasing nonlinearity and a large number of degrees of freedom. Since the FOL approach learns a parametric mapping of the solution space, rather than solving a new nonlinear system at every query, the resulting surrogate model can substantially reduce evaluation costs after training. Consequently, for highly nonlinear forward or optimization problems, the relative speedup compared to conventional FEM solvers is expected to increase, making such applications a particularly promising direction for future work.}

\newpage
\noindent
\textbf{Data Availability}:
\textcolor{black}{The data supporting the findings of this study are openly available and can be accessed via the following link: \href{https://github.com/RezaNajian/folax}{FOLAX}.}
\\ \\ 
\textbf{Acknowledgements}:
The authors would like to thank the Deutsche Forschungsgemeinschaft (DFG) for the funding support provided to develop the present work in the project Cluster of Excellence “Internet of Production” (project: 390621612). 
\\ \\ 
\textbf{Author Statement}:
S.R.: Conceptualization, Supervision, Software, Writing - Review \& Editing. R.N.A.: Conceptualization, Methodology, Software, Writing - Review \& Editing. K. T.: Investigation, Software. A. M.: Software, Review \& Editing. M. K.: Supervision, Review \& Editing. M. A.: Funding, Supervision, Review \& Editing.
\\ \\ \\
\appendix
\section{Derivation of FEM residuals} 
\label{app:A}
In Section \ref{sec:governing_equation}, we derived the weak form of the steady-state diffusion equation. Utilizing the standard finite element method, the temperature field $T$, conductivity field $k$ as well as their first spatial derivatives, are approximated as
\begin{equation}
T= \sum {N}_i T_i =\boldsymbol{N}^T_e \boldsymbol T_e, \quad \nabla T =\sum {B}_i T_i =\boldsymbol{B}_e\boldsymbol T_e, \quad 
k= \sum {N}_i k_i =\boldsymbol{N}^T_e \boldsymbol k_e. 
\end{equation}
Here, $T_i$ and $k_i$ are the nodal values of the temperature and conductivity field of node $i$ of element $e$, respectively. Moreover, matrices $\boldsymbol N_e$ and $\boldsymbol B_e$ store the elemental shape functions and their spatial derivatives. For a quadrilateral 2D element, they are written as
\begin{equation}
\boldsymbol N^T_e=\left[
\begin{matrix}
N_1    &\cdots & N_4 \\
\end{matrix}
\right],\quad 
\boldsymbol B_e=\left[
\begin{matrix}
N_{1,x} &\cdots & N_{4,x}  \\
N_{1,y} &\cdots & N_{4,y}\\
\end{matrix}
\right],
\end{equation}

\begin{equation}
\boldsymbol T^T_e=\left[
\begin{matrix}
T_1    &\cdots & T_4 \\
\end{matrix}
\right],\quad 
\boldsymbol k^T_e=\left[
\begin{matrix}
k_1    &\cdots & k_4
\end{matrix}
\right].
\end{equation}

The notation $N_{i,x}$, and $N_{i,y}$ represent the derivatives of the shape function $N_i$ with respect to the coordinates $x$ and $y$, respectively. To compute these derivatives, we utilize the Jacobian matrix $\boldsymbol J = \partial \boldsymbol X / \partial \boldsymbol \xi$, a standard procedure in any finite element subroutine \cite{bathe, hughes}. Here, $\boldsymbol X = [x, y]$ and $\boldsymbol \xi = [\xi, \eta]$ represent the physical and parent coordinate systems, respectively. Finally, one can derive the so-called \textit{discrete residual} form of the weak formulation in Eq.~\ref{eq:weakformthermal} as
\begin{align}
\label{eq:dis_residual}
\boldsymbol r = \sum\int_{\Omega_e} [\boldsymbol B_{e}]^T k~[\boldsymbol B_{e} \boldsymbol T_e] ~dV - \int_{\Omega_t}[\boldsymbol  N_{e}]^T k~[\boldsymbol B_{e} \boldsymbol T_e]~\boldsymbol  n~dS -\int_{\Omega_{e}} [\boldsymbol  N_{e}]^T  Q~dV.
\end{align}

The above formulation is shown for the 2D case, which is primarily reported in the results of this work. 
Extending this to 3D is straightforward.


\newpage
\section{Details on sample preparation}
\label{sec:app_sample}
By providing the network with a sufficiently diverse set of inputs, the performance of the deep learning model will improve. Here, we will briefly discuss our strategies to explore each design space. 

\subsection{Free parameter design for learning based on material property distribution}

The material parameters listed in Table~\ref{tab:par} are selected for calculation in section \ref{sec:mat_prop}. 
Note that in all the calculation, the temperature values remain within the range of $0$ to $1$. However, alternative parameters and boundary values can be chosen, and the model variables can be normalized to ensure they fall within the range of $-1$ to $1$. 
\begin{table}[H]
\centering
\caption{Model input parameters for the thermal problem discussed in section~\ref{sec:mat_prop}.}  
\label{tab:par}
\begin{footnotesize}
\begin{tabular}{ l l }
\hline
      & value / unit    \\
\hline
Heat conductivity phase 1  ($k_{\text{mat}}$)  & $1.0$~W/mK  \\
Heat conductivity Phase 2 ($k_{\text{inc}}$)  & $0.01$~W/mK  \\
Heat source/sink ($Q$)  & $0.0$~W/m$^3$  \\
Applied temperature ($T_L$,$T_R$)  & ($1.0$~K, $0.0$~K) or ($1.0$~K, $0.1$~K)\\
\hline
\end{tabular}
\end{footnotesize}
\end{table} 

\begin{figure}[H] 
  \centering
  \includegraphics[width=0.99\linewidth]{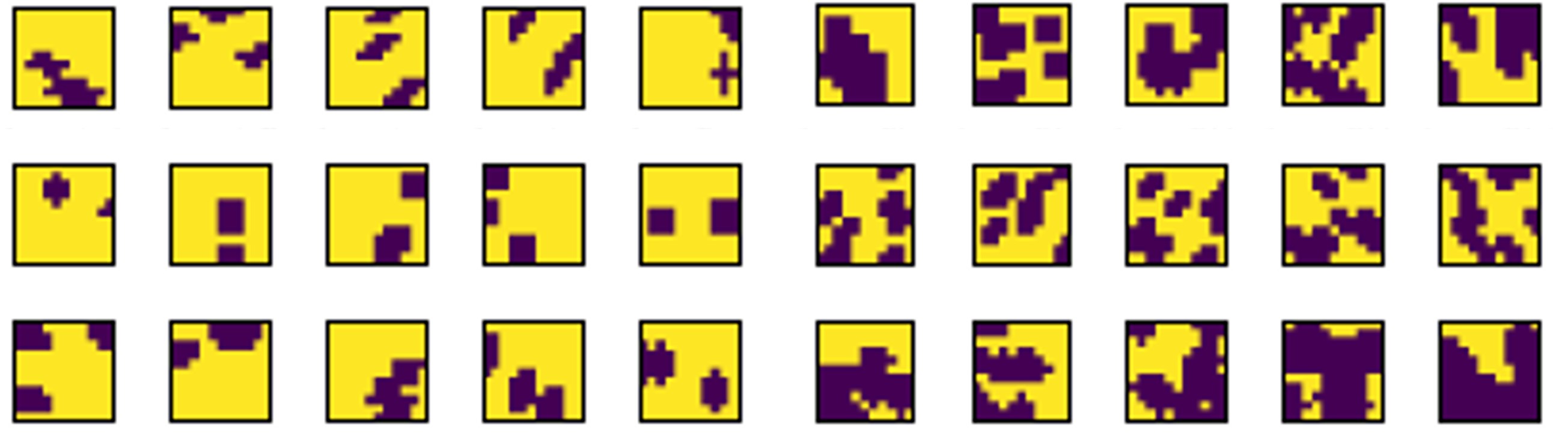}
  \caption{Examples showcasing created morphologies for a two-phase microstructure used in section~\ref{sec:mat_prop}.}
  \label{fig:samples}
\end{figure}

\begin{figure}[H] 
  \centering
  \includegraphics[width=0.99\linewidth]{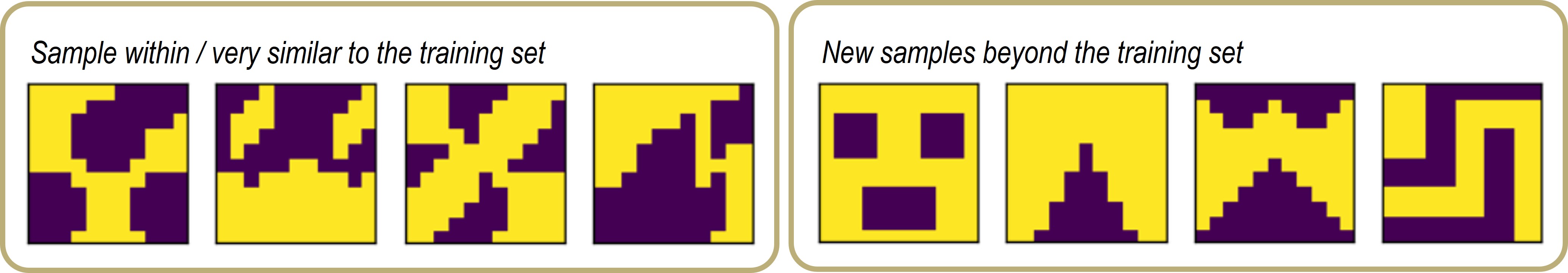}
  \caption{Unseen samples used for testing the neural network after training. The results of this study is discussed in section \ref{sec:mat_prop}. }
  \label{fig:test}
\end{figure}

\begin{figure}[H] 
  \centering
  \includegraphics[width=0.99\linewidth]{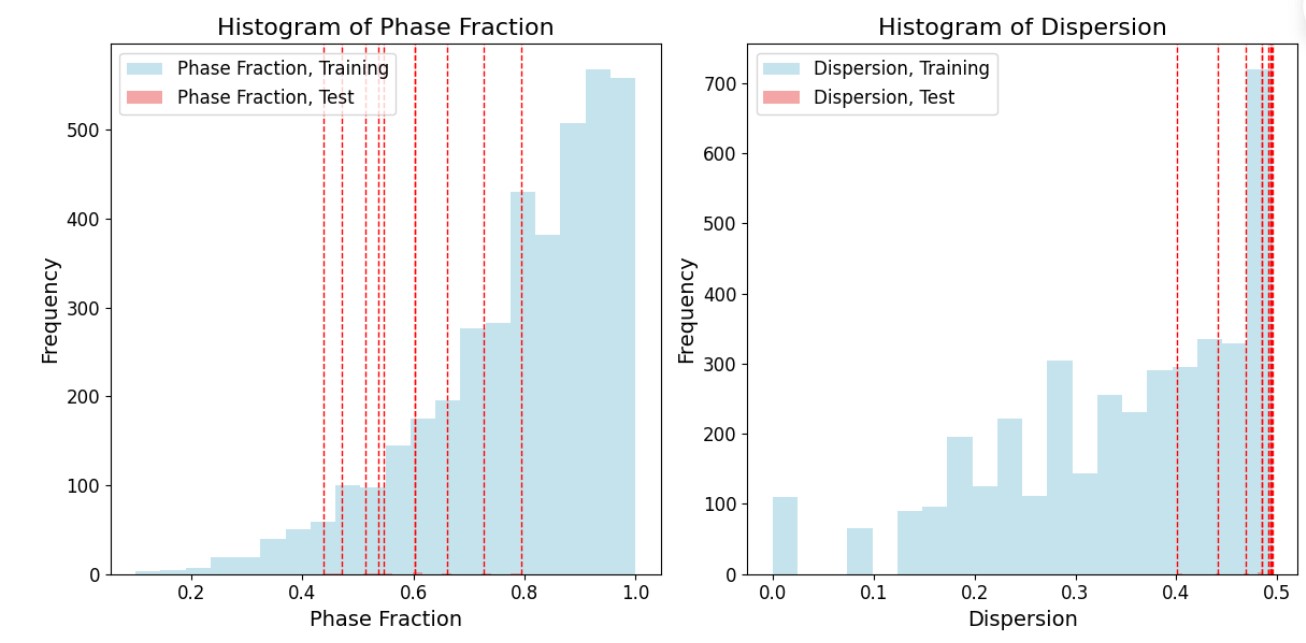}
  \caption{Histogram of the phase fraction as well as dispersion values of the created samples for the studies in section \ref{sec:mat_prop}.  }
  \label{fig:hist}
\end{figure}

Based on the description in Section 2.2.1, we generate $4000$ samples. 
A subset of the generated samples is presented in Fig.~\ref{fig:samples}. Additionally, some samples are chosen for testing (see Fig.~\ref{fig:test}). These test cases are entirely novel and some are deliberately selected for their symmetrical aspects, ensuring they are beyond the scope of the training dataset. The corresponding histograms for volume fraction and dispersion value are depicted in Fig.~\ref{fig:hist}. 

\subsection{Fourier-based parametrization for learning based on material property distribution }
\begin{figure}[H] 
  \centering
  \includegraphics[width=0.99\linewidth]{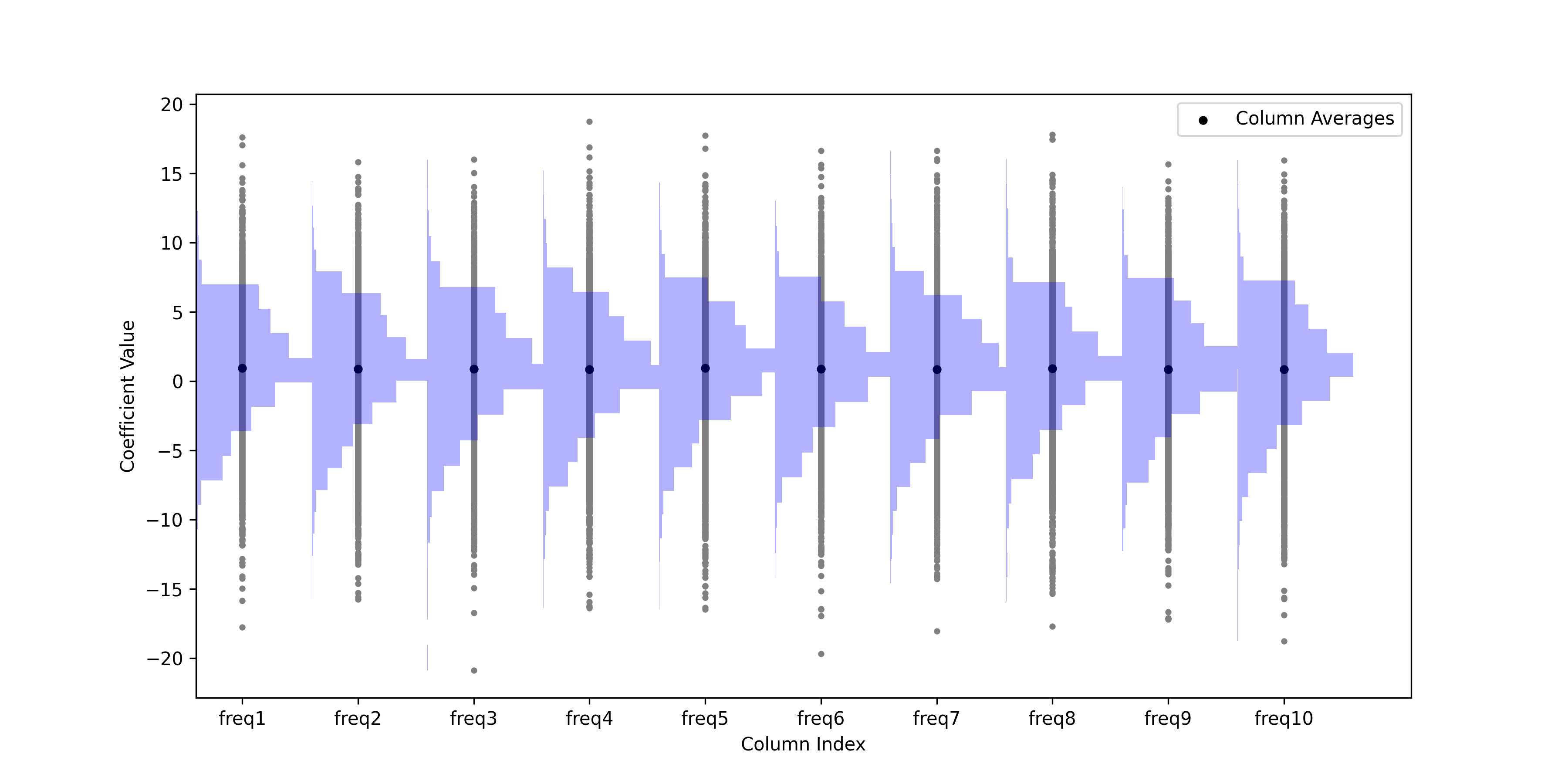}
  \caption{ Distribution of the 8000 random values for components of each frequency term. These are used in section~\ref{sec:mat_prop}. The parameter $\text{freq}_i$ represents the $i$-th coefficient corresponding to the $i$-th frequency. }
  \label{fig:coeffs_matrix}
\end{figure}

\begin{figure}[H] 
  \centering
  \includegraphics[width=0.99\linewidth]{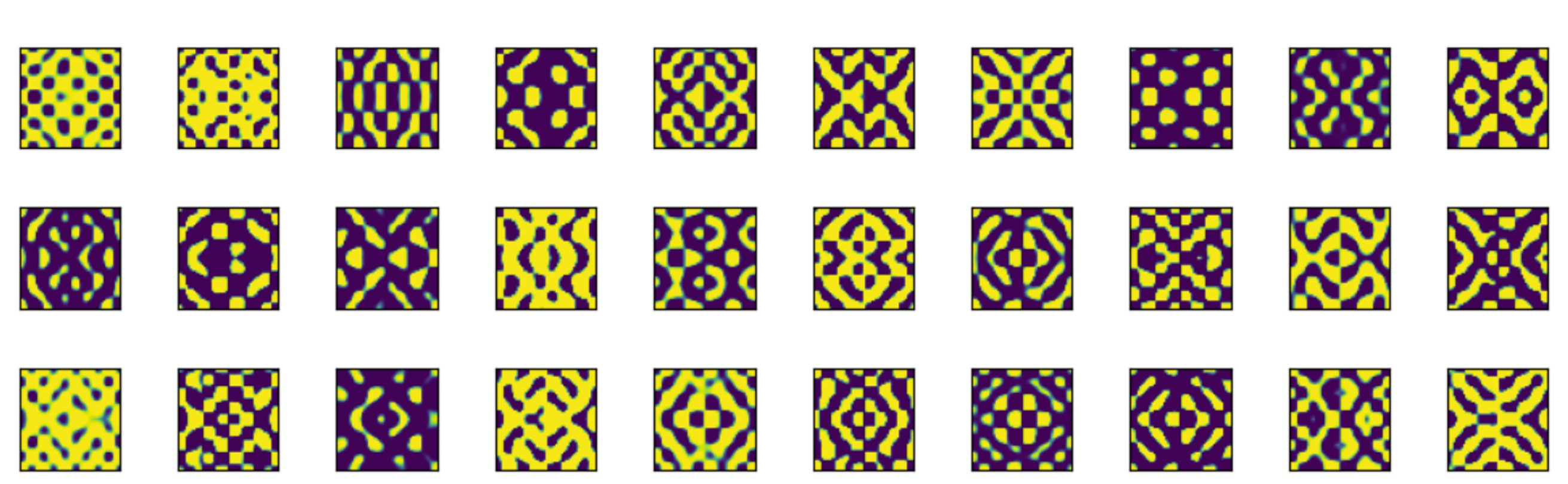}
  \caption{Examples showcasing created morphologies for a two-phase microstructure used in section~\ref{sec:mat_prop}.}
  \label{fig:samples_ff}
\end{figure}

The input samples used for training are generated by combining different random distributions for ten inputs. A summary of all the $8000$ selected samples is shown in Fig.~\ref{fig:coeffs_matrix}. See Fig.~\ref{fig:samples_ff} for a few examples of samples obtained by the Fourier-based parameterization.

\begin{figure}[H] 
  \centering
  \includegraphics[width=0.7\linewidth]{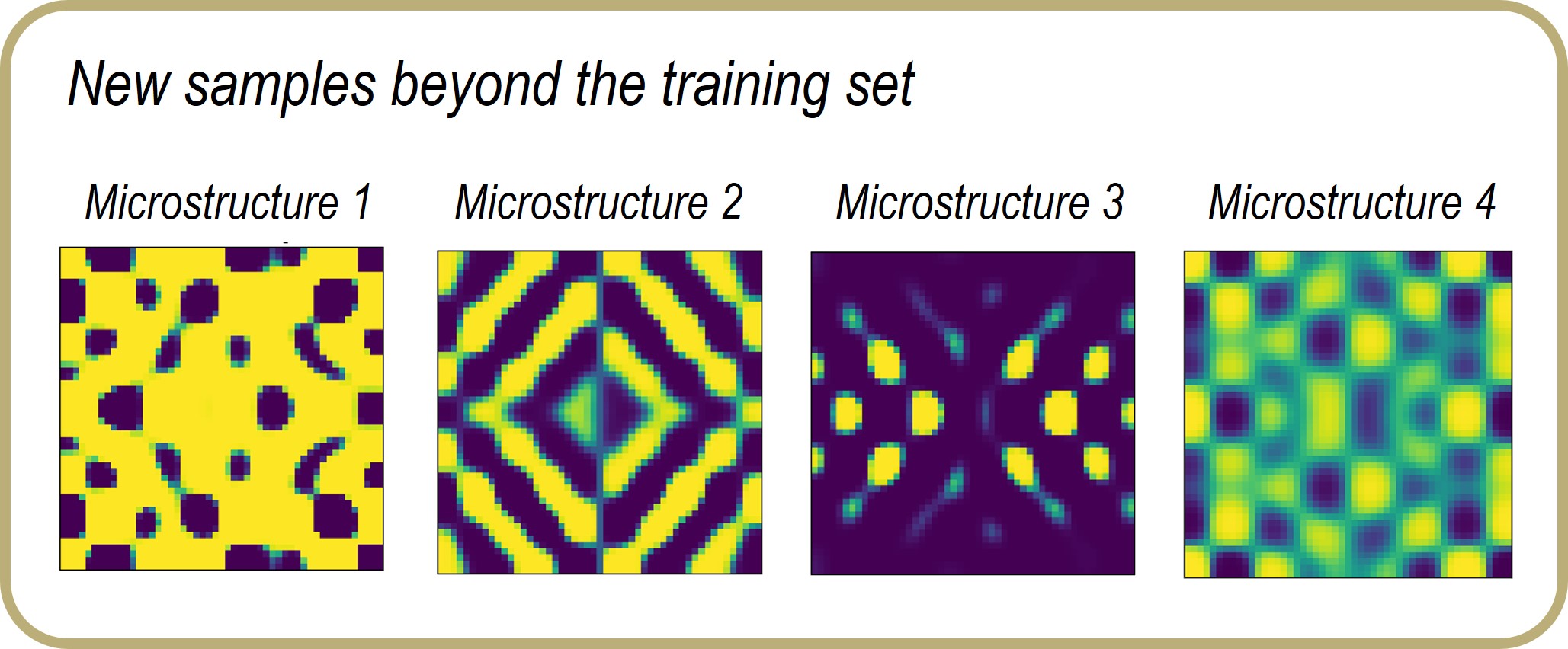}
  \caption{Unseen samples used in section~\ref{sec:mat_prop} for testing the neural network after training. }
  \label{fig:test_f}
\end{figure}

A few unseen test cases were also selected to check the performance of the trained network beyond the training set. In some test cases, the matrix material is highly conductive, while in others, the matrix acts as an insulating material with highly conductive inclusions. The distribution of these frequencies is randomly generated and deliberately selected to be uncommon in the training set (i.e., $5<\text{freq}_i<10$). This allows us to test the performance of the pre-trained parametric FOL.
For a given image of the microstructure, using the Fourier-based parameterization, one must first perform an inverse Fourier analysis to find the corresponding coefficients.
See Table \ref{Table:Fourier_coeff} for the chosen amplitudes of the test cases and Fig.~\ref{fig:test_f} for the corresponding conductivity maps.

\begin{table}[H]
  \centering
  \caption{Selected test cases and corresponding Fourier coefficients used in section~\ref{sec:mat_prop}.}
   \begin{tabular}{lll} 
      \hline
          Study type &   Value ($\text{freq}_i, i=1 \cdots 10$)\\
      \hline
       Microstructure 1 (Figs.~\ref{fig:66_plot_mesh_vec_data}~\ref{fig:66_plot_mesh_vec_grad_data}~\ref{fig:66_Qx_i2_sens_comparisons}) & $\left[5.3, 6.0, 7.7, 5.1, 5.1, 6.8, 5.5, 8.3, 8.1, 7.5 \right]$\\
       Microstructure 2 (Fig.~\ref{fig:73_plot_mesh_vec_data})  & $\left[0.7, -0.5, -0.0, 0.3, 0.9, 1.6, -0.2, 0.9, -0.3, -1.3 \right]$\\
       Microstructure 3 (Fig.~\ref{fig:28_plot_mesh_vec_data}) & $\left[-1.7, 0.7, -0.8, 0.6, 0.3, 0.5, -0.8, -0.9, 1.8, -0.6 \right]$\\
       Microstructure 4 (Fig.~\ref{fig:mesh})   & $\left[-3.6, 0.8, 0.5, 2.0, 3.8, 0.0, -0.8, 2.6, 0.3, -0.3 \right]$\\
      \hline
      \end{tabular}
      \label{Table:Fourier_coeff}
  \end{table}

\newpage
\color{black}
\subsection{Fourier-based parametrization for learning based on source term distribution }
For parametric learning over the source term with constant Dirichlet boundary conditions on all four edges, we utilized the same strategy with slightly different values for frequencies to investigate further non-periodic inputs. 
The material parameters for this study are listed in Table~\ref{tab:par_Q}.
\begin{table}[H]
  \centering
  \caption{\textcolor{black}{Model input parameters for the thermal problem discussed in section~\ref{sec:source_prop}.} }  
  \label{tab:par_Q}
  \begin{footnotesize}
  \begin{tabular}{ l l }
  \hline
        & value / unit    \\
  \hline
  Heat conductivity  ($k$)  & $1.0$~W/mK  \\
  Heat source/sink, max value ($Q_{max}$)  & $-1.0$~W/m$^3$  \\
  Heat source/sink, max value ($Q_{min}$)  & $-10.0$~W/m$^3$  \\
  Applied temperature ($T_b$)  & $1.0$~K \\
  \hline
  \end{tabular}
  \end{footnotesize}
\end{table} 

\textcolor{black}{For the sake of briefness, we avoid extra visualization and similar explanations for the above study.
Similar to the previous case, we use sigmoidal projection to generate the heat-sink profiles according to
\begin{align}
\label{eq:sigmoid_Q}
\begin{aligned}
    Q(x,y) &= (Q_{max}-Q_{min})\cdot\text{Sigmoid}\left(\beta(Q_f-0.5)\right) + Q_{min}, \\
    Q_f(x,y) &= \sum_i^{n_{sum}} [  D_i \cos{(f_{x,i}~x)} \cos({f_{y,i}~y)}].
\end{aligned}
\end{align}
See also explanations after Eq.~\ref{eq:foruier_eq} and results in section~\ref{sec:source_prop}. } 

\subsection{Samples for learning based on boundary conditions }
\begin{figure}[H] 
  \centering
  \includegraphics[width=0.99\linewidth]{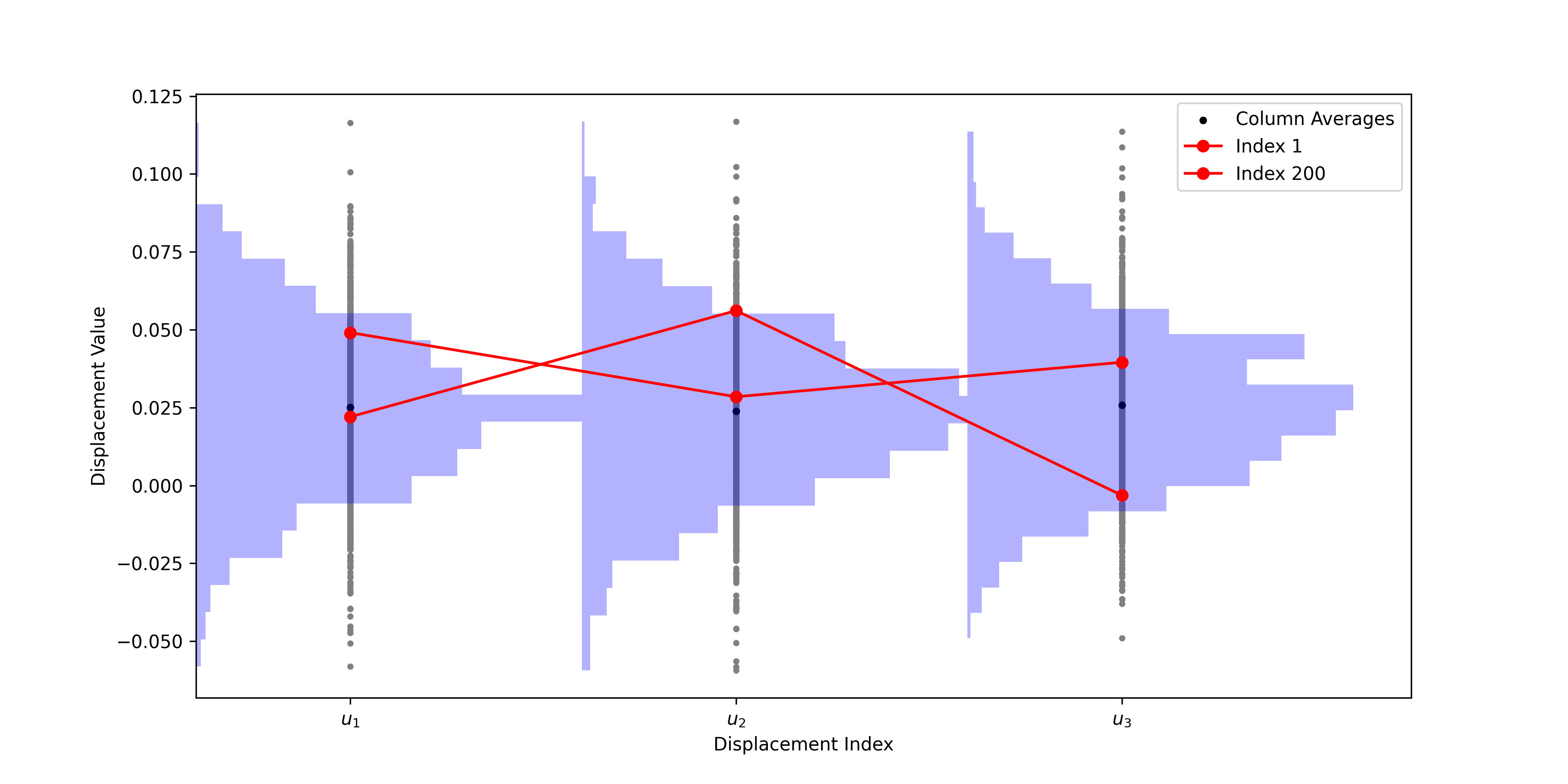}
  \caption{ \textcolor{black}{Distribution of the 1000 random values for components of the displacement vector. These are used in Section~\ref{sec:BCs_prop}.} }
  \label{fig:coeffs_matrix_bc}
\end{figure}

For the results presented in Section~\ref{sec:BCs_prop}, we varied the components of the applied displacement vector on the top surface according to normalized random distributions. 
About $1000$ samples were created and used for the training, and their distributions are shown in Figs.~\ref{fig:coeffs_matrix_bc}.

\color{black}

\color{black}
\newpage
\section{Further details on networks hyperparameters}
\label{sec:NN_det}
In what follows, we provide the details of the selected hyperparameters for the different network architectures investigated in this work, as part of the comparison between various operator-learning approaches combined with the FOL framework. These hyperparameters were determined through exploratory tuning and yield highly promising results; however, we do not claim them to be fully optimal.
\begin{table}[h!]
  \centering
  \begin{tabular}{l l}
  \hline
  \textbf{Hyperparameter} & \textbf{Value} \\
  
  \hline
  \textbf{Branch network} & \\
  \quad Input size & 52 $\times$ 52\\
  \quad Hidden layers & [128, 128, 128, 128] \\
  \quad Output size & 128 \\
  \quad Activation & swish \\
  \hline
  \textbf{Trunk network} & \\
  \quad Input size & 3 \\
  \quad Hidden layers & [128, 128, 128, 128] \\
  \quad Output size & 128 \\
  \quad Activation & swish \\
  \hline
  \textbf{DeepONet} & \\
  \quad Activation function name & swish \\
  \quad batch size & 100\\
  \quad bias & 0. \\
  \hline
  \textbf{Optimizer} & \\
  \quad Type & Adam\\
  \quad learning rate & $1 \times 10^{-4}$ \\
  \hline
  Number of epochs & 10000 \\
  \hline
  \end{tabular}
  \caption{DeepONet AD-based and FOL-based Hyperparameters}
  \end{table}

  \begin{table}[h!]
  \centering
  \begin{tabular}{l l}
  \hline
  \textbf{Hyperparameter} & \textbf{Value} \\
  \hline
  \textbf{FNO Model} & \\
  \quad Number of nodes in the first dimension & 12 \\
  \quad Number of nodes in the second dimension & 12 \\
  \quad Number of channels to which the input is lifted & 32 \\
  \quad Number of fourier stages & 4 \\
  \quad last projection dimension & 128 \\
  \quad Output scale factor & 0.001 \\
  \quad Activatio function & GELU\\
  \quad Grids & 52 $\times$ 52 \\
  \quad batch size & 100\\
  \quad Number of epochs & 1000 \\
  \hline
  \textbf{Optimizer} & \\
  \quad Type & Adam \\
  \quad Learning rate & $1 \times 10^{-3}$ \\
  \hline
  \end{tabular}
  \caption{Fourier Neural Operator (FOL-FNO) Hyperparameters}
  \end{table}

  \begin{table}[h!]
  \centering
  \begin{tabular}{l l}
  \hline
  \textbf{Hyperparameter} & \textbf{Value} \\
  \hline
  \textbf{FOL-MLP Model} & \\
  \quad Input size & 10 \\
  \quad Hidden layers & [300, 300] \\
  \quad Activation function & SIREN \\
  \quad Prediction gain & 5 \\
  \quad Output size & 52 $\times$ 52 \\
  \hline
  \textbf{Optimizer} & \\
  \quad Type & Adam \\
  \quad Learning rate & $1 \times 10^{-4}$ \\
  \hline
  \end{tabular}
  \caption{FOL-MLP Hyperparameters}
  \end{table}

\begin{figure}[H] 
  \centering
  \includegraphics[width=0.99\linewidth]{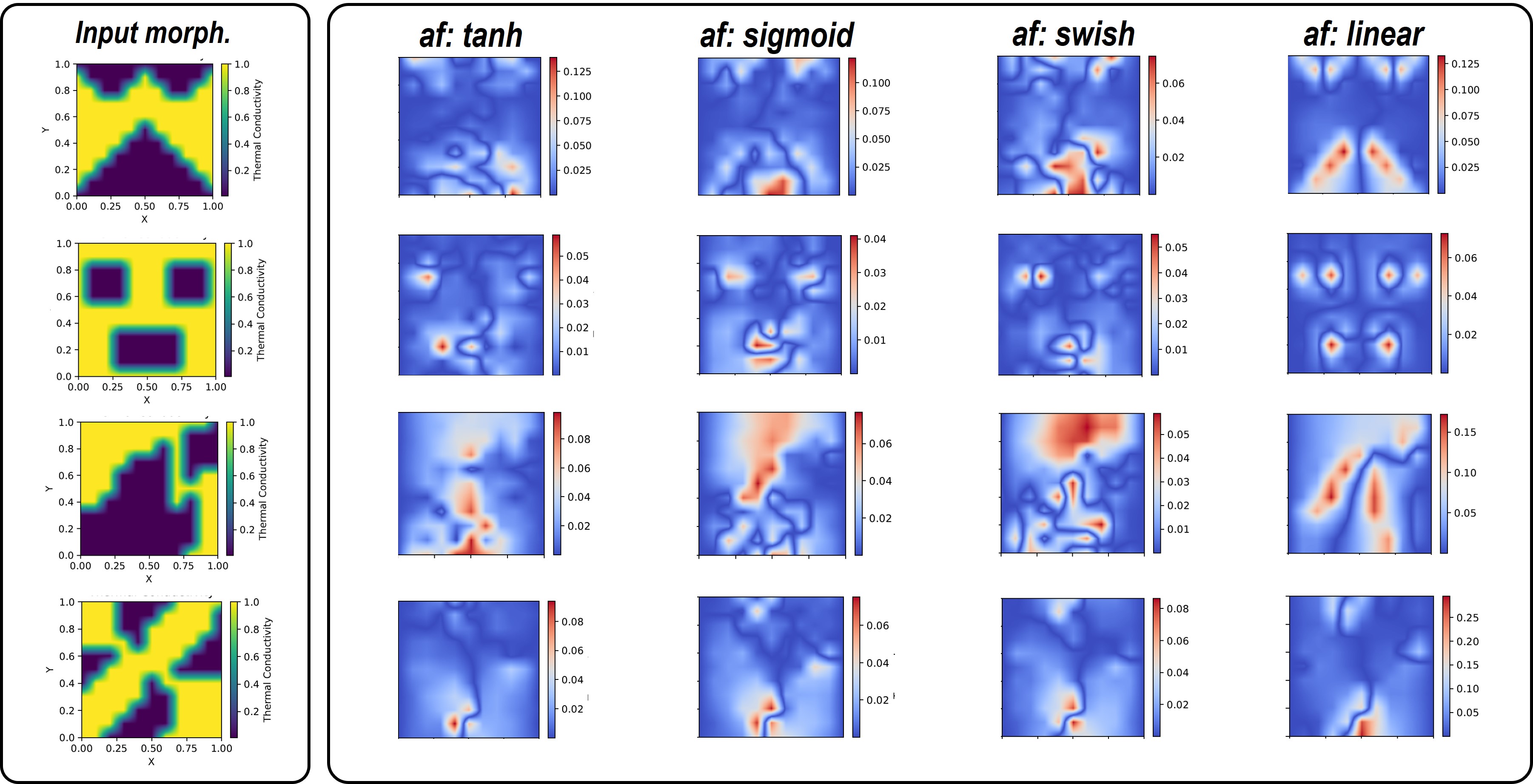}
  \caption{Influence of the type of activation function on the obtained results.}
  \label{fig:af}
\end{figure}

In Fig.~\ref{fig:af}, we look into the influence of various choices for the activation function on the predictions. To avoid repetition, only the pointwise error for each choice of the activation function is shown in each row and column of Fig.~\ref{fig:af}. Except for the linear case, which is an oversimplified case, the errors using other types of activation functions remain low.

\begin{figure}[H] 
  \centering
  \includegraphics[width=0.8\linewidth]{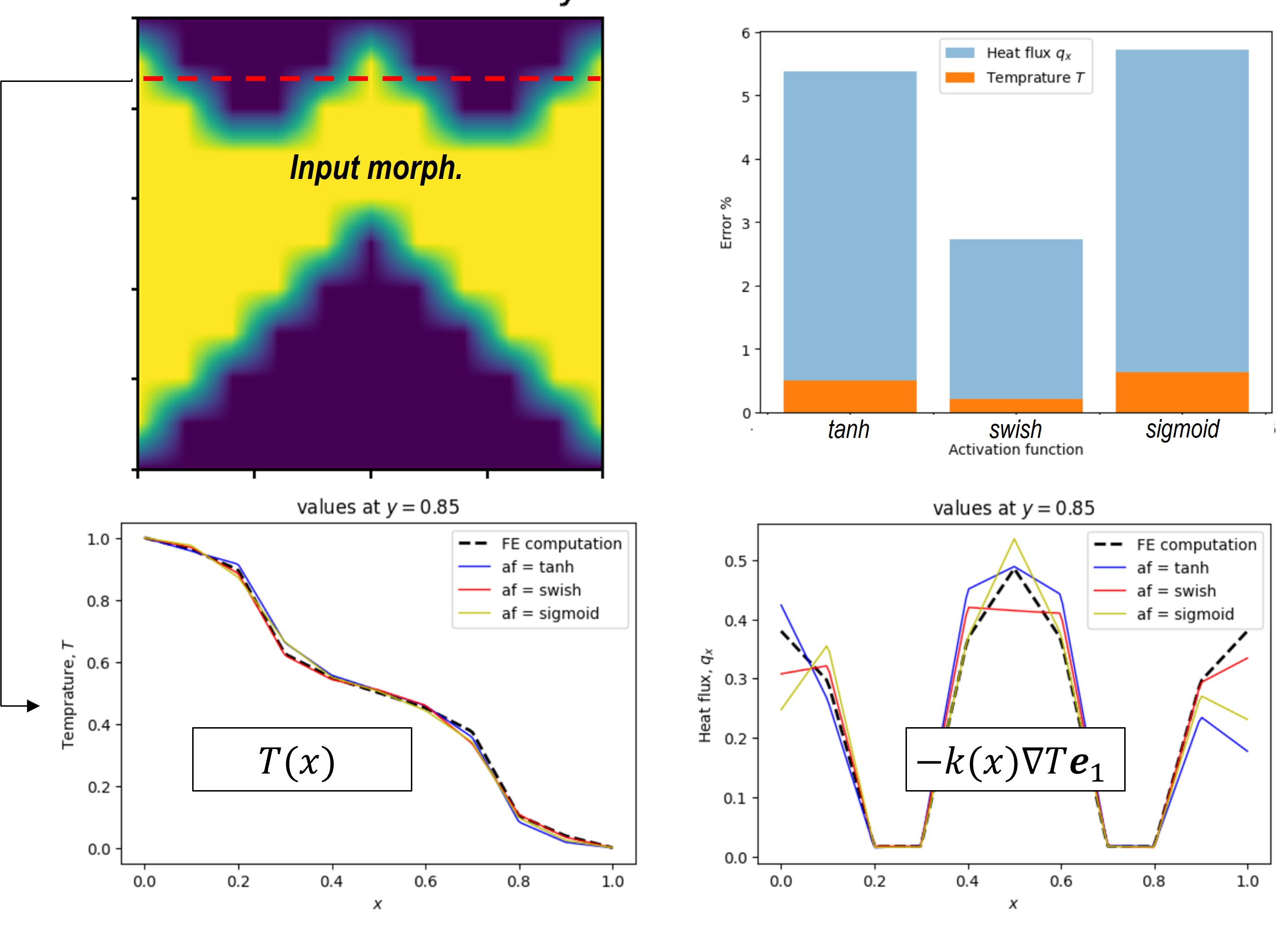}
  \caption{ Evaluation of averaged errors for different choice of activation functions. }
  \label{fig:err}
\end{figure}

In total, the tanh and swish functions showed the best performance. For a better and more quantitative comparison, the cross-section of the predictions and averaged errors are reported in Fig.~\ref{fig:err}. To avoid repetition, the error values in this case are calculated based on $Err = \frac{| \langle T_{NN} \rangle - \langle T_{FE} \rangle|}{| \langle T_{FE} \rangle|}\times 100$, where we have $\langle T_{NN} \rangle = \frac{1}{N^2}\Sigma_i T_{NN}(i)$. This is motivated by the idea of homogenization, where we average the obtained temperature and heat flux to obtain the overall material properties. The lower this homogenized error is, the higher the accuracy will be for the homogenization of properties

Interestingly, the errors for the homogenized temperature are below $1\%$, which also holds for many other test cases not reported here. On the other hand, the average error for the flux in x-direction is higher, which is reasonable since we only predict the temperature profile and any slight fluctuations in the predicted temperature influence its spatial derivatives. The errors for the heat flux using the swish option are still acceptable and remain between $2\%$ to $3\%$. One should note that ideas from mixed formulations, where we predict the heat flux as an output, can also be integrated with the current formulation to enhance the predictive ability of the neural network for higher-order gradients of the solutions (see investigations in \cite{REZAEI2022PINN, Faroughi2023}).

\color{black}

\bibliography{Ref}

\end{document}